\documentclass[aos]{imsart}

\RequirePackage{amsthm,amsmath,amsfonts,amssymb}
\RequirePackage[authoryear]{natbib}
\RequirePackage[colorlinks,citecolor=blue,urlcolor=blue]{hyperref}
\RequirePackage{graphicx}
\usepackage{bm}
\usepackage{multirow}
\usepackage{booktabs}
\usepackage{algorithm}
\usepackage{algorithmic}
\usepackage{xcolor}

\startlocaldefs
\theoremstyle{plain}
\newtheorem{coro}{Corollary}[section]
\newtheorem{theorem}{Theorem}[section]

\newtheorem{lemma}[theorem]{Lemma}

\theoremstyle{remark}
\newtheorem{assump}{Assumption}[section]
\newtheorem{defi}{Definition}

\newtheorem{remark}{Remark}
\newtheorem*{assump*}{Assumption*}

\def\bx{\bm{x}}

\def\bI{\mathbb{I}}

\def\bS{\mathbb{S}}
\def\E{\mathbb{E}}

\def\mA{\mathcal{A}}
\def\CB{\mathcal{B}}
\def\CC{\mathcal{C}}

\def\CH{\mathcal{H}}

\def\CS{\mathcal{S}}
\def\CM{\mathcal{M}}
\def\CT{\mathcal{T}}

\def\s{{\bf s}}
\def\bs{\bm{s}}
\def\bu{\bm{u}}

\def\bxi{\bm{\xi}}
\def\cov{\mbox{cov}}
\def\var{\mbox{var}}
\def\BSigma{\pmb \Sigma}
\def\BTheta{\pmb \Theta}
\def\BK{\mbox{\pmb{K}}}
\def\BA{\mbox{\pmb{A}}}
\def\BB{{\pmb{B}}}
\def\BP{{\pmb{P}}}
\def\trans{^{\rm T}}

\def\Bdelta{{\pmb{\delta}}}
\def\BDelta{{\pmb{\Delta}}}
\def\BE{{\pmb{\varepsilon}}}
\def\Beta{{\pmb{\eta}}}
\def\Bdelta{{\pmb{\delta}}}
\def\BX{{\pmb{X}}}
\def\BD{{\pmb{D}}}

\def\CX{{\mathcal{X}}}
\def\Bmu{{\pmb{\mu}}}
\DeclareMathOperator*{\argmax}{arg\,max}

\newcommand{\blue}[1]{\textcolor{blue}{#1}}

\endlocaldefs

\begin{document}

\begin{frontmatter}
\title{Change-Point Detection and Support Recovery for Spatially Indexed Functional Data}
\runtitle{Spatial Functional Change Detection}
\begin{aug}
\author[A]{\fnms{Fengyi}~\snm{Song}\ead[label=e1]{sauntbai@gmail.com}},
\author[A]{\fnms{Decai}~\snm{Liang}\ead[label=e2]{liangdecai@nankai.edu.cn}}
\and
\author[A]{\fnms{Changliang}~\snm{Zou}\ead[label=e3]{nk.chlzou@gmail.com}}
\address[A]{School of Statistics and Data Science, Nankai University \printead[presep={,\ }]{e1,e2,e3}}
\end{aug}

\begin{abstract}
Large volumes of spatiotemporal data, characterized by high spatial and temporal variability, may experience structural changes over time. Unlike traditional change-point problems, each sequence in this context consists of function-valued curves observed at multiple spatial locations, with typically only a small subset of locations affected. This paper addresses two key issues: detecting the global change-point and identifying the spatial support set, within a unified framework tailored to spatially indexed functional data. By leveraging a weakly separable cross-covariance structure---an extension beyond the restrictive assumption of space-time separability---we incorporate functional principal component analysis into the change-detection methodology, while preserving common temporal features across locations. A kernel-based test statistic is further developed to integrate spatial clustering pattern into the detection process, and its local variant, combined with the estimated change-point, is employed to identify the subset of locations contributing to the mean shifts. To control the false discovery rate in multiple testing, we introduce a functional symmetrized data aggregation approach that does not rely on pointwise $p$-values and effectively pools spatial information. We establish the asymptotic validity of the proposed change detection and support recovery method under mild regularity conditions. The efficacy of our approach is demonstrated through simulations, with its practical usefulness illustrated in an application to China's precipitation data.
\end{abstract}


\end{frontmatter}

\section{Introduction}\label{sec: intro}
\subsection{Motivation and Background}
With the rapid advancement of technology, large-scale spatiotemporal data at high frequencies are increasingly available in many fields, such as ecology, epidemiology, and environmental science \citep[e.g.,][]{cressie2015statistics, liu2017functional, yan2018real}.
Motivated by an empirical analysis of China's precipitation data in Section \ref{sec: real data}, we investigate the change detection of annual patterns over a spatial region, followed by the support recovery of locations where these changes occur.
Suppose that for each year $i$, we observe precipitation data $X_i(\s_j;t)$ across times $t\in \CT$ at spatial locations $\s_j \in \CS$.
Our first concern is the following change-point model:
\begin{equation*}\label{cp model}
    X_{i}(\s_j;t) =
    \begin{cases}
\mu_0^* (\s_j;t)+ \varepsilon_i(\s_j;t), & \text { for } i=1, \ldots, \tau^*; \\
\mu_1^* (\s_j;t)+ \varepsilon_i(\s_j;t), & \text { for } i=\tau^*+1, \ldots,n,
\end{cases}
\end{equation*}
where $\tau^*$ is an unknown change point, $\mu_0^*(\cdot;\cdot)$ and $\mu_1^*(\cdot;\cdot)$ represent the spatiotemporal mean functions before and after the change, and $\varepsilon_i(\cdot;\cdot)$ is a mean-zero random field with some spatiotemporal covariance structure.
After detecting a change-point, we then focus on identifying the subset of spatial locations that contribute to the change, as the number of affected sites is typically small relative to the entire domain.
Therefore, the second goal is to identify the non-zero support set
\begin{equation*}\label{sp model}
    \left\{ \s_j\in \CS: \E \{ X_{\tau^*+1}(\s_j;\cdot)-X_{\tau^*}(\s_j;\cdot) \} \not= 0 \right\},
\end{equation*}
while controlling the associated error rate.
In particular, it is anticipated that incorporating spatial structures, such as clustering or correlation patterns, into the models could improve the efficacy of signal detection and the interpretability of scientific findings.

Functional data analysis (FDA) has received growing attention due to its effectiveness in feature extraction and dimension reduction for data observed or measured over a continuous domain. Modern scientific applications increasingly involve functional data as a basic measurement structure across diverse settings, giving rise to the so-called {\em second-generation functional data} \citep{wang2016functional, koner2023second} and the corresponding challenges in statistical inference.
Taking the observed dataset in our study as an example, for each year $i$, the measurement $X_i(\s_j;\cdot)$ at location $\s_j$ is treated as a functional curve across $t\in\CT$, and the annual collection over a set of locations $\s_j\in \CS$ forms replicated spatial functional curves, commonly referred to as {\em spatially indexed functional data}.
Change-point analysis for functional data has been extensively studied in these years, with various applications for channel profile monitoring \citep{paynabar2016change}, temperature shift detection \citep{berkes2009detecting}, and mortality rates analysis \citep{li2024detection} etc.
Most existing methodologies have been developed within the scope of univariate or dependent functional data \citep{hormann2010weakly, aston2012detecting, aue2018detecting}, with classical techniques---such as the cumulative sum (CUSUM) procedure---being widely adopted in combination with FDA tools \citep{zhang2011testing, li2024detection}.
For spatially indexed functional data, \cite{gromenko2017detection} proposed a change detection approach for the mean function of the spatiotemporal process under the assumption of space-time separability, a well-known concept in conventional spatial statistics \citep[e.g.][]{cressie2015statistics}.
This assumption allows the spatial and temporal covariance structure to be estimated separately, and the functional principal component analysis (FPCA) can then be exploited to depict the temporal pattern.

Our second objective---the support recovery for massive spatial components--falls within the framework of spatial multiple testing, with relevant applications involving the change identification of environmental or climate fields \citep{sun2015false}, functional neuroimaging \citep{brown2014incorporating}, and disease mapping \citep{gao2023spatial}, etc. 
For instance, the study of weather risk attribution forecast typically covers a large number of events, namely the systematic study, instead of the targeted study that examines one single event \citep{risser2019spatially}, leading to simultaneous analyses on a list of events.
The false discovery rate \citep[FDR;][]{benjamini1995controlling}, defined as the expected fraction of false discoveries among all discoveries, has become a popular and practical criterion for large-scale hypothesis testing. While the classical Benjamini-Hochberg (BH) method has been shown to be valid for controlling the FDR under various dependence conditions, \citep[e.g.][]{benjamini2001control}, a variety of methods have been proposed to improve power by leveraging specific characteristics of the data at hand, such as spatial clusters or local sparsity \citep{perone2004false, cao2015changepoint, yun2022detection}.  
Recently \cite{cai2022laws} proposed an adaptive weighting $p$-value approach by incorporating the local sparsity structure of spatial signals, which provides a convenient and suitable tool for a wide range of spatial multiple testing problems.

\subsection{Challenges and Connections to Existing Works}
The analysis of cross-covariance structures plays a critical role in multivariate or spatial functional data, particularly in the formulation of an appropriate FPCA procedure \citep{paynabar2016change, zhang2021unified}.  
For change-point detection, similar approaches have been adopted by \cite{berkes2009detecting} for univariate functional data and generalized by \cite{gromenko2017detection} for spatially indexed functional data based on the space-time separability, which decomposes the full covariance into a product of a purely spatial covariance and a purely temporal covariance.
Though heavily used in spatial statistics, the separability assumption has been shown to be unrealistic for many real applications through hypothesis testing procedures \citep[e.g.][]{aston2017tests}.
As a result, it remains both challenging and valuable to bring up more flexible and reasonable covariance structures that can serve as a valid foundation for subsequent inference tasks, such as dimension reduction and change detection.

The second challenge involves effectively integrating the spatial clustering pattern into the detection approach for massive datasets.
In real-world applications, it is generally expected  
that a location and its adjacent neighbors are more likely to belong to similar regions, whether or not they experience changes, with nearby locations being more heavily affected than those that are spatially distant.
\cite{gromenko2017detection} proposed averaging variability across diverse locations using a spatial weighting function, but this method does not account for the clustering pattern among spatial neighborhoods.  
In other fields, such as large-scale sequential surveillance, research has demonstrated that aggregating local information and utilizing spatial clustering can improve the efficiency of anomaly detection \citep{yan2018real, ren2022large}. 
When multiple testing is performed on spatial signals, incorporating the relationships between individual locations and spatial clusters into the inferential process is also commonly recommended to increase statistical power \citep[e.g.][]{benjamini2007false, sun2015false}.
Although not directly applicable to our dataset, where observations at each spatial location are function-valued objectives, these insights suggest the potential for developing a more powerful procedure for detecting spatial structures of interest.

Spatial support recovery presents another challenge not typically encountered in conventional spatiotemporal statistics. While global change-point tests for entire random fields have been extensively investigated \citep[e.g.][]{li2016comparison}, it is often more appealing if one can further identify the specific subsets where changes occur.
This necessitates a finite approximation strategy at the location level for inference within a continuous spatial domain,  inevitably leading to issues of multiplicity or high dimensionality due to the large number of spatial sites.
To control the number of incorrect rejections at an acceptable level, the FDR criterion has been widely adopted and extended to spatial signals, with many existing methods relying on a two-component mixture prior distribution for individual $p$-values.
For example, \cite{sun2015false} proposed an oracle procedure that utilizes hyperparameter estimation to represent the posterior probability of this model, while \cite{yun2022detection} employed an EM algorithm to fit the density of $p$-values based on temporally stationary spatiotemporal fields. However, in the spatiotemporal context of interest, the assumption of such a pre-specified model for $p$-values may be questionable.
Furthermore, in comparison to global change-detection procedures, 
accurately deriving the asymptotic properties of local test statistics that incorporate clustering information poses significant challenges, rendering a $p$-value-based approach less feasible. Alternatively, one may consider some $p$-value-free approaches, such as the knockoff filter \citep{barber2015controlling} and the symmetrized data aggregation (SDA) approach \citep{du2021false, chen2021data}, 
which, to our knowledge, have not yet been adapted to the spatial multiple testing problem under consideration.

\subsection{Our Contributions}
In this article, we adopt a global-to-local perspective for spatially indexed functional data, addressing both change-point detection and spatial multiple testing within a unified framework.
\begin{itemize}
\item
We leverage a weakly separable structure for spatiotemporal modeling, which offers a more flexible and empirically realistic cross-covariance configuration compared to the conventional separability assumption \citep[e.g.][]{gromenko2017detection}.
Building on this structure, we develop an efficient spatial FPCA procedure via a differencing covariance estimator, which preserves the structural change information and serves as the foundation for the subsequent inference tasks. 
\item
For global change detection, we construct an asymptotically valid hypothesis testing procedure via a CUSUM procedure integrated with the above FPCA method.
The change-point is then consistently estimated through a novel kernel aggregation strategy that enhances detection capability by exploiting spatial clustering patterns.
\item 
To identify the specific locations with mean shifts, we propose a functional SDA approach based on a variant of the kernel-based statistics, ensuring valid FDR control and power improvement in multiple testing.
In particular, we derived uniform convergence rates for the quantities based on the FPCA and change-point estimators, providing a cornerstone for establishing the symmetry property of our ranking statistics.
\item
Our method is model-free in several aspects.
First, the proposed change detection procedure avoids strong structural assumptions---such as stationarity or Gaussianity---by leveraging replicated spatiotemporal observations.
Second, the SDA-based multiple testing rule does not rely on the conventional two-component mixture model for location-wise $p$-values \citep[e.g.][]{cai2022laws}, resulting in a leaner framework for FDR control.
Furthermore, the established theoretical guarantees hold broadly across diverse spatial sampling schemes, enhancing the method’s practical applicability.
\end{itemize}

To the best of our knowledge, this work is the first to integrate global change detection and local multiple testing within a coherent framework, 
providing a more appealing and comprehensive approach to spatiotemporal change-point modeling. 
Simulation studies demonstrate the superiority of the proposed detection procedure over existing ones in terms of finite-sample performance, and the application to  China's precipitation data further illustrates the practical utility and interpretability of our methods.

\subsection{Organization}
The rest of this article is organized as follows.
In Section \ref{sec: model}, we present the foundational change-point model for spatially indexed functional curves under the weakly separable structure.
Based on the proposed FPCA estimators in Section \ref{sec: FPCA}, we develop a global change-point detection procedure, accompanied by an asymptotic investigation on the test statistics, in Section \ref{sec: cp detect}. 
This is followed by an SDA-based support recovery procedure, along with a justification of the FDR validity in Section \ref{sec: sup recov}.
We illustrate the empirical performance of the proposed methods through a simulation study in Section \ref{sec: simulation}, and a real example using China's precipitation data in Section \ref{sec: real data}.
Additional technical details and numerical results are provided in the Appendix.

\subsection{Notations}
We let $\lambda_{\min }(\BA)$ and $\lambda_{\max }(\BA)$ denote the smallest and largest eigenvalues of a square matrix $\BA$, and denote $\BA(j,k)$ as the $(j,k)$-th element of $\BA$.
For a set $\CM$, let $| \CM |$ be its cardinality. 
For a vector $\bx = (x_1,\dots,x_p)\trans$, let $\|\bx \|$ be the $l_2$ norm and $\| \bx \|_\infty = \max_{j=1,\dots,p}|x_j|$ represent the $l_{\infty}$ norm. 
Denote $\left\langle \cdot, \cdot \right\rangle_p$ as the inner product between two $p$-dim vector.
For any two functions $f(t)$ and $g(t) \in L^2(\CT)$, where $L^2(\CT)$ denotes the space of square-integrable functions on $\CT$, we denote the inner product $\langle f(\cdot),g(\cdot) \rangle_{\CT}=\int_\CT f(t)g(t) dt$ and the corresponding norm $\|f(\cdot)\|_\CT=\langle f(\cdot),f(\cdot) \rangle_\CT^{1/2}$.

\section{The basic spatiotemporal change-point model and FPCA approach}\label{sec: model}
Let $X_i(\s_j;t)$ be the functional observation across $t\in \CT$ taken at the spatial location $\s_j\in\CS$ on day or year $i$, 
where $\CS$ and $\CT$ represent the spatial and time domains, respectively.
In this article, we treat the temporal records across $\CT$ as fully observed functional data, while the spatial observations are available at discrete locations $\{\s_{j}\}_{j=1}^{p}$.
Moreover, $X_i(\s_j;t)$ is observed from the latent random process $X_i(\bs;t)$.
\begin{assump}\label{assump: L2}
     The process $X_i(\bs;t)$ is a spatiotemporal random field in $L^2(\CS \times \CT)$, i.e. $\E \int_{\CS\times \CT} X_i^2(\bs;t) d\bs dt < \infty$.
\end{assump}
Under Assumption \ref{assump: L2}, for each $i=1,\ldots,n$, $X_i(\bs;t)$ can be expressed as 
\begin{equation*}\label{model: st L2}
X_i(\bs;t)=\mu_i(\bs;t)+\varepsilon_i(\bs; t), \quad i=1,\ldots,n,
\end{equation*}
where $\mu_i(\bs;t)$ is a deterministic mean function, and $\varepsilon_i(\bs; t)$ is a zero-mean random effect.

\begin{assump}\label{assump: mean and error}
Each $\mu_i(\bs;t)$ is a smooth function on $\CS \times \CT$.
Moreover, $\{\varepsilon_i(\bs;t)\}_{i=1}^n$ are i.i.d spatiotemporal random fields with continuous covariance functions defined as  $C(\bs,t;\bs',t')=\E \left\{\varepsilon_i(\bs; t)\,\varepsilon_i(\bs'; t') \right\}$.
\end{assump}
Under Assumptions \ref{assump: L2} to \ref{assump: mean and error}, the covariance function $C(\cdot,\cdot;\cdot,\cdot)$ can also be viewed from the perspective of operators in Hilbert-Schmidt spaces; see Appendix \ref{sec: proof consis} for more details.
Considering that at some time $\tau^*$, a change has occurred in the mean function $\mu_i$, where we abbreviate $\mu_i(\bs;t)$ as $\mu_i$ for simplicity. 
Our detection problem involves two main steps.
The first step is formulated as the following hypothesis test:
\begin{equation*}\label{test: cp}
\mathcal{H}_0: \mu_1=\cdots=\mu_n \ \mbox{versus}\ \ 
\mathcal{H}_1: \mu_1=\cdots=\mu_{\tau^*}\neq \mu_{\tau^*+1}=\cdots=\mu_{n} \mbox{ for some }1\leq \tau^*<n,
\tag{P1}
\end{equation*}
 where the equalities are in the $L^2(\CS \times \CT)$ sense, followed by the estimation of the occurring time $\tau^*$. In this paper, we focus on the one-change-point problem, although additional change points can be further detected using the proposed methods as needed. 
Additionally, the problem in \eqref{test: cp} implies that we are testing for a change in the mean function, under the assumption that the covariance structures remain unchanged.
This approach is common in change-point detection, as allowing both the mean and variance to change can complicate the analysis, even for scalar variables
\citep{horvath2012inference}.

After detecting the global change-point, the next step is to identify the spatial locations where these changes occur. 
Specifically, let $\bS = \{\s_1,\dots,\s_p\}$ represent the set of discretely observed locations, and denote $\mA=\{\s_j\in\bS:\mu_{\tau^*}(\s_j;\cdot)\neq \mu_{\tau^*+1}(\s_j;\cdot)\}$ the subset of $\bS$ with mean changes at $\tau^*$,
{where the equalities are in the $L^2(\CT)$ sense.}
The focus here is on detecting this breaking set in space, as it is often believed that only a subset of spatial locations contributes to the changes, {meaning that $|\mathcal{A}|$ is small relative to $p$.}
One can formulate an analogous model for temporal support recovery if the data exhibit similar patterns in the time domain.
Define $\theta_j = \bI\{ \s_j \in \mA \}$, where $\bI$ is an indicator function and $\theta_j=0/1$ corresponds to a null/nonnull variable,
we consider the multiple hypothesis tests:
 \begin{equation*}\label{test: mcp}
     \mathcal{H}_{0}^j: \theta_j=0  \ \mbox{versus}\ \  \mathcal{H}_{1}^j: \theta_j=1, \mbox{ for } j=1,\dots,p.
     \tag{P2}
 \end{equation*}
Let $\delta_j$ be the binary test decision at location $\s_j$, where $\delta_j=1$ if $\mathcal{H}_{0}^j$ is rejected and $\delta_j=0$ otherwise. 
The false discovery proportion (FDP) and true discovery proportion (TDP) are defined as follows:
\begin{equation}\label{FDP TDP defi}
\mbox{FDP} =  \frac{\sum_{j=1}^p (1-\theta_j) \delta_j} {\max \{ \sum_{j=1}^p \delta_j,1 \} }, \quad \mbox{TDP} = \frac{\sum_{j=1}^p \theta_j\delta_j}{\max \{ \sum_{j=1}^p \theta_j,1 \}}.
\end{equation}
Our objective is to accurately recover the informative set $\mA$ while controlling $\mbox{FDR} = \E(\mbox{FDP})$, and simultaneously achieving a high level of power, defined as $\E(\mbox{TDP})$.

\subsection{Weakly separable covariance structure}\label{sec: weak sepa}
As the covariance of spatiotemporal processes may be quite intricate, we aim to develop the methodology under some flexible covariance structure; i.e. the following concept of weak separability \citep{liang2023test}.
\begin{defi}\label{def: WS}
   A spatiotemporal process $X_i(\bs;t)$ in $L^2(\CS\times\CT)$ is weakly separable if there exists some orthonormal basis $\{\psi_r(\cdot)\}_{r=1}^\infty$ in $L^2(\CT)$ such that
\begin{equation}\label{WS model}
X_i(\bs;t) = \mu_i(\bs,t)+ \sum_{r=1}^{\infty}\xi_{ir}(\bs)\psi_r(t)
\end{equation}
holds almost surely in $L^2(\CS \times \CT)$,
where 
$\{\xi_{ir}(\cdot)\}_{r=1}^\infty$ are mutually uncorrelated spatial processes in $L^2(\CS)$, i.e.  for any $r\not= r'$, $\cov\{\xi_{ir}(\bs),\xi_{ir'}(\bs')\}=0$ for any $\bs$ and $\bs'$ in $\CS$.
\end{defi}

The expansion \eqref{WS model} entails that $\xi_{ir}(\bs) = \langle \varepsilon_i(\bs;\cdot),\psi_r(\cdot) \rangle_\CT$ for any $\bs\in\CS$.
Moreover, it follows from Definition \ref{def: WS} that the covariance function 
of $X_i(\bs;t)$ can be decomposed as 
\begin{equation*}\label{WS cov}
C(\bs,t;\bs',t') = \sum_{r=1}^\infty \sigma_r(\bs,\bs') \psi_r(t)\psi_r(t'),
\end{equation*}
where $\sigma_r(\bs,\bs')=\E \{ \xi_{ir}(\bs)\xi_{ir}(\bs') \}$ is the $r$th eigenvalue of the {covariance function $C_{\bs,\bs'}(\cdot,\cdot)= C(\bs,\cdot;\bs',\cdot)$}, and $\psi_r(\cdot)$ is the corresponding eigenfunction.
Following the common perspective of functional principal component analysis (FPCA), the model \eqref{WS model} generalizes the Karhunen-Lo\`{e}ve (KL) expansion for spatial functional fields, therefore we also call $\{\psi_r(\cdot)\}_{r=1}^\infty$ and 
$\{\xi_{ir}(\cdot)\}_{r=1}^\infty$ 
as the functional principal components (FPCs) and FPC scores, respectively.  

It can be seen from the definition that, the proposed weak separability generalizes the conventional concept of separability (also referred to as {\em{strong separability}} in the sequel for a more clear comparison), and provides a flexible yet parsimonious representation for the spatially indexed functional data of concern. For each spatial location $\s_j$, the observed functional process {$X_i(\s_j;\cdot)$} can be projected onto a common basis $\{ \psi_r(\cdot)\}_{r=1}^\infty$,
resulting in a sequence of FPC scores $\{\xi_{ir}(\s_j)\}_{r=1}^\infty$.
Let $\bm{\xi}_{ir}= \left(\xi_{ir}(\s_1),\dots,\xi_{ir}(\s_p)\right)\trans$, the weak separability implies that provided with such a basis system, 
$\{\bm{\xi}_{ir}\}_{r=1}^\infty$ are mutually uncorrelated for any $r\ne r'$, i.e. $\E \{ \xi_{ir}(\s_j) \xi_{ir'}(\s_k) \}=0 $ for any $j,k=1,\dots p$. On the other hand, for each $r$, the $p$-variate vector $\bm{\xi}_{ir}$ may be correlated with some covariance structure 
\begin{align*}
    \sigma_r(\s_j,\s_k) = \E \{ \xi_{ir}(\s_j) \xi_{ir}(\s_k) \}, \quad j,k=1,\dots,p,
\end{align*}
and thus can be modeled using techniques in traditional spatial or multivariate statistics. This lays the foundation for the following FPCA procedure and the testing approaches in Sections \ref{sec: cp detect} and \ref{sec: sup recov}.

\subsection{Spatial FPCA under the change-point model}\label{sec: FPCA}
To develop the change detection procedure for spatial functional data, we first need to estimate the eigenvalues and eigenfunctions, which relies on the proper estimation of the covariance function.
Define $\BX_i(t)=(X_i(\s_1;t),\dots,X_i(\s_p;t))\trans$ and 
$\BE_i(t)=(\varepsilon_i(\s_1;t),\dots,\varepsilon_i(\s_p;t))\trans$,
both of which can be regarded as $p$-dim functional processes with some cross-covariance structures implied by $C(\cdot,\cdot;\cdot,\cdot)$.
Under the weakly separable model \eqref{WS model}, it can be observed that
\{$\psi_r(\cdot)$\} are, unique up to a sign, the eigenfunctions of the marginal covariance function
\begin{equation*}
    H(t,t') = \frac{1}{p} \, \E \left\{ \left\langle \BE_i(t),\BE_i(t')  \right\rangle_p \right\} =  \frac{1}{p}\sum_{j=1}^p C(\s_j,t;\s_j,t').
\end{equation*}
Moreover, the eigenbasis of $H(t,t')$ is shown to be optimal in the sense that it retains the largest amount of total variability among the $p$-dim vectors 
$ \langle \BX_i(\cdot),\tilde{\psi}_r(\cdot) \rangle_\CT  $
for any orthonormal basis $\{\tilde\psi_r(\cdot)\}_{r=1}^\infty$ \citep{zapata2021partial}.
This motivates us to estimate $\psi_r(\cdot)$ based on the eigendecomposition of an estimator of $ H(\cdot,\cdot)$.
For this purpose, a naive approach may be performed based on $\widetilde H(t,t')=p^{-1}\sum_{j=1}^p \widetilde  C_{jj}(t,t')$, where $\widetilde C_{jj}(t,t')=n^{-1}\sum_{i=1}^n \{ X_{i}(\s_j;t)-\bar X_j(t)\}\{ X_{i}(\s_j;t')-\bar X_j(t')\} $ is the sample covariance function of $X(\s_j;\cdot)$ and $\bar X_j(t)=n^{-1} \sum_{i=1}^n X_i(\s_j;t)$.
However, under the alternative hypothesis in \eqref{test: cp}, such an estimated covariance function may be inconsistent due to the bias in the sample mean estimates.
To address this issue, we consider the marginal covariance estimator via differencing \citep{paynabar2016change}, say
\begin{equation}\label{hatH}
 \widehat{H}(t,t')=\frac{1}{2 p (n-1)}\sum_{i=1}^{n-1} \left\langle \left\{ \BX_{i+1}(t)-\BX_{i}( t) \right\}, \left\{ \BX_{i+1}(t')-\BX_{i}(t') \right\} \right\rangle_p,
\end{equation}
following a similar spirit as the typical robust estimation for the covariance matrix.
We will show in the proof of Theorem \ref{thm: consistency of estimators} that $\widehat H(\cdot,\cdot)$ is a consistent estimator of $H(\cdot,\cdot)$ under both $\mathcal{H}_0$ and $\mathcal{H}_1$ in \eqref{test: cp}.

Based on the above marginal covariance estimators, 
we then obtain the estimates $\{\hat\lambda_r$, $\hat\psi_r(\cdot)\}$ as the eigenvalues and eigenfunctions of $\widehat H(\cdot,\cdot)$.
One can see from Theorem \ref{thm: consistency of estimators} that $\hat\psi_r(\cdot)$ and $\hat\lambda_r$ are respectively consistent to $\psi_r(\cdot)$ and $\lambda_r:=p^{-1}\sum_{j=1}^p \lambda_{r,j}$, where $\lambda_{r,j} := \E\{\xi_{ir}^2(\s_j)\}$ reflects the heteroskedasticity across different locations $\{\s_j\}_{j=1}^p$.
Besides the functional components that depict the temporal variation, 
we propose to estimate the cross spatial covariance $\sigma_r(\s_j,\s_k)$, i.e., the $p\times p$ cross-covariance matrix $\BSigma_r=\left(\sigma_r(\s_j,\s_k)_{j,k=1,\dots,p}\right)$, by
\begin{equation}\label{hat Sigma_r}
\widehat\BSigma_r = \frac{1}{2(n-1)} \sum_{i=1}^{n-1} \left\langle  \left\{ \BX_{i+1}(\cdot)-\BX_{i}(\cdot) \right\},\hat\psi_r(\cdot) \right\rangle_\CT \left\langle \left\{ \BX_{i+1}(\cdot)-\BX_{i}(\cdot) \right\}, \hat\psi_r(\cdot) \right\rangle_\CT\trans,
\end{equation}
of which the off-diagonal element $\widehat\BSigma_r(j,k)$ and diagonal element $\hat\lambda_{r,j}:=\widehat\BSigma_r(j,j)$ yield the cross-covariance and variance estimates respectively for $\sigma_r(\s_j,\s_k)$ and $\lambda_{r,j}$.
Moreover, define the spatial correlation $\rho_r(\s_j,\s_k) =\sigma_r(\s_j,\s_k)/ \sqrt{\lambda_{r,j}\lambda_{r,k}}$, 
we propose the corresponding estimator $\hat\rho_r(\s_j,\s_k) = \widehat\BSigma_r(j,k)/ \sqrt{\widehat\BSigma_r(j,j)\widehat\BSigma_r(k,k)}$.
The convergence results for the developed FPCA estimators are established in Theorem \ref{thm: consistency of estimators}.

\begin{assump}\label{assump: moment}(Moments)
   The multivariate functional process $\BE_i(\cdot)$ satisfies that 
 for each $j=1,\dots,p$, $\E \| \,\BE_i(\s_j;\cdot) \|_\CT^4  < \infty $ (which means $\E \{\int_\CT\BE_i^2(\s_j;t)dt \} ^2<\infty$).
\end{assump}

\begin{assump}\label{assump: eigengap}(Eigen-gaps)
    For some integer $R>0$, the eigenvalues \{$\lambda_r$\} satisfy $\lambda_1 > \lambda_2 > \dots \lambda_R > 0$. Moreover, there exist constants $c_1,c_2>0$ such that for each  $1 \leq r \leq R$, $c_1 \leq \lambda_{\min }\left(\BSigma_r\right) < \lambda_{\max }\left(\BSigma_r\right) \leq c_2$.
\end{assump}

Assumption \ref{assump: moment} is a standard moment assumption for the weak convergence of (cross) covariance operators in Hilbert-Schmidt spaces \citep[e.g.][]{bosq2000linear, Hsing2015Theoretical}.
Assumption \ref{assump: eigengap} states the distinctness of the eigenvalues, which is common in FDA and ensures the consistency of eigen-estimators in Theorem \ref{thm: consistency of estimators}.
Without loss of generality, the eigendecompositions of $\widehat H$ and $H$ are ordered in terms of the decreasing value of $\{\hat\lambda_r\}$ and $\{\lambda_r\}$ in what follows, thereby the expansion of \eqref{WS model} is ordered in the same way.    
Assumption \ref{assump: eigengap} also implies that each of $\mathbf{\Sigma}_r$ is positive definite.

\begin{theorem}\label{thm: consistency of estimators}
Suppose that the spatiotemporal process $X_i(\cdot;\cdot)$ is weakly separable under Assumptions \ref{assump: L2}--\ref{assump: eigengap}.
Then for each $1 \leq r \leq R$ and as $n \rightarrow \infty$, we have
\begin{align*}
\hat\lambda_{r} \stackrel{p} \rightarrow \lambda_{r},   \quad 
\hat\psi_r(\cdot) \stackrel{p} \rightarrow  \psi_r(\cdot), 
\quad  \widehat\BSigma_r \stackrel{p} \rightarrow \BSigma_r,
\end{align*}
where $\hat\psi_r(\cdot) \stackrel{p} \rightarrow  \psi_r(\cdot)$ means $\| \hat\psi_r(\cdot)- \psi_r(\cdot)\|_\CT= o_p(1)$, and $ \widehat\BSigma_r \stackrel{p} \rightarrow \BSigma_r$ means the element-wise convergence of $\widehat\BSigma_r$ in probability.
\end{theorem}

Theorem \ref{thm: consistency of estimators} establishes the consistency of the FPCA estimator resulting from the covariance estimator $\widehat H(\cdot,\cdot)$. Notably, the convergence results hold under both the null and alternative hypotheses in \eqref{test: cp}.
In Appendix B.1 we provide a more precise convergence rate for $\hat\psi_r(\cdot)$, showing that $\| \hat\psi_r(\cdot)- \psi_r(\cdot)\|_\CT=O_p(n^{-1/2})$.
Furthermore, when the dependency among the covariance functions at different locations is negligible, the convergence rate improves to  
$\| \hat\psi_r(\cdot)- \psi_r(\cdot)\|_\CT = O_p\left( (pn)^{-1/2} \right)$. This result suggests that the estimation errors decrease as the number of locations $p$ increases, indicating that estimation efficiency is enhanced by pooling information across different locations, benefiting from the weakly separable structure.

\section{Change-point detection based on FPCA}\label{sec: cp detect}
Building on the FPCA estimators developed in Section \ref{sec: FPCA},
we propose a change-point detection method tailored for the spatially indexed functional data under study. 
In this section, we focus first on the scenario where the number of locations $p$ is fixed, as the primary objective is to test for a global change-point across all locations.

\subsection{Test statistics and null distribution}\label{sec: CP model}
Recall that at each location $\s_j$, we observe the functional objects $X_{i}(\s_j; \cdot)$ across $i=1,\dots,n$.
To formulate the change-point model, for any $1\leq \tau<n$, we define the standardized CUSUM difference:
\begin{align*}\label{deltat}
{\widehat\Delta}_{\tau}(\s_j; \cdot) = \frac{1}{\sqrt{n}} \left\{ \sum\limits_{i=1}^{\tau} X_{i}(\s_j; \cdot) - \frac{\tau}{n}\sum\limits_{i=1}^n X_{i}(\s_j; \cdot) \right\},
\end{align*}
where we observe that 
$$ \sum_{i=1}^\tau X_{i}(\s_j; \cdot) - \frac{\tau}{n}\sum_{i=1}^n X_{i}(\s_j; \cdot) = \frac{\tau(n-\tau)}{n}\left\{ \frac{1}{\tau}\sum _{i=1}^{\tau}X_{i}(\s_j; \cdot)-\frac{1}{(n-\tau)}\sum_{i=\tau+1}^{n}X_{i}(\s_j; \cdot)\right\}. $$
Treating ${\widehat\Delta}_{\tau}(\s_j; \cdot)$ as an element in $L^2(\CT)$, we then project it onto the FPCs for $r=1,\dots,R$, yielding the scaled statistics
\begin{equation}\label{eta_r,s}
 {\hat{\eta}}_{\tau,r}(\s_j) = 
 \hat\lambda_{r,j}^{-1/2} \left\langle{{\widehat\Delta}}_{\tau}(\s_j; \cdot), {\hat\psi}_{r}(\cdot) \right\rangle_{\CT},
\end{equation}
where $\hat\lambda_{r,j}$ and $\hat{\psi}_{r}(\cdot)$ are the consistent FPCA estimates obtained from the procedure in Section \ref{sec: FPCA}. 
By utilizing this dimension-reduction technique, we can evaluate the functional elements ${\widehat\Delta}_{\tau}(\s_j; \cdot)$ through the aggregation of the first few components of  $\{\hat{\eta}_{\tau,r}(\s_j)\}_{r=1}^\infty$, say 
$\sum_{r=1}^R \hat{\eta}_{\tau,r}(\s_j)$,
which captures a substantial portion of the variability in ${\widehat\Delta}_{\tau}(\s_j; \cdot)$, owing to the weak separability property. 

As is typically observed for spatial signals, a location and its adjacent neighbors are more likely to fall in a similar type of region, either occurring changes or not. Meanwhile, the number of spatial locations affected by an abrupt break is usually not large.
To incorporate this local clustering pattern among functional curves into the change-detection approach, we proposed the following kernel-based statistic:
\begin{equation}\label{kernel statistic}
{\hat{\eta}}_{\tau,r}^{(h)} =\mathop{\sum\sum}_{1\leq j,k\leq p}K_h(\s_j-\s_k)\hat{\eta}_{\tau,r}(\s_j)\hat{\eta}_{\tau,r}(\s_k),
\end{equation} 
where $K_h(\cdot)=K(\cdot/h)/h^2$, with $h>0$ representing a bandwidth that depends on the spatial observations.
Here $K(\cdot)$ is a nonnegative, symmetric and bounded kernel function from $\mathbb{R}^2$, which integrates to 1 and has bounded derivatives.
From the perspective of nonparametric regression, this approach measures the quantity
$\E_{\bs} \{\E (\hat\eta_{\tau,r}^2 | \bs)\}$
where $\E_{\bs}$ represents the expectation with respect to some sampling density $f_{\bs}$ of the spatial points $\bs \in \bS$, and $\E(\hat\eta_{\tau,r}^2|\bs)$ denotes the expectation of 
$\E \{\hat\eta_{\tau,r}^2(\bs)\}$ conditional on $\bs$ \citep{ren2022large}. The kernel statistic \eqref{kernel statistic} can be viewed as employing this approach by utilizing a uniform sampling density for all spatial points.
Rather than relying solely on the statistic at $\s_j$, it aggregates the quantity \eqref{eta_r,s} from neighboring locations around $\s_j$, thereby enhancing signal detection capabilities by pooling information from adjacent areas.

By aggregating the statistic \eqref{kernel statistic} across different FPC directions, we
define $Q_{h}(\tau)=\sum_{r=1}^R\hat{\eta}_{\tau,r}^{(h)}$. 
The test statistic can then be constructed as follows:
\begin{equation}\label{Qh}
Q_h^{max} = \max_{1\leq\tau<n}Q_{h}(\tau), \quad \mbox{ or } \quad
Q_h^{sum} = \frac{1}{n}\sum_{1\leq\tau<n}Q_{h}(\tau).
\end{equation}
The change-point test can be conducted using either $Q_h^{{max}}$ or $Q_h^{{sum}}$, which can be respectively regarded as the Kolmogorov-Smirnov or Cram\'{e}r-von Mises type of functionals based on the functional cumulative sum process. 
\cite{gromenko2017detection} noted that the convergence rate for Kolmogorov-Smirnov-type statistics may be slow, and therefore focused on the latter approach. By contrast, \cite{paynabar2016change} proposed using a simulation-based approach to approximate the distribution of the ``max'' statistic. In this paper, we focus on both the ``max'' and ``sum'' statistics, presenting their asymptotic null distribution in Theorem \ref{thm: asy null of Qh}, and investigating their numerical performance via simulation in Section \ref{sec: simulation}.

\begin{theorem}\label{thm: asy null of Qh}
Under the assumptions in Theorem \ref{thm: consistency of estimators} and the
null hypothesis $\mathcal{H}_0$ in \eqref{test: cp},
\begin{align}\label{convergence of Qh}
Q_h^{{max}} & \stackrel{d }{\rightarrow} \sup_{0<x<1}  \sum_{r=1}^R {\BB_{r}\trans(x) {\mbox{\BK}}_h \BB_{r}(x)}, 
\quad Q_h^{{sum}}  \stackrel{d }{\rightarrow}       \sum_{r=1}^R \int_0^1 {\BB_{r}\trans(x) {\mbox{\BK}}_h \BB_{r}(x)}   dx ,
\end{align}
where ${\BK}_h = \left(K_h(\s_j-\s_k)_{j,k=1,\dots,p} \right)$ is the $ p\times p$ kernel matrix, $\BB_{r}(\cdot) = \{B_{1r}(\cdot),\dots,B_{pr}(\cdot)\}\trans$ are $p$-dimensional Brownian bridges with covariance matrix $\BP_r= \left( \rho_r(\s_j,\s_k)_{j,k=1,\dots,p}\right)$, i.e. $\BP_r^{-1/2} \BB_r(\cdot)$ is a vector of independent standard Brownian bridges.
\end{theorem}

\begin{remark}
The convergence properties in Theorem \ref{thm: asy null of Qh} hold when the number of locations $p$ is fixed, regardless of the shape or sampling scheme of the spatial domain.
This contrasts with typical requirements for modeling spatiotemporal data without replicates, where consistent estimators often necessitate specific asymptotic regimes such as increasing-domain asymptotics. Given the availability of realizations for $\BX_i(\cdot)$ across $i=1,\dots,n$, the results in \eqref{convergence of Qh} also appear to be independent of the strength of spatial dependence.
However, as is well-acknowledged in
multivariate change-detection literature \citep[e.g.][]{guanghui2023changepoint}, 
the (second-order) correlation among different elements can not be too strong for identifying the mean breaks.
This phenomenon is similarly observed in our spatiotemporal setup, where the convergence in \eqref{convergence of Qh} tends to be slow if the spatial correlation is too strong.  
Our simulation results indicate that the proposed detection procedure performs well under a range of scenarios with moderate spatial dependency.
\end{remark}

To approximate the quantiles of the asymptotic distribution on the right-hand side of (\ref{convergence of Qh}), one can compute the Monte Carlo distribution based on multivariate Brownian bridges with the covariance structure $\BP_r$.
A typical approach involves incorporating spatial correlation estimates $\hat\rho_r(\cdot,\cdot)$, as provided in Section \ref{sec: FPCA}, to obtain the corresponding $\widehat\BP_r$. 
Alternatively, if the spatial correlation can be further assumed to be stationary and isotropic, as is common in spatial statistics, $\BP_r$ can be estimated through some smoothing procedures.
{For example, assuming $\rho_r(\cdot)$ is a univariate correlation function that depends only on pairwise distances,
one can employ a nonparametric regression approach on the observations $\{\tilde{d},\hat\rho_r(\tilde{d})\}$, where $\hat\rho_r(\tilde{d})=\hat\rho_r(\s_j,\s_k)$, and $\tilde{d}=\|\s_j-\s_k\|$ denotes the distance between locations $\s_j$ and $\s_k$.
Following \cite{gromenko2017detection}, we use B-splines to obtain a nonparametric estimator $\hat\rho_r(\cdot)$ and consequently the matrix $\widehat\BP_r$, by plugging $\hat\rho_r(\cdot)$.}
Note that this approach allows for varying variances across different locations, providing a more general framework than the typical assumption of spatial stationarity.

\subsection{The proposed change-detection procedure}\label{sec: cd proce}
Despite the null distribution presented in Theorem \ref{thm: asy null of Qh}, our primary interest lies in detecting the change-point $\tau^*$, specifically understanding the behavior of the test under the alternative hypothesis.
As is typical in change-point detection, we assume that $\lim_{n\rightarrow \infty} \tau^*/n  =\theta_0 \in (0,1)$.
According to Theorem \ref{thm: asy null of Qh}, a large value of $Q_h(\tau)$ exceeding a certain threshold would suggest rejecting the null hypothesis. Define that $\BDelta(\cdot) = \{\Delta(\s_1;\cdot),\dots,\Delta(\s_p;\cdot)\}\trans$, 
where $\Delta(\s_j;\cdot) = \mu_{\tau^*+1}(\s_j;\cdot) - \mu_{\tau^*}(\s_j;\cdot)$ denotes the functional process of the mean shift at location $\s_j$.
The following theorem establishes the rate of divergence for the test statistics under the alternative.

\begin{theorem}\label{thm: asy alter of Qh}
Under the assumptions in Theorem \ref{thm: consistency of estimators} and the alternative $\mathcal{H}_1$ in \eqref{test: cp}, 
if there exists at least one $r=1,\dots,R$ such that $ \BDelta_r := \langle {\BDelta}(\cdot), \psi_{r}(\cdot) \rangle_\CT \neq 0$, 
then $n^{-1} Q_h^{max}= O_p(1)$ and $n^{-1} Q_h^{sum} = O_p(1)$,
with the explicit formulas for $O_p(1)$ provided in \eqref{Q_h lim} of Appendix.
\end{theorem}

Theorem \ref{thm: asy alter of Qh} demonstrates that the developed tests are consistent provided the change is not orthogonal to the subspace spanned by the eigenfunctions.
Notably, the established $n$-rate of divergence for $Q_h^{max}$ and $Q_h^{sum}$ 
remains valid when $p \rightarrow \infty$, since the consistency of the proposed FPCA estimators is unaffected by the dimension of $p$.
Upon rejection of the null hypothesis, the change-point can be estimated by
\begin{equation}\label{est tau}
\hat{\tau}= \argmax_{1\leq\tau<n} Q_h(\tau).
\end{equation}

\begin{coro}\label{thm: cp consis}
Under the assumptions in Theorem \ref{thm: asy alter of Qh}, we have $|\hat{\tau} - \tau^*| = O_p(1)$.
\end{coro}

To implement the change-point test based on the derived methodology, selecting an appropriate bandwidth $h$ for the statistic $Q_h^{max}$ or $Q_h^{sum}$ is crucial.
While the weak convergence results in \eqref{convergence of Qh} hold for any $h>0$ theoretically (for a fixed $p$), the convergence rate generally improves when $h$ is smaller.
On the other hand, choosing a relatively large $h$ may incorporate more non-zero components in $\BDelta_r$, enhancing the power of the test by capturing broader spatial signals. These considerations suggest that the test can perform well across a reasonable range of $h$.
From a data-driven perspective, we propose to determine the bandwidth by first fitting a pilot model for the correlation function $\rho(\cdot)$. 
This can be estimated by averaging the nonparametric estimates $\hat\rho_r(\cdot)$ across different components $r$.
The bandwidth $h$ can then be selected based on the range parameter of the fitted spatial correlation or by identifying the minimum distance at which the fitted correlation drops below a threshold $\varrho$.
The rationale behind this approach is that neighboring locations with relatively strong dependence are more likely to belong to the informative set of signals.
Additionally, it is expected that the resulting change-point will be consistent across various values of $\varrho$. 

Another important implementation issue is the selection of the truncation number $R$ for the eigenfunctions. 
It is well recognized in FDA that, the chosen parameter $\widehat R$ should not be too large, as this can lead to increasingly unstable FPC estimates.
On the other hand, as indicated by Theorem \ref{thm: asy alter of Qh}, an excessively small $\widehat R$ may fail to capture sufficient signals across the components $\BDelta_r$, thereby reducing the ability to detect changes. To provide practical guideline, we adopt the fraction of variance explained (FVE) criterion, defined in our setting as 
$$ \mbox{FVE}(R)=\left(\sum_{r=1}^R \hat\lambda_r\right)/\left(\sum_{r=1}^\infty \hat\lambda_r\right), $$ where \{$\hat\lambda_r$\} are the estimated eigenvalues from Section \ref{sec: FPCA}.
Common thresholds for FVE include 80\%, 90\% or 95\%, though our simulation results suggest that the proposed method remains robust across a wide range of FVE values.

\begin{algorithm}
\begin{algorithmic}[1]
    \caption{The detection procedure for the global change-point} 
    \label{alg: cp detect}
    \STATE Conduct the FPCA procedure in Section \ref{sec: FPCA}, and
    \begin{itemize}
        \item obtain the FPCs $\{\hat\psi_r(\cdot)\}$ using $\widehat H(\cdot,\cdot)$ as defined in \eqref{hatH}.
        \item calculate the  covariance matrix $\widehat{\BSigma}_r$ by \eqref{hat Sigma_r} with the corresponding correlation matrix $\widehat{\BP}_r$.
    \end{itemize}
    \STATE Determine the truncation number $\widehat{R}$ according to a pre-specified FVE level.
    \STATE Fit a nonparametric correlation function $\hat\rho(\cdot)$ based on $\widehat{\BP}_r$, and select the bandwidth by $h = \inf\{d>0:\hat \rho(d)<\varrho\}$ for a specified $\varrho$. Then calculate the kernel matrix ${\BK_h}$.
    \STATE For each $\tau = 1,\dots,n$,
    \begin{itemize}
        \item  obtain ${\hat{\eta}}_{\tau,r}(\s_j)$ based on the FPC estimates by \eqref{eta_r,s} across $r = 1,\dots,\widehat{R}$ for each $\s_j$.
        \item  calculate the quadratic form 
        ${\hat{\eta}}_{\tau,r}^{(h)} = {\hat{\Beta}_{\tau,r}}\trans {\BK_h} \hat{\Beta}_{\tau,r}$ with $\hat{\Beta}_{\tau,r} = \{{\hat{\eta}}_{\tau,r}(\s_1),\dots,{\hat{\eta}}_{\tau,r}(\s_p)\}\trans$. 
    \end{itemize}
    Then calculate the statistic $Q_h^{max}$ or $Q_h^{sum}$ by \eqref{Qh} based on $Q_h(\tau) = \sum_{r=1}^{\widehat{R}} {\hat{\eta}}_{\tau,r}^{(h)} $.
    \STATE For each $r = 1,\dots,\widehat{R}$ and $0<x<1$, generate the Monte Carlo samples of $\tilde\BB_r(x) = \widehat{\BP}_r^{1/2}\BB_r^*(x)$, where $\BB_r^*(x) = \{B_{1r}^*(x),\dots,B_{pr}^*(x)\}\trans$ is a standard $p$-dimensional Brownian bridge. Then calculate the samples of
    \begin{equation*}
        \tilde F_{Q_h}(x) = \sum\limits_{r=1}^{\widehat{R}} {\tilde\BB_{r}\trans(x) {\mbox{\BK}}_h \tilde\BB_{r}(x)}.
    \end{equation*}
    \begin{itemize}
    \item For the test statistic $Q_h^{max}$, obtain the $p$-value by using the Monte Carlo distribution of $\sup_{0<x<1} \tilde F_{Q_h}(x)$.
    Conduct the test similarly for $Q_h^{sum}$.
    \end{itemize}
    \STATE When the test is rejected, plot the function $Q_h(\tau)$ and obtain the change-point $\hat\tau$ by \eqref{est tau}.
\end{algorithmic}
\end{algorithm}

We illustrate the proposed change-point detection procedure in Algorithm \ref{alg: cp detect}.
It is noteworthy that valid change-point detection can also be performed using the non-kernel statistics 
$$ Q_0^{max} = \max_{1\leq\tau<n} Q_{0}(\tau) \mbox{ or } Q_0^{sum} = {n}^{-1}\sum_{1\leq\tau<n} Q_{0}(\tau),$$ 
where $Q_{0}(\tau)=\sum_{r=1}^{\widehat{R}} \sum_{j=1}^p {\hat{\eta}}^{\,2}_{\tau,r}(\s_j)$ defines the statistic $Q_h(\tau)$ for $h=0$.
This procedure employs the conventional sum of squares approach, whereas the test statistic $Q_0^{sum}$ coincides with the statistic $\hat\Lambda_1$ proposed by \cite{gromenko2017detection} under uniform weighting across all locations.
It is important to note that $\hat\Lambda_1$ is derived under the assumption of strong separability.
In contrast, the proposed test based on $Q_h$ efficiently incorporates the spatial clustering information through a kernel approach, resulting in a substantial increase in power.
We compare the empirical test performance of $Q_h$, $Q_{0}$ and those proposed by \cite{gromenko2017detection} through the simulation study in Section \ref{sec: simulation}.

\section{Support recovery based on sample splitting}\label{sec: sup recov}
After detecting the change-point using the procedure described in Algorithm \ref{alg: cp detect}, the next objective is to develop a valid support recovery method for the problem (\ref{test: mcp}). 
A natural approach is to utilize the statistic in (\ref{eta_r,s}) with the estimated change-point $\hat\tau$, followed by calculating the $p$-values for all locations, denoted as $\{p_j;j=1,\dots,p\}$,
and then applying a multiple testing procedure, such as the BH procedure or the 
LAWS method proposed by \cite{cai2022laws}.
The LAWS method incorporates local weights for the point-wise $p$-values by assuming 
that each location has a prior probability of being an alternative, say $\pi(\s_j) = \Pr\left( \theta_j = 1 \right)$.
However, for the spatial sampling scheme under consideration, this assumption may be questionable. 
To address this limitation, we propose a $p$-value-free procedure that adaptively and effectively leverages the spatial structure, based on a combination of the proposed FPCA method and the SDA approach in \cite{du2021false}.

\subsection{Construction of ranking statistics and FDR thresholding}\label{sec: rank statistic}
The SDA method involves constructing a series of ranking statistics with global symmetry property through sample splitting, followed by selecting a data-driven threshold to control the FDR. 
Adhering to this framework, we first partition the dataset into two subsets, both used to construct statistics that evaluate the evidence against null hypothesis.
Specifically, we adopt an order-preserved sample-splitting approach by dividing the full sample $\CX=\{ \BX_i(\cdot): i=1,\dots,n \} $ into
\begin{equation}\label{data split}
  \mathcal{X}_O:=\left\{\BX_{2 i-1}(\cdot): i=1, \ldots, m\right\} \text { and } \mathcal{X}_E:=\left\{\BX_{2 i}(\cdot): i=1, \ldots, m\right\} \text {, }  
\end{equation}
where we assume that $n$ is even for convenience. 
As suggested by \cite{chen2021data},
this splitting strategy helps preserve the original change-point structure
with minimal variation between the training and validation samples.
The statistics on $\CX_O$ are constructed using (\ref{eta_r,s}) with the estimated change-point $\hat\tau$, denoted as $\hat{\eta}_{\hat\tau, r}^O\left(\s_j\right)$ for $j=1,\dots,p$; while for $\CX_E$, we propose the kernel-weighted statistic:
\begin{equation}\label{NW est}
    \tilde{\eta}_{\hat\tau, r}\left(\s_j\right)=\frac{\sum_{k}K_h(\s_k-\s_j)\hat{\eta}_{\hat\tau, r}\left(\s_k\right)}{\sum_{k}K_h(\s_k-\s_j)},
\end{equation}
where $K_h(\cdot)$ is the kernel function as defined in \eqref{kernel statistic}.
Following a similar spirit to \eqref{kernel statistic}, this statistic
aggregates information from the neighboring locations of $\s_j$, resulting in enhanced signals at each spatial point.
The corresponding statistic calculated using the sub-sample $\CX_E$ is denoted as $\tilde\eta_{\hat\tau, r}^E\left(\s_j\right)$.

To combine information from both data splits and across different FPCA directions, we define the detection statistic for each $j$ as:
\begin{equation}\label{FSDA statistic}
    W_j=\sum_{r=1}^{R} \hat{\eta}_{\hat\tau, r}^O\left(\s_j\right) \tilde{\eta}_{\hat\tau, r}^E \left(\mathbf{s}_j\right).
\end{equation}
This statistic fulfills the symmetry property under $\mathcal{H}_{0}^j$ and tends to take large positive values under $\mathcal{H}_{1}^j$, as discussed further in Section \ref{sec: intui}.
Based on the computed $W_j$ values across all locations, 
we determine a threshold $L>0$ according to the following rule:
\begin{equation}\label{thresh}
    L=\inf \left\{t>0: \frac{1+\#\left\{j: W_j \leq-t\right\}}{\#\left\{j: W_j \geq t\right\} \vee 1} \leq \alpha\right\},
\end{equation}
where $\alpha$ denotes the target FDR level. If the set is empty, we simply 
let $L=+\infty$.
The desired support set is then recovered by $\CM=\{j: W_j\ge L\}$.

To implement the multiple testing statistic in \eqref{FSDA statistic}, a similar FVE criterion for the truncation number $\widehat{R}$ can be employed as in the change-detection statistic $Q_h$.
For bandwidth selection, we typically use a common $h$ in $\tilde\eta_{\hat\tau, r}^E\left(\s_j\right)$ across all locations $\s_j$, following the same approach outlined in Algorithm \ref{alg: cp detect}.
To ensure the theoretical guarantees for the kernel estimator in Assumption \ref{assump: kernel}, a relatively large threshold value of $\rho$ is determined for practical applicability.
Remarkably, our method also accommodates varying $h$ values at each $\s_j$ to account for spatial inhomogeneity or differing clustering patterns, based on the researchers' knowledge of the local signal magnitude.
We summarize the proposed functional SDA (fSDA) procedure as follows.

\begin{algorithm}
\begin{algorithmic}[1]
    \caption{The fSDA procedure for support recovery}
    \label{alg: FSDA}
    \STATE Obtain the estimated change-point $\hat{\tau}$ as outlined by Algorithm \ref{alg: cp detect}.
    \STATE Split the data into $\mathcal{X}_O$ and $\mathcal{X}_E$ as \eqref{data split} and conduct the FPCA procedure on each split separately.
    \STATE Determine the truncated number $\widehat{R}$ 
    {using the same FVE criterion in Algorithm \ref{alg: cp detect}.}
    \STATE  For each $j=1,\dots,p$, 
    \begin{itemize}
        \item calculate  $\hat{\eta}_{\hat{\tau}, r}^O\left(\mathbf{s}_j\right)$ by \eqref{eta_r,s} across $r=1,\dots,\widehat{R}$ based on the FPCA estimates on $\mathcal{X}_O$.
        \item calculate $\tilde{\eta}_{\hat{\tau}, r}^E\left(\mathbf{s}_j\right)$ by \eqref{NW est} across $r=1,\dots,\widehat{R}$ based on the FPCA estimates on $\mathcal{X}_E$ and the kernel function $\BK_h$ in Algorithm \ref{alg: cp detect}.
        \item obtain the test statistic $W_j$ by \eqref{FSDA statistic}.
    \end{itemize}
    \STATE Rank $W_j$'s and obtain the threshold $L$ by \eqref{thresh} for a target $\alpha$. Finally, output the support set 
    $$ \CM=\{j: W_j\ge L\}. $$
\end{algorithmic}
\end{algorithm}

\subsection{An intuitive explanation of the fSDA procedure}\label{sec: intui}
As seen from the procedure in Section \ref{sec: rank statistic}, the statistics $W_j$’s play a central role in the proposed fSDA approach. 
To explain why this works, let $T_{j,r}^O = \hat{\eta}_{\hat\tau, r}^O(\s_j)$, which \blue{asymptotically follows $N(0,1)$ under the null hypothesis}, and let $T_{j,r}^E=\tilde{\eta}_{\hat\tau, r}^E (\s_j )$.
For each $r$, the product $ W_{j,r}=T_{j,r}^O T_{j,r}^E$ aggregates evidence from the two groups.
When the signal magnitude at location $\s_j$, defined as $\Delta_{j,r}:= \langle \Delta(\s_j;\cdot),\psi_r(\cdot) \rangle_\CT $ is large, both $T_{j,r}^O$ and $T_{j,r}^E$ tend to exhibit large absolute values with the same sign. 
This results in a large positive $W_{j,r}$, and thus a large $W_j$, which aids in ranking and selecting informative locations.
On the other hand, negative $W_j$'s, typically corresponding to the null cases, are useful for thresholding. 
Intuitively, $W_j$ is, at least asymptotically symmetric with mean zero for $j\in\mathcal{A}^c$,
due to the asymptotic normality of $T_{j,r}^O$ and the independence between $\CX_O$ and $\CX_E$.
The key idea is to leverage the following symmetry property:
\begin{equation}\label{sym}
    \sup _{0 \leq t \leq c_p}\left|\frac{\sum_{j\in\mathcal{A}^c} \bI\left(W_j \geq t\right)}{\sum_{j\in\mathcal{A}^c} \bI\left(W_j \leq-t\right)}-1\right| = o_p(1)
\end{equation}
for some upper bound {$c_p$}. 
Under the selection rule in \eqref{thresh}, this implies that the number of false discoveries, $\#\{j\in\mathcal{A}^c: W_j \geq t\}$, is asymptotically equal to $\#\{j\in\mathcal{A}^c: W_j \leq-t\}$, which is conservatively approximated by $\#\{j: W_j \leq-t\}$.
Therefore, the ratio in \eqref{thresh} provides an overestimate of FDP, leading to conservative control of FDR.
Moreover, since most elements in $\{j: W_j\le -L\}$ are expected to belong to the null set for a suitably chosen $L$, the discrepancy between the fraction in \eqref{thresh} and the actual FDP is typically small in practice, ensuring that the empirical FDR level closely matches $\alpha$.
In summary, the proposed procedure approximates the FDP based on the symmetry of the ranking statistics $W_j$'s, rather than relying on asymptotic normality---which may be difficult to achieve due to the kernel aggregation approach.

\begin{figure}
    \centering
    \includegraphics[width=1\linewidth]{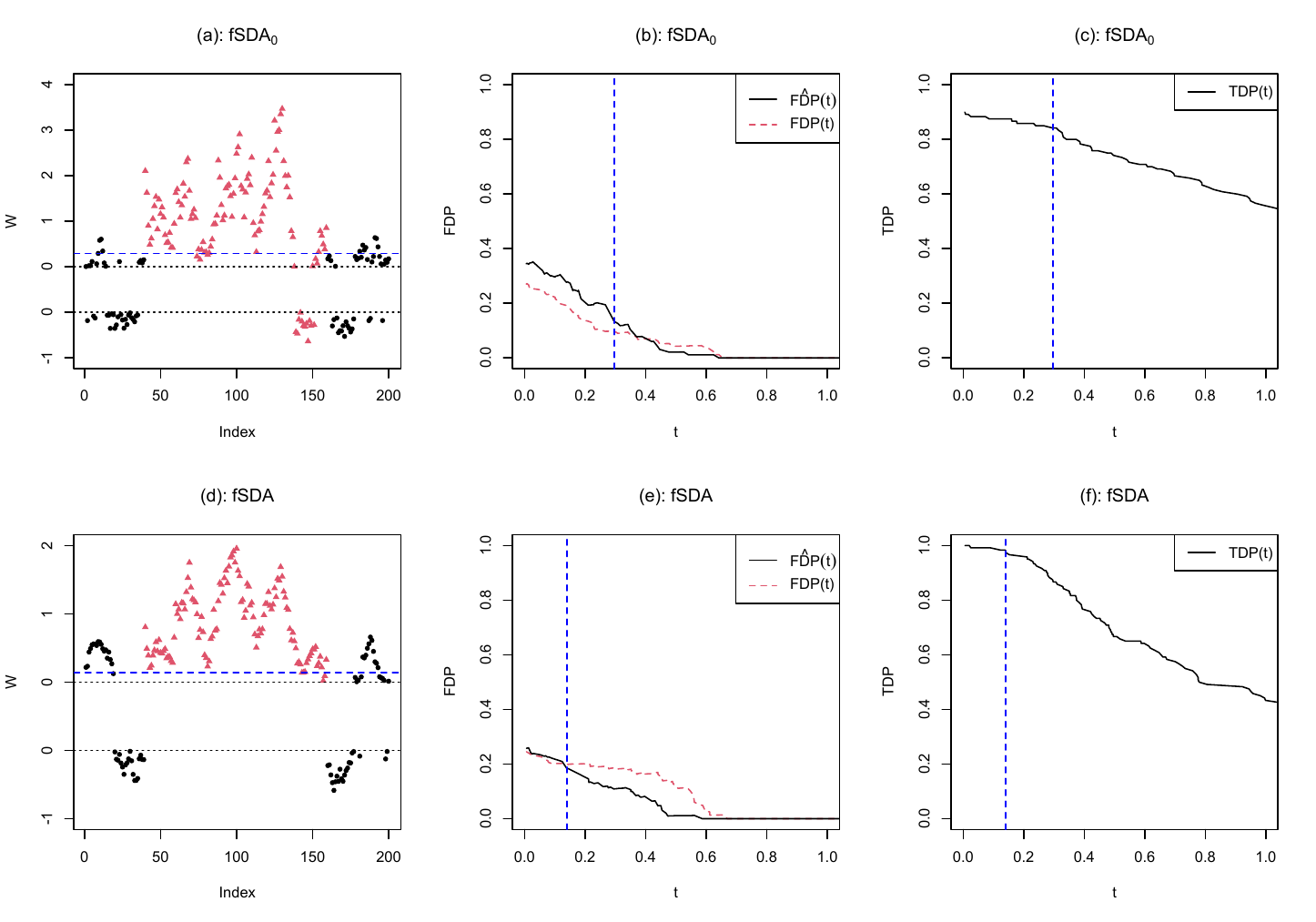}
    \caption{\small (a) Scatterplot of the 200 $W_j$'s obtained by $\text{fSDA}_0$ with black dots and red triangles denoting nulls and true signals, respectively. 
    A vertical space is added to the middle of the plot to better contrast positive and negative $W_j$'s, and the blue dashed line represents the threshold $L$. 
    (b) The corresponding estimate of FDP curve (against $t$) along with the true FDP for $\text{fSDA}_0$. 
    (c) The TDP curve (against $t$) for $\text{fSDA}_0$. 
    (d)-(f) The scatterplot, the corresponding FDP estimate, and the TDP for the fSDA method.
    }
    \label{symmetry compar}
\end{figure}

Similar to the change-detection procedure in Algorithm \ref{alg: cp detect}, the proposed multiple testing approach can also be conducted based on a non-kernel version of $W_j$, computed directly from $\hat{\eta}_{\hat\tau, r}(\s_j)$ on both $\CX_O$ and $\CX_E$. 
The corresponding ranking and thresholding process can be performed following the same steps as outlined in Algorithm \ref{alg: FSDA}, and we refer to this non-kernel variant as $\text{fSDA}_0$.
Figure \ref{symmetry compar} illustrates a comparison between the operation of $\text{fSDA}_0$ and the kernel-based fSDA.
The data are generated from the model \eqref{simu setup} in Section \ref{sec: simulation} under scenario (i), where the spatial domain is a one-dimensional regular grid on $\CS=[0,1]$ with $p=200$ locations.
The alternative mean function is set as $\mu_1(\bs,t)=0.5 \, \bI(\bs\in\Omega_s)$ with $\Omega_s=[0.2,0.8]$, resulting in a non-null set of size $|\mA| = 120$.
Panels (a) and (d) show scatterplots of $W_j$'s for $\text{fSDA}_0$ and fSDA,  respectively, where black dots represent null locations and red triangles denote true signals. 
For both methods, the true signals predominantly appear above zero, while the uninformative $W_j$'s values (nulls) are roughly symmetrically distributed around the horizontal axis. 
This symmetry  enables accurate estimation of the FDP, as defined by the fraction in \eqref{thresh} and illustrated in Panels (b) and (e), 
demonstrating the effectiveness of both methods in controlling FDR. 

On the other hand, the proposed fSDA approach achieves a substantial improvement in power over $\text{fSDA}_0$. 
As shown in Figure \ref{symmetry compar}, several true signals fall below the threshold in Panel (a), with some even falling below zero.
In contrast, the kernel aggregation in \eqref{NW est} enables fSDA to effectively leverage spatial structure among neighboring locations,
resulting in stronger positive values and observable clustering patterns among true signals, as illustrated in Panel (d).
This refinement leads to a notable increase in TDP, as defined in \eqref{FDP TDP defi}, with Panel (f) showing consistently higher TDP values compared to Panel (c).
Specifically, at an FDR level of 0.2, fSDA achieves a TDP of 0.98 with a threshold $L$ = 0.14, significantly outperforming $\text{fSDA}_0$, which yields a TDP of 0.84 with $L$ = 0.30.
It is also noteworthy that, by utilizing an order-preserved strategy that harnesses the full sample information, both fSDA and $\text{fSDA}_0$ effectively mitigate the power loss commonly associated with conventional sample-splitting methods.

\subsection{Theoretical property}\label{sr theory}
In this section, we establish the asymptotic validity of the FDR control for the proposed fSDA procedure described in Algorithm \ref{alg: FSDA}. As is typical in multiple testing frameworks, we require that both $n$ and $p$ tend to infinity. Note that the test statistics $W_j$'s are constructed based on the estimated change-point $\hat\tau$, which remains consistent to $\tau^*$ as $p\rightarrow\infty$ according to Theorem \ref{thm: asy alter of Qh}.
Specifically, we assume the following:

\begin{assump}\label{assump: p}(Order of location number to sample size)
    {$p=o\left\{\exp \left(n^{1 / 4}\right)\right\}$}.
\end{assump}

\begin{assump}\label{assump: sr moment}(Moments)
There exists $\gamma>0$ such that $\E\{\exp ( s\, \xi_{ir}(\s_j) ) \} \leq \exp(s^2\gamma^2\lambda_{r,j}/2)$ for all $j=1,\ldots,p$, $r= 1,\ldots,R$ and $s\in\mathbb{R}$, i.e.
the standardized FPC scores $\xi_{ir}(\s_j)/\lambda_{r,j}^{1/2}$ are sub-Gaussian random variables with variance proxy $\gamma^2$.
Furthermore, there exists a constant $M$ independent of $p$ such that $\sup_{j=1,\dots,p}\sum_{r=1}^{\infty}\lambda_{r,j}<M$.
\end{assump}

Assumption \ref{assump: p} constrains the growth rate of $p$ relative to $n$ to facilitate technical derivations.
Specifically, we apply the moderate deviation result for the two-sample $t$-statistic \citep{cao2007moderate} to establish the uniform convergence of $\Pr(\eta_{j,r}^*>x)$ within $x\in \left(0,o(n^{1/6})\right)$,
where $\eta_{j,r}^*:= \lambda_{r, j}^{-1/2} \langle \widehat{\Delta}_{\tau^*}(\s_j;\cdot),\psi_r(\cdot)\rangle_\CT$ denotes the oracle version of $\hat\eta_{\hat\tau,r}(\s_j)$ constructed using the true FPC elements. Assumption \ref{assump: sr moment} is imposed to control the tail probability of $\eta_{j,r}^*$, 
and it additionally requires that the covariance function $C(\s_j,\cdot;\s_j,\cdot)$ be uniformly bounded in $j$.
As shown in \cite{zapata2021partial}, this assumption ensures the derivation of concentration inequalities for estimators obtained via the FPCA approach. 
Moreover, we establish uniform bounds for the quantities $|\hat{\lambda}_{r,j}^{-1/2}-\lambda_{r,j}^{-1/2}|$, $\|\hat\psi_r(\cdot)-\psi_r(\cdot)\|_\CT$, and $\left\langle \widehat\Delta_{\hat\tau}(\s_j;\cdot) -  \widehat\Delta_{\tau^*}(\s_j;\cdot),\psi_r(\cdot) \right\rangle_\CT$ in Lemma \ref{lem: bound for est}, which provide more explicit concentration results than those in Theorem \ref{thm: consistency of estimators} and facilitate the derivation of the uniform convergence rate for $W_j$.

\begin{assump}\label{assump: signal}(Signals)
Define that $\Bdelta_j^*=(\delta_{j,1}^*,\dots,\delta_{j,R}^*)\trans$ where $\delta_{j,r}^*=\lambda_{r, j}^{-1/2}\Delta_{j,r}$, and $\beta_p=\{1\leq j\leq p:\sqrt{n}\|\Bdelta_j^*\|/\sqrt{\log p}\to\infty\}.
$ 
Assume that $b_p=|\beta_p|\to\infty$ as $n,p\to\infty$.
\end{assump}

\begin{assump}\label{assump: dependence}(Dependence)
For each $j=1,\dots,p$, define that $\CC_j = \{1\leq k \leq p: \left|\sum_{r=1}^{R}\rho_r(\s_j,\s_k)\right| \geq C_\rho(\log n)^{-2-v}\}$, where $C_\rho>0$ is a large constant, and $v>0$ {is some small constant}. Assume that $|\CC_j| \leq l_p$ for some $l_p$ with $l_p/b_{p}\to 0$ as $n,p \rightarrow \infty$.
\end{assump}

\begin{assump}\label{assump: kernel}(Bandwidth)
For each $j$, the bandwidth $h$ used in \eqref{NW est} satisfies that $ \sup_{k: |k-j|\leq h} \sqrt{n} \| \Bdelta_{k}^*- \Bdelta_{j}^* \| = O(\sqrt{\log p})$.
\end{assump}

Assumption \ref{assump: signal} requires that the number of identifiable effect sizes, defined in terms of the standardized signal $\Bdelta_j^*$, does not become too small as $p\rightarrow \infty$.
This is a standard requirement in FDP control methods. As demonstrated by \cite{liu2014phase}, if a multiple testing approach controls the FDP with high probability, then the number of true alternatives must diverge as the number of tests goes to infinity.
Assumption \ref{assump: dependence} allows $\xi_{ir}(\s_j)$ to be correlated across all locations, provided that the number of strong correlations does not grow too rapidly \citep{du2021false}. 
Since spatial correlation typically diminishes as the distance between locations increases, this assumption seems natural and can be satisfied if, for instance, the spatial random field is $m$-dependent
(i.e. observations are independent if their distance exceeds $m$).
Assumption \ref{assump: kernel} concerns the bandwidth condition for the proposed kernel aggregation approach. It is approximately satisfied when $nh^2=O(\log p)$, given the smoothness of the mean function $\mu_i$.
From the perspective of estimation accuracy, this assumption ensures that $\tilde{\eta}_{\hat\tau, r}\left(\s_j\right)$ serves as a reasonable approximation of $\hat{\eta}_{\hat\tau, r}\left(\s_j\right)$, similarly to Assumption 2 in \cite{du2021false}.

\begin{remark}
We highlight that the obtained result for FDR control remains generally valid under various spatial sampling schemes, including both increasing and infill domain frameworks, as long as Assumptions \ref{assump: dependence}--\ref{assump: kernel} hold.
As discussed in \cite{cai2022laws}, an essential technical challenge in the context of the two-component mixture prior is to establish the uniform convergence of $\sum_{j=1}^p \bI(p^w_j\leq t, \theta_j = 0)/\sum_{j=1}^p \pi(\s_j)\,t$, where $p_j^w$ denotes the weighted $p$-value. Achieving this convergence relies on an accurate estimation of $\hat\pi(\cdot)$ under an infill-domain asymptotic regime.
In contrast, our proposed fSDA procedure is independent of any $p$-value model; its validity primarily depends on the asymptotic symmetry property \eqref{sym}. This property can typically be satisfied under the dependence structure specified in Assumption \ref{assump: dependence}, ensuring uniform convergence of $\sum_{j\in\mathcal{A}^c}\bI(W_{j} \geq t)/\sum_{j\in\mathcal{A}^c}\bI(W_{j} \leq -t)$,
as shown in the proof of \eqref{sym property} in Appendix. Alternatively, one may impose other assumptions on the dependence structure, such as the mixing conditions on $W_j$'s or ${\xi}_{ir}(\cdot)$ \citep[e.g.][]{yun2022detection}, to achieve similar outcomes.
\end{remark}

\begin{theorem}\label{thm: FDR control}
Suppose the assumptions in Theorem \ref{thm: asy alter of Qh} and Assumptions \ref{assump: p}--\ref{assump: kernel} hold, then for any $\alpha \in (0,1)$, the FDP of the fSDA method satisfies
    \begin{equation*}\label{FDP control}
        {\operatorname{FDP}_w(L)}:=\frac{\#\left\{j: W_j \geq L, j\in\mathcal{A}^c\right\}}{\#\left\{j: W_j \geq L\right\} \vee 1} \leq \alpha+o_p(1).
    \end{equation*}
    It follows that $\limsup_{(n,p)\to\infty}\operatorname{FDR}_w \leq \alpha$.
\end{theorem}

Theorem \ref{thm: FDR control} confirms the asymptotic validity on FDR control of the proposed support recovery approach. 
In contrast to existing multiple testing methods, our fSDA procedure incorporates both kernel smoothing and sample-splitting, with test statistics constructed from FPCA estimators under a spatiotemporal change-point model. These components introduce additional challenges for both theoretical analysis and computational implementation. To better present the theoretical results in Theorem \ref{thm: FDR control}, we summarize the key uniform convergence results in Lemmas \ref{lem: sym for true W}--\ref{lem: bound for est} in Appendix \ref{appA}, which serve as the foundation for deriving the convergence rate of the ranking statistic $W_j$.
The next result further demonstrates that our fSDA procedure is capable of not only maintaining FDR control, but also guaranteeing the full recovery of identifiable effect sizes.

\begin{coro}\label{thm: ident sr}
Under the assumptions in Theorem \ref{thm: FDR control}, we have
\begin{equation*}
\limsup_{(n,p)\to\infty}\Pr\left(\CM\supseteq\beta_p\right)=1.
\end{equation*}
\end{coro}

The detailed proofs of Theorem \ref{thm: FDR control} and Corollary \ref{thm: ident sr} are provided in Appendices \ref{B5}--\ref{B6} of Appendix and the numerical performance of the proposed method is assessed through simulations in Section \ref{sec: simulation}.

\section{Monte Carlo simulation}\label{sec: simulation}
\subsection{Data generating process.}
For $i=1,\dots,n$, we generate the spatiotemporal data $X_i(\cdot;\cdot)$ as follows:
\begin{equation}\label{simu setup}
X_i(\s_j;t)=\mu_i(\s_j;t)+\sum_{r=1}^{R}\xi_{ir}(\s_j)\psi_r(t), \quad j=1,\dots,p, \quad  t\in\CT,
\end{equation}
where $R=6$ and the eigenfunctions are $ \psi_r(t)= \sqrt{2}\cos\{r\pi t\}$ for odd $r$, and $\psi_r(t)=\sqrt{2}\sin\{(r-1)\pi t\}$ for even $r$. The time domain is set as $\CT = [0,1]$ with 100 equally spaced time points.
The following three scenarios of the spatial domain are investigated, with the locations $\{\s_j\}_{j=1}^p$ being generated: 
\begin{longlist}
\item at a 1-dim regular grid $\{\frac{1}{p},\frac{2}{p},\dots,1\}$ on $\CS=[0,1]$; 
\item at a 2-dim regular grid $\{\frac{1}{\sqrt{p}},\frac{2}{\sqrt{p}},\dots,1\} \otimes \{\frac{1}{\sqrt{p}},\frac{2}{\sqrt{p}},\dots,1\}$ on $\CS=[0,1]\times [0,1]$; 
\item from a homogeneous Poisson process on $\CS=[0,1]\times [0,1]$ with $p$ (irregular) points.
\end{longlist}
For each $i$ and $r=1,\dots,R$, the FPC score
 $\xi_{ir}(\cdot)$ is generated from a mean-zero Gaussian random field with the covariance structure
$\sigma_r(\bs,\bs') = \omega_r \mbox{M}(\|\bs-\bs'\|;\nu_r,\phi_r)$,
where $\omega_r = 4 r^{-1.6}$ and 
$\mbox{M}(\cdot; \nu, \phi)$ denote the isotropic Mat\'ern covariance with the smoothing parameter $\nu$ and range parameter $\phi$. 
Following \cite{liang2023test} and to reflect the magnitude of weak separability,  we set $\nu_1=1$ and $\nu_r=0.5$ for $r=2,\dots,6$, whereas $(\phi_1,\phi_2,\phi_3,\phi_4) = (0.1,0.05,0.075,0.02)$ and $\phi_5=\phi_6=0$, i.e. no spatial correlation for the high-order FPC scores. 
The true change-point is set as $\tau^*=n/2$.
We define the mean function before the change-point as $\mu_i(\bs;t) = \mu_0(\bs;t)=0$ for $1 \le i \le \tau^*$, and
the alternative mean as
\begin{align}\label{alter mean}
    \mu_i(\bs;t) = \mu_1(\bs;t)= \delta\, K_{\Omega}(|\bs-\s_0|), \quad \mbox{ for } i = \tau^*+1,\dots,n,
\end{align} 
 where $K_{\Omega}(\cdot)$ is a univariate Gaussian kernel function with the support $\Omega=\{\bs\in\CS: | \bs- \s_0 | \leq r_s\} $ and $\s_0$ being the center of $\CS$.
According to the definition in \eqref{alter mean}, the parameters $\delta$ and $r_s$ reflect the strength and magnitude of the spatial signal, respectively.
We first implement the global change-point detection outlined in Algorithm \ref{alg: cp detect}. For the alternative case with $\delta>0$, the support recovery procedure is then performed following Algorithm \ref{alg: FSDA}.

For space considerations, we only present the test results under scenario (iii), i.e. the irregular 2-dimensional case, which most closely resembles the structure of China's precipitation data analyzed in Section \ref{sec: real data} and is more representative of practical applications.
Our simulations also show satisfactory results for the other two scenarios, demonstrating the robustness of our method across different spatial sampling schemes.
We also investigate the impact of spatial covariance on the test performance by varying the range parameters, with more simulation results provided in Appendix \ref{E1}.

\subsection{Results for change-point detection}
We implement the proposed change-point detection approach using either the ``max'' or ``sum'' statistic of $Q_h$ and $Q_0$. 
The truncation number $R$ and bandwidth $h$ are determined based on the procedure outlined in Section \ref{sec: cd proce}. The test results are shown to be robust across FVE values in the range $[0.8,0.95]$ and $\varrho$ values in $[0.01,0.1]$, with the latter yielding an average $h$ value in $[0.18,0.47]$. For clarity and brevity, we present the results under FVE$=90\%$ and $\varrho=0.05$. We compare the performance of our methods with the statistics $\widehat\Lambda_1$ and $\widehat\Lambda_2$ from \cite{gromenko2017detection} (referred to as Gro) under the same FVE criterion. Here, $\widehat\Lambda_1$ can be considered as a counterpart to $Q_0^{sum}$ under the assumption of {strong separability}, while $\widehat\Lambda_2$ is a non-normalized version of $\widehat\Lambda_1$. The empirical size and power are assessed through 1000 and 200 runs respectively, conducted at a 0.05 significant level.

Table \ref{tab: size} compares the empirical Type I errors using the proposed statistics with those using $\widehat\Lambda_1$ and $\widehat\Lambda_2$ across various sample sizes and numbers of locations. The proposed tests using $Q_0$ and $Q_h$ exhibit overall stable sizes for both ``sum'' and ``max'' types across different values of $p$ and $n$.
In contrast, the test using $\widehat{\Lambda}_2$ fails to control the Type I errors, while the test using $\widehat{\Lambda}_1$ appears to be oversized when $p$ is relatively small. 
{For the proposed kernel statistic $Q_h$, the empirical sizes tend to be closer to the nominal level as the sample size $n$ increases from 100 to 200,} with little difference observed between the ``sum'' and ``max'' statistics.

\begin{table*}
    \caption{Empirical size (\%) of the change-point tests under different $(p,n)$-settings.}
    \label{tab: size}
    \centering
    \begin{tabular}{c c ccc ccc}
    \toprule
   \multirow{2}{*}{ Method } & \multirow{2}{*}{Type} & \multicolumn{3}{c}{$n=100$} & \multicolumn{3}{c}{$n=200$} \\
    \cmidrule(r){3-5} \cmidrule(l){6-8}
    & & $p=20$ & $p=50$ & $p=100$ & $p=20$ & $p=50$ & $p=100$ \\
    \midrule
    \multirow{2}{*}{$Q_0$} & sum & 5.0 & 5.4 & 4.5 & 5.5 & 5.6 & 4.5 \\
    & max & 4.5 & 3.4 & 4.4 & 5.2 & 4.3 & 4.0 \\
    \multirow{2}{*}{$Q_h$} & sum & 5.2 & 5.8 & 5.5 & 5.6 & 5.3 & 5.0 \\
    & max & 6.0 & 6.3 & 6.2 & 5.5 & 5.7 & 5.6 \\
    \multirow{2}{*}{Gro} & $\widehat{\Lambda}_1$ & 7.2 & 6.9 & 5.7 & 6.9 & 6.7 & 5.5 \\
    & $\widehat{\Lambda}_2$ & 7.8 & 10.1 & 9.8 & 7.9 & 9.1 & 8.1 \\
    \bottomrule
    \end{tabular}
\end{table*}

Table \ref{tab: power} presents the empirical power results for $n=100$ under various scenarios of spatial signals, with the highest power in each scenario highlighted in bold. The results for $n=200$, which show a clear improvement in power, are deferred to the Appendix \ref{E1} due to space constraints.
It can be seen that the test using $Q_h$, whether employing the ``sum'' or ``max'' statistic, generally outperforms other methods across different settings. As the spatial magnitude $r_s$ increases from 0.4 to 0.6, the proportion of locations with mean changes rises significantly, from 48.8\% to 94.6\%, resulting in a rapid increase in the power of $Q_h$.
In contrast, the power of $Q_0$ and $\widehat\Lambda_1$ remains relatively unchanged, indicating the superiority of $Q_h$ as the signal magnitude grows. Additionally, all methods exhibit improved performance as $p$ increases, with $Q_h$ demonstrating the most significant enhancement in power. This aligns with the intuition that denser locations allow kernel aggregation to integrate more information from the neighborhood, underscoring the advantages of incorporating spatial clustering patterns into inference.

\begin{table}
    \caption{Empirical power (\%) of the change-point tests under different signal settings when $n=100$.} 
    \label{tab: power}
    \centering
    \begin{tabular}{c c c ccc ccc}
    \toprule
    & \multirow{2}{*}{Method} & \multirow{2}{*}{Type} & \multicolumn{3}{c}{$r_s = 0.4$} & \multicolumn{3}{c}{$r_s = 0.6$} \\
    \cmidrule(r){4-6} \cmidrule(l){7-9}
    & & & $\delta = 0.2$ & $\delta = 0.3$ & $\delta = 0.4$ & $\delta = 0.2$ & $\delta = 0.3$ & $\delta = 0.4$ \\
    \midrule
    & \multirow{2}{*}{$Q_0$} & sum & 19.0 & 31.5 & 49.0 & 16.0 & 33.0 & 53.5 \\
    & & max & 12.5 & 30.5 & 46.5 & 14.0 & 28.0 & 52.0 \\
    \multirow{2}{*}{$p=20$} & \multirow{2}{*}{$Q_h$} & sum & \textbf{20.0} & 35.5 & \textbf{53.5} & \textbf{25.0} & 45.0 & \textbf{67.0} \\
    & & max & \textbf{20.0} & \textbf{38.5} & 51.5 & 23.5 & \textbf{46.5} & 65.0 \\
    & \multirow{2}{*}{Gro} & $\widehat{\Lambda}_1$ & 19.0 & 28.5 & 46.5 & 17.0 & 27.5 & 48.0 \\
    & & $\widehat{\Lambda}_2$ & 9.0 & 11.0 & 20.0 & 10.0 & 11.0 & 27.0 \\
    \midrule
    
    & \multirow{2}{*}{$Q_0$} & sum & 23.0 & 47.0 & 68.5 & 17.0 & 42.5 & 69.5 \\
    & & max & 14.5 & 40.0 & 65.0 & 9.5 & 44.0 & 68.5 \\
    \multirow{2}{*}{$p=50$} & \multirow{2}{*}{$Q_h$} & sum & \textbf{31.5} & \textbf{59.5} & \textbf{78.0} & \textbf{30.0} & 73.0 & 84.0 \\
    & & max & 30.0 & 57.5 & 77.5 & 28.0 & \textbf{75.0} & \textbf{86.0} \\
    & \multirow{2}{*}{Gro} & $\widehat{\Lambda}_1$ & 20.5 & 44.0 & 65.5 & 17.5 & 39.5 & 68.0 \\
    & & $\widehat{\Lambda}_2$ & 17.0 & 30.0 & 32.0 & 12.0 & 28.0 & 34.5 \\
    \midrule
    
    & \multirow{2}{*}{$Q_0$} & sum & 24.5 & 51.5 & 79.5 & 15.5 & 52.5 & 82.5 \\
    & & max & 14.0 & 48.5 & 80.0 & 10.5 & 49.0 & 81.0 \\
    \multirow{2}{*}{$p=100$} & \multirow{2}{*}{$Q_h$} & sum & \textbf{36.0} & \textbf{76.0} & 91.5 & \textbf{44.5} & 88.0 & \textbf{98.5} \\
    & & max & 30.0 & 71.0 & \textbf{92.0} & 37.5 & \textbf{89.5} & 98.0 \\
    & \multirow{2}{*}{Gro} & $\widehat{\Lambda}_1$ & 22.5 & 47.5 & 78.0 & 19.5 & 54.5 & 80.5 \\
    & & $\widehat{\Lambda}_2$ & 19.5 & 31.5 & 42.0 & 17.5 & 26.5 & 45.0 \\
    \bottomrule
    \end{tabular}
\end{table}

We further assess the performance of the proposed change-point estimation approach in terms of the mean and standard deviation. As shown in Table \ref{tab: consistency}, the change detection using $Q_h$ exhibits the smallest estimation errors across nearly all settings, particularly for smaller $\delta$ and larger $p$. 
When the signal strength $\delta$ increases to 1, all methods except $\widehat{\Lambda}_2$ achieve 100\% power with significantly improved accuracy in change-point estimation, and $Q_h$ continues to demonstrate greater efficiency than $Q_0$ and $\widehat{\Lambda}_1$.
The additional result with $r_s=0.6$, provided in Appendix \ref{E1}, also illustrate the superior consistency of $Q_h$ compared to the other methods for larger signal magnitudes. 

\begin{table}
    \caption{Simulation study on the consistency of the change-point detection procedures with the mean and standard deviation (in parentheses) of $|\hat{\tau} - \tau^*|$'s with $r_s = 0.4$ and $n=100$. }
    \label{tab: consistency}
    \centering
    \begin{tabular}{cccccccc}
    \toprule
    \multirow{2}{*}{Method} & \multicolumn{3}{c}{$\delta = 0.4$} & & \multicolumn{3}{c}{$\delta = 1$}\\
    \cmidrule(r){2-4} \cmidrule(r){6-8}
    & $p=20$ & $p=50$ & $p=100$ & & $p=20$ & $p=50$ & $p=100$\\
    \midrule
    
    $Q_0$ & $\mathbf{3.48_{(3.62)}}$ & $2.52_{(3.08)}$ & $2.38_{(2.63)}$ & & $\mathbf{0.72}_{(1.28)}$ & $0.52_{(1.05)}$ & $0.22_{(0.60)}$\\
    $Q_h$ & $3.55_{(3.73)}$ & $\mathbf{2.43_{(2.76})}$ & $\mathbf{1.95_{(2.54)}}$ & & $\mathbf{0.72_{(1.24)}}$ & $\mathbf{0.50_{(1.01)}}$ & $\mathbf{0.21_{(0.54)}}$\\
    $\widehat{\Lambda}_1$ & $4.12_{(3.88)}$ & $2.76_{(3.21)}$ & $2.58_{(2.69)}$ & & $0.74_{(1.34)}$ & $0.59_{(1.08)}$ & $0.24_{(0.60)}$\\
    $\widehat{\Lambda}_2$ & $5.22_{(4.67)}$ & $4.92_{(4.56)}$ & $4.31_{(4.48)}$ & & $1.83_{(2.53)}$ & $1.53_{(2.30)}$ & $0.81_{(1.17)}$\\
    \bottomrule
    \end{tabular}
\end{table}

\subsection{Results for support recovery}
\begin{figure}
\centering
\includegraphics[width=1\linewidth]{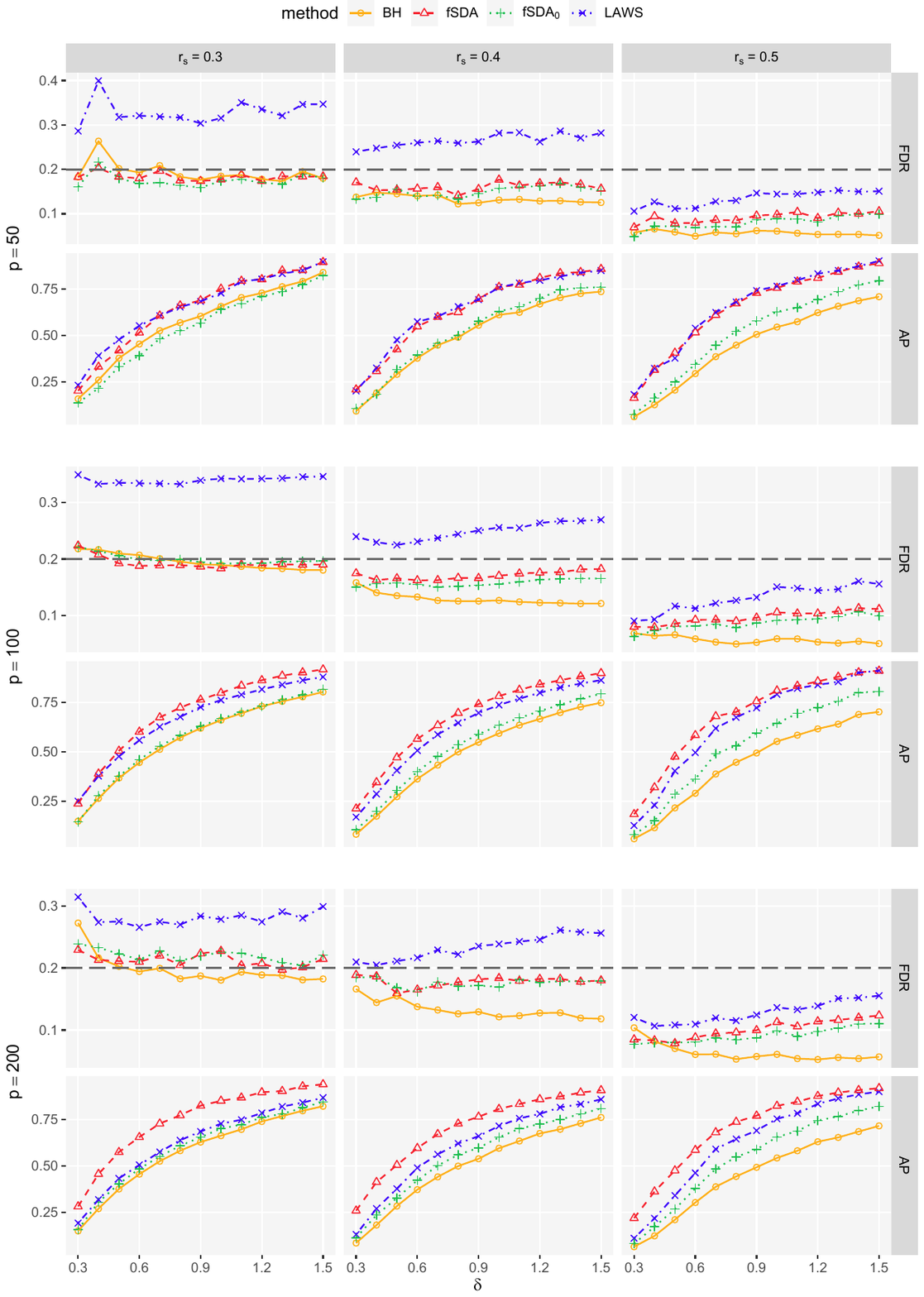}
\caption{The empirical FDR and AP comparison for $n=100$, with the dashed line representing the nominal level $\alpha=0.2$.}
\label{simu: fdr1}
\end{figure}

{\label{SR simu}}
After estimating the change point for each simulated dataset under the alternatives, we assess the performance of support recovery using the following methods: (i) BH procedure,  (ii) LAWS procedure, (iii) the $\text{fSDA}_0$ procedure without kernel aggregation, and (iv) The full fSDA procedure.
Both the BH and LAWS methods employ a $p$-value screening approach based on the asymptotic normality of $\hat\eta_{\hat\tau,r}(\s_j)$ from \eqref{eta_r,s}.
For LAWS, a nonparametric estimator for the prior probability  $\pi(\s_j)$ is employed with the bandwidth selected by the R-package ``kedd''. 
In our fSDA method, the truncation number $R$ and bandwidth $h$ are determined according to Algorithm \ref{alg: FSDA}. 
Under a nominal level at $\alpha=0.2$, these methods are compared based on empirical FDR and average power, defined as $\mbox{AP}=\E(\mbox{TDP})$, across 200 replications. 
Figure \ref{simu: fdr1} illustrates the multiple testing results for $p=50, 100$, and $200$ with $n=100$. Additional results with $n=200$ are provided in Appendix \ref{E1}, showing a similar FDR pattern along with increased power.

From Figure \ref{simu: fdr1}, it is evident that all methods, except LAWS, control the FDR at the nominal level in most settings, with BH being more conservative due to overestimating false positives. 
The FDR for all methods tends to decrease as the spatial magnitude $r_s$ increases, whereas LAWS exhibits substantial FDR inflation when $r_s = 0.3$ and $0.4$. 
When $r_s=0.5$, corresponding to an $80\%$ proportion of the informative set, even LAWS is able to control the FDR. 
It is also noteworthy that inflated FDR levels are observed for both BH and LAWS when the values of $r_s$ and $\delta$ are relatively small.
This deficiency may stem from inaccurate estimation of the $p$-values when signals are weak, further demonstrating the advantage of the proposed $p$-value-free approach.

In terms of average power, the fSDA method consistently outperforms the other approaches across different $r_s$ and $\delta$, 
with its superiority becoming more pronounced as $p$ increases.
Even for $p=50$, the power of fSDA remains comparable to LAWS, which fails to control the FDR when $r_s$ is less than 0.4.
While $\text{fSDA}_0$ also controls the empirical FDR across various scenarios, its power is notably inferior to those of fSDA.
These findings are similar to those from in Table \ref{tab: power}, demonstrating that the kernel statistic effectively aggregates spatial signals, particularly when stations are more densely located.
Additionally, all methods exhibit higher power as the signal strength $\delta$ increases, though varying the magnitude $r_s$ has a less noticeable impact on power. In conclusion, the proposed fSDA method offers the most powerful approach while maintaining effective FDR control, and the results demonstrate robustness across different values of $n$ and $p$.

\section{Empirical application to China's precipitation data}\label{sec: real data}

\begin{figure}
    \centering
    \includegraphics[width=0.8\linewidth]{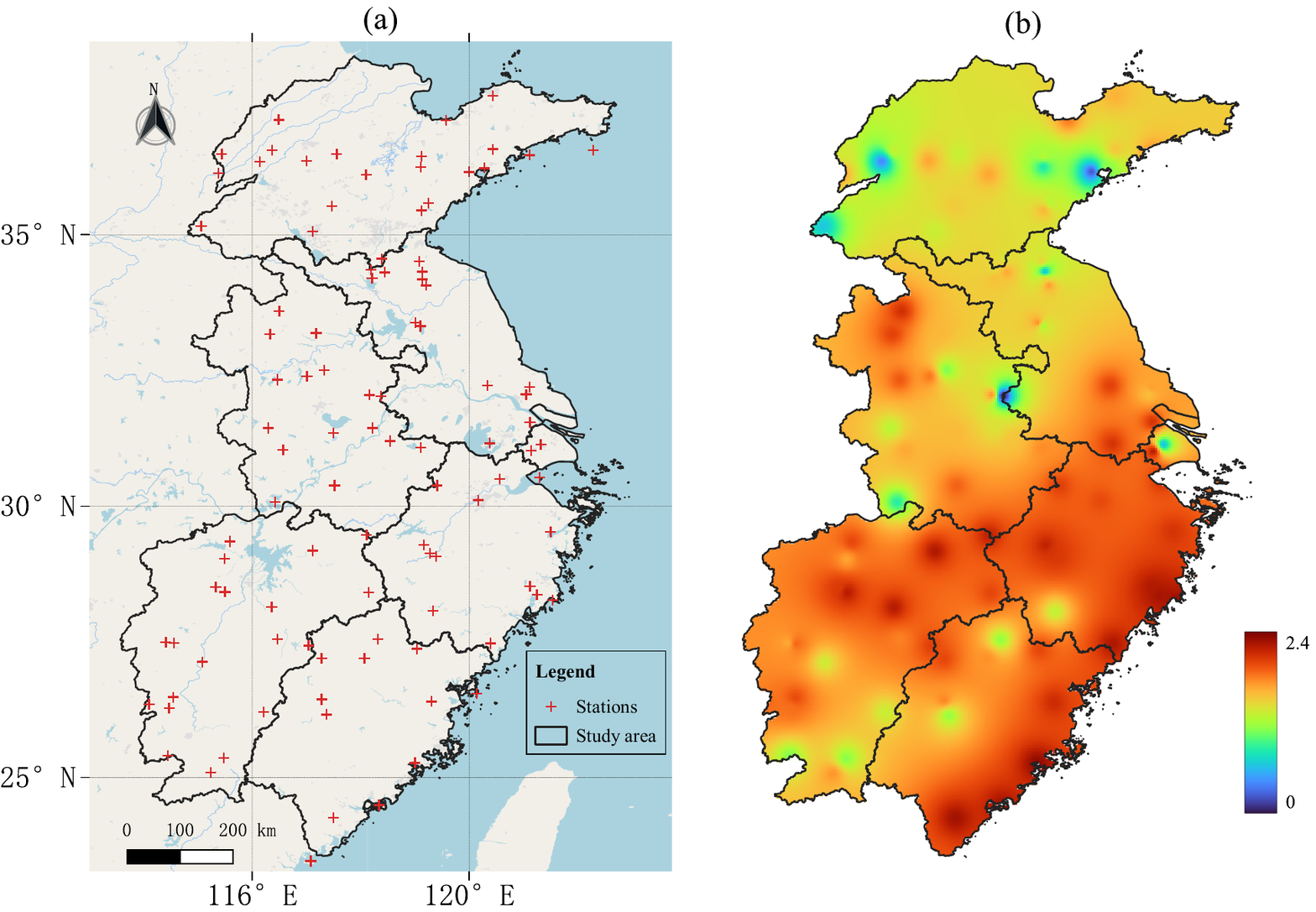}
    \caption{(a) Illustration of 100 meteorological stations in Eastern China. (b) A heatmap of the estimated spatial field of $\|\hat{\mu}_0(\s;\cdot) - \hat{\mu}_1(\s;\cdot)\|_\CT$ before and after 1988.}
    \label{fig: realstations}
\end{figure}

The study of precipitation plays an important role in the fields of climatology, meteorology, and hydrology \citep[e.g.][]{brunsell2010, liu2019change}. 
Precipitation data in China, influenced by Asian monsoon and complex terrain, exhibit intricate spatial and temporal patterns with notable seasonal, interannual, and regional variations \citep{ng2021changes,zhang2019identifying}. The data analyzed in this study were provided by the National Meteorological Information Center, China Meteorological Administration (CMA), consisting of daily precipitation records from over 1500 meteorological stations across mainland China from 1961 to 2013. Following standard environmental practices, we aggregate the daily observations into half-month intervals, resulting in $12\times 2=24$ records per year. Due to the vast and complex landscape in China, we select a subset of $p=100$ meteorological stations in Eastern China for this analysis. 
Panel (a) of Figure \ref{fig: realstations} visualizes the geographical locations of the selected subregion, including the provinces of Shandong, Jiangsu, Anhui, Shanghai, Zhejiang, Jiangxi, and Fujian, which comprise one of the areas with the highest precipitation levels in China.

To mitigate the impact of heavy-tailed distributions, we apply a logarithmic transformation to the original half-monthly precipitation data $Y_i(\s_j; t_m)$ for year $i$ at time point $t_m$, $m = 1, \dots, 24$:
$$ X_i(\bs_j;t_m) = \log_{10}\{Y_i(\bs_j;t_m)+1\},$$ as recommended by \cite{gromenko2017detection}. The transformed temporal records at each location are then projected onto the functional time domain $\CT = [0,1]$ using cubic smoothing splines with six degrees of freedom.
Based on the resulting pre-smoothed spatially index functional dataset $\{X_i(\bs_j;t):i=1961,\dots,2013; j=1,\dots,p; t\in\CT\}$,
our objective is to apply the proposed methodology to detect a potential change-point year and subsequently identify the specific spatial locations where significant mean changes occur.

Before proceeding with the change-detection procedure, it is crucial to evaluate the space-time separability structure of the dataset.
Additional results in Appendix \ref{E2} demonstrate that the hypothesis of strong separability is rejected for our data, whereas the assumption of weak separability is not rejected across different FVE levels. Given the validity of the weakly separable model, we first apply the proposed FPCA procedure outlined in Algorithm \ref{alg: cp detect}.
At a $90\%$ level of FVE, the selected number of eigenfunctions is $\widehat{R}=5$. 
Using the bandwidth selection rule, we set $h=487.1$km, slightly larger than the mean (351.7km) and the median (323.8km) of the pairwise distances among all spatial locations.
It is also noteworthy that the change-detection procedure yields similar results over a reasonable range, specifically with $4 \le \widehat{R} \le 8$ and $300 \mbox{km} \le h \le 700 \mbox{km}$.

We perform the change-point test using both ``max'' and ``sum'' types of $Q_h$, as well as $Q_0$.
As shown in Table \ref{real pvalue}, all tests using $Q_h$ and $Q_0$ reject the null hypothesis, with $Q_h$ yielding slightly smaller $p$-values compared to $Q_0$.
In contrast, the statistics $\widehat\Lambda_1$ and $\widehat\Lambda_2$ from \cite{gromenko2017detection} fail to detect a change-point.
We then plot the function $Q_h(\tau)$ in Figure \ref{fig: real Qh} of Appendix \ref{E2} which identifies the estimated change-year as $\hat\tau=1988$.
This finding is consistent with the results of \cite{liu2019change}, which indicated that change-points in precipitation extremes increasingly occurred around 1990 in the southeastern region of the Hu Huanyong line, a vast and well-known area fully encompassing our selected locations.
Further tests conducted separately on the two segmented periods, as summarized in Table \ref{real pvalue seg}, confirm that no additional change-points are present apart from the year 1988.

\begin{table}
    \centering
    \caption{The $p$-values of change-point tests for China precipitation data.}
    \label{real pvalue}
    \begin{tabular}{cccccccccc}
    \toprule
    & \multicolumn{2}{c}{Gro} & & \multicolumn{2}{c}{$Q_0$} & & \multicolumn{2}{c}{$Q_h$} & \\
    \cmidrule(r){2-3} \cmidrule(r){5-6} \cmidrule(r){8-9}
    & $\widehat{\Lambda}_1$ & $\widehat{\Lambda}_2$ & & sum & max & & sum & max & \\
    \midrule
    & 0.370 & 0.385 & & 0.035 & 0.005 & & 0.025 & $<0.001$ & \\
    \bottomrule
    \end{tabular}
\end{table}

\begin{table}
    \centering
    \caption{The $p$-values of change-point test for different segment of years.}
    \label{real pvalue seg}
    \begin{tabular}{cccccccccc}
    \toprule
    \multirow{2}{*}{Segment}& \multicolumn{2}{c}{Gro} & & \multicolumn{2}{c}{$Q_0$} & & \multicolumn{2}{c}{$Q_h$} & \\
    \cmidrule(r){2-3} \cmidrule(r){5-6} \cmidrule(r){8-9}
    & $\widehat{\Lambda}_1$ & $\widehat{\Lambda}_2$ & & sum & max & & sum & max & \\
    \midrule
    1961-1988 & 0.915 & 0.885 & & 0.495 & 0.675 & & 0.890 & 0.945 & \\
    1989-2013 & 0.715 & 0.755 & & 0.520 & 0.665 & & 0.665 & 0.750 & \\
    \bottomrule
    \end{tabular}
\end{table}

To figure out the locations that contributed to the changes in $\hat\tau=1988$, we begin by exploring the spatial distribution of the mean difference.
Specifically, an ordinary kriging interpolation approach is applied to the quantity $\|\hat{\mu}_0(\s;\cdot) - \hat{\mu}_1(\s;\cdot)\|_\CT$ with an exponential variogram, where 
\begin{equation}\label{mean diff}
    \hat{\mu}_0(\s;t) = \hat{\tau}^{-1}\sum_{i=1}^{\hat{\tau}} X_{i}(\s;t) \mbox{ and }\hat{\mu}_1(\s;t) = (n-\hat{\tau})^{-1}\sum_{i=\hat{\tau} + 1}^n X_{i}(\s;t)
\end{equation}
represent the estimated mean functions for the periods before and after the detected change-point, respectively.
As indicated in Panel (b) of Figure \ref{fig: realstations}, aside from a few areas where the kriging estimates of this mean-difference is nearly zero, most of Eastern China experienced notable changes in the amount of precipitation,
with a more pronounced increasing trend in the southern part of the selected region. While this method provides a rough and intuitive sketch of the spatial patterns, it falls short of rigorously identifying the specific subsets of locations where significant changes have occurred. 

\begin{figure}
    \centering
    \includegraphics[width=0.8\linewidth]{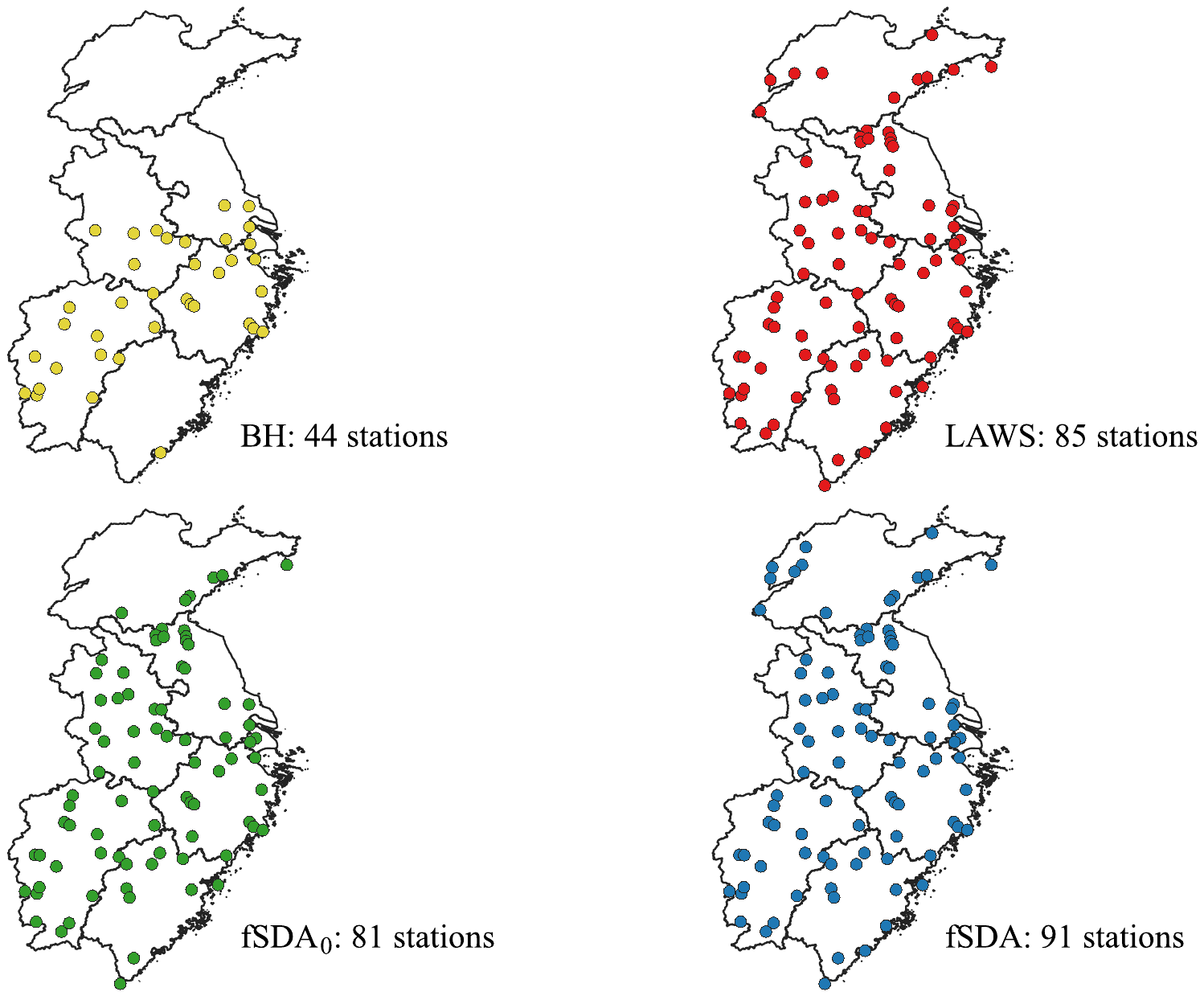}
    \caption{The recovery sets of meteorological stations using various multiple testing approaches.}
    \label{fig: realchange}
\end{figure}

To accurately identify the regions contributing to the observed changes, we perform the multiple testing procedure as outlined in \eqref{test: mcp} across these 100 stations.
Using an FDR level of $\alpha = 0.2$, we implement the proposed fSDA approach following Algorithm \ref{alg: FSDA} and compare it with the BH, LAWS, and fSDA$_0$ methods. As shown in the detection results in Figure \ref{fig: realchange}, fSDA recovers the largest set of contributing stations, while BH identifies the fewest.
All methods highlight a majority of stations in Zhejiang, Jiangxi, and the southern part of Anhui and Jiangsu, areas that display the most substantial differences in precipitation, as shown in Figure \ref{fig: realstations}(b).
In comparison to BH, the other three methods identify additional stations in Fujian, a highly informative area in the heatmap of Figure \ref{fig: realstations}(b).
Both fSDA$_0$ and fSDA detect over 80 stations, with fSDA capturing more stations located contiguously in northwest Shandong.
The result from LAWS, which detects 85 stations, shows some disparity compared to fSDA, particularly in Shandong and northern Anhui. Given the superior FDR control of fSDA as demonstrated by our simulation study, we recommend its discovery set as a more accurate identification of the changing regions. Our detection results align with the notable north-south disparity around the Yangtze River observed by \cite{liu2019change} and pinpoint specific locations exhibiting significantly increasing trends.
However, further research, including a more detailed environmental analysis of the specific precipitation changes at each location, is necessary to deepen our understanding of these findings.

\begin{appendix}
\section{Main Lemmas}\label{appA}
In this section, we provide several key lemmas that characterize the convergence properties of the proposed detection statistics and serve as fundamental components in the proof of the main theorems. 
We begin with a weak convergence result for the i.i.d. sum of FPC scores.
\begin{lemma}\label{lem: CLT}
Suppose that Assumptions \ref{assump: L2}--\ref{assump: eigengap} hold. Then
the FPC scores $\{ \xi_{ir}(\bs_j): i=1,\dots,n; \,r=1,\dots,R; \,j=1,\dots,p \}$ satisfy the functional central limit theorem (with respect to the Skorokhod space $D^{pR}[0,1]$) that 
\begin{equation}\label{functional CLT}
n^{-1 / 2} \sum_{1\leq i \leq nx } \mbox{vec}\left(\begin{array}{ccc}
 \xi_{i1}(\s_1) & \ldots & \xi_{iR}(\s_1) \\
\vdots & \ddots & \vdots \\
 \xi_{i1}(\s_p) & \ldots & \xi_{iR}(\s_p) \\
\end{array}\right) \stackrel{d} \rightarrow \mathbf{W}(x)
\end{equation}
for $x \in (0,1)$,
where $\mathbf{W}(x)$ is a $pR$-dimensional vector of Brownian motions with block-diagonal covariance matrix  $\BSigma=\mbox{diag}\left(\BSigma_1,\dots,\BSigma_R\right)$.
\end{lemma}

Define that $\Beta_j^* = (\eta_{j,1}^*,\dots,\eta_{j,R}^*)\trans$ where $\eta_{j,r}^*=\lambda_{r,j}^{-1/2}\left\langle \widehat\Delta_{\tau^*}\left(\s_j;\cdot\right), \psi_r(\cdot)\right\rangle_\CT$ denotes the counterpart of $\hat\eta_{j,r}:=\hat\eta_{\hat\tau,r}(\s_j)$ using the true FPC scores.
Define similarly that $\tilde\Beta_j^* = (\tilde{\eta}_{j,1}^*,\dots,\tilde{\eta}_{j,R}^*)\trans$ where $\tilde{\eta}_{j,r}^*={\sum_{k}K_h(\s_k-\s_j)\eta_{k,r}^*}/{\sum_{k}K_h(\s_k-\s_j)}$. 
The next lemma describes the moment and uniform bounds for $\Beta_j^*$ and  $\tilde\Beta_j^*$.
\begin{lemma}\label{lem: bound for eta*}
Suppose that the process $\BX_i(\cdot)$ is weakly separable under Assumption \ref{assump: eigengap}. \\
(a) The covariance matrix of $\Beta_j^*$, denoted as $\BSigma_j^*$, is diagonal with the entries equal to $\theta_0(1-\theta_0)(1-2\theta_0)$, where we assume $\theta_0<1/2$ without loss of generality. \\
(b) Suppose Assumption \ref{assump: sr moment} hold. Then for each {$1\le j \le p$},
\begin{align*}
    &\operatorname{Pr}\left(\left\|\Beta_j^* -\sqrt{n} \theta_0(1-\theta_0)\Bdelta_j^* \right\| > \sqrt{C_\eta\log p} \right) = o(1/p)
\end{align*}
for some large $C_\eta>0$, and the same argument holds for $\tilde\Beta_j^*$ under Assumption \ref{assump: kernel}.
\end{lemma}

Define $W_j^*:= \sum_{r=1}^R \eta_{j,r}^{*O}\tilde\eta_{j,r}^{*E}$, which denotes the counterpart of the statistic \eqref{FSDA statistic} based on $\eta_{j,r}^*$ and  $\tilde{\eta}_{j,r}^*$.  
Lemma \ref{lem: sym for true W} provides the asymptotic symmetry property of $W_j^*$.
\begin{lemma}\label{lem: sym for true W}
Suppose Assumptions \ref{assump: eigengap}, \ref{assump: p}--\ref{assump: sr moment} and \ref{assump: dependence}--\ref{assump: kernel} hold. Then we have 
\begin{equation}\label{sym property true W}
    \sup _{0 \leq t \leq t^*}\left|\frac{\sum_{j\in\mathcal{A}^c} \bI\left(W_j^* \geq t\right)}{\sum_{j\in\mathcal{A}^c} \bI\left(W_j^* \leq-t\right)}-1\right| = o_p(1),
\end{equation}
for some upper bound $t^*$ with $t^*\leq C_\eta \log p$.
\end{lemma}

Lemma \ref{lem: bound for est} establishes the uniform bounds for the estimators used in $W_j$.
\begin{lemma}\label{lem: bound for est}
Suppose that $\BX_i(\cdot)$ is weakly separable under Assumptions \ref{assump: L2}--\ref{assump: eigengap} and Assumption \ref{assump: sr moment}. \\
(a) For each $1\le r \le R$, we have
\begin{equation*}
\max_{1\leq j \leq  p}\left|\hat{\lambda}_{r,j}^{-1/2}-\lambda_{r,j}^{-1/2}\right| =  O_p\left(\left(\log p /n\right)^{1/2}\right).
\end{equation*}
(b) For each $r=1,\dots,R$, we have
\begin{equation*}
 \max_{1\leq j \leq  p} \left|\left\langle \widehat\Delta_{\hat\tau}(\s_j) -  \widehat\Delta_{\tau^*}(\s_j),\psi_r \right\rangle_\CT\right| =  O_p\left(\left(\log p /n\right)^{1/2}\right).
\end{equation*}
(c) Define that $\mA^c_h=\mA^c \cup \{k:|k-j|\leq h \mbox{ where } j\in \mA^c \}$ and suppose Assumption \ref{assump: kernel} holds. Then for each $r=1,\dots,R$,
\begin{equation*}
\max_{j\in\mA^c_h}\left|\hat\eta_{j,r}-\eta_{j,r}^*\right| = O_p\left(\log p /n^{1/2}\right).
\end{equation*}
(d) Under Assumption \ref{assump: p}, we have
\begin{equation*}
\max_{j\in\mA^c}\left| W_j - W_j^* \right| = o_p \left( (\log p)^{-1/2} \right)
\end{equation*}
\end{lemma}

\setcounter{section}{1}

\section{Technical Details}

\subsection{Proof of Theorem \ref{thm: consistency of estimators}}\label{sec: proof consis}
We begin by introducing some standard definitions for the covariance operators in Hilbert spaces. Let $\CB_{HS}\{ L^2(\CT)\}$ denote the space of Hilbert-Schmidt operators on $L^2(\CT)$, which is itself a Hilbert space with the inner product $\langle \CC_1,\CC_2\rangle_{HS}=\sum_{r=1}^\infty \langle \CC_1 e_r,\CC_2 e_r\rangle$ for any orthonormal basis $\{e_r\}$ in $L^2(\CT)$, and the induced norm $\Vert \cdot \Vert_{HS}$. For any $f_1,f_2\in L^2(\CT)$, $f_1\otimes f_2$ denotes the operator in  $\CB_{HS}\{ L^2(\CT)\}$ defined by $(f_1\otimes f_2)(g)=\langle f_1,g\rangle f_2$ for any $g\in L^2(\CT)$. Let $(L^2(\CT))^p$ denote the space of $p$-dim functions with inner product $\langle\bm{f},\bm{g}\rangle_{L_p}=\sum_{j=1}^{p}\int_{\CT} f_j(t)g_j(t)dt$ for $\bm{f},\bm{g}\in (L^2(\CT))^p$,
and the norm $\|\cdot\|_{L_p}=\langle \cdot,\cdot \rangle_{L_p}^{1/2}$. 
Recall that $\BE_i=\{\varepsilon_i(\s_1),\dots,\varepsilon_i(\s_p)\}\trans$ is a $p$-dim mean-zero functional process in $(L^2(\CT))^p$, 
where $\varepsilon_i(\s_j,\cdot)$ is abbreviated as $\varepsilon_i(\s_j)$ for the objective in $L^2(\CT)$.
Define $\CC_{jk}=\E\{\varepsilon_i(\s_j) \otimes \varepsilon_i(\bs_k)\}$ for any $j,k=1,\ldots,p$, which is the covariance operator for $\varepsilon_i(\s_j) $ and $\varepsilon_i(\s_k) $ in $\CB_{HS}\{ L^2(\CT)\}$. Consequently, the covariance function $C(s_j,t;s_k,t')$, as defined in Section \ref{sec: weak sepa}, can be regarded as the covariance kernel of the operator $\CC_{jk}$.
Moreover, the marginal covariance function $H(t,t')$ can be seen as the kernel of the marginal covariance operator defined by 
\begin{equation*}
	\CH =p^{-1}\sum_{j=1}^{p}\CC_{jj}.
\end{equation*}
We refer to Chapters 7.2 and 7.3 of \cite{Hsing2015Theoretical} for more details on the (cross) covariance operators in Hilbert space.

When the $p$-dim functional process $\BE_i$ is weakly separable, as defined in \eqref{WS model}, the covariance operator of $\CC_{jj}$ entails the eigen-decomposition
$\CC_{jk}= \sum_{r=1}^{\infty} \sigma_{r}(\s_j,\s_k) \, \psi_{r}\otimes \psi_{r}$, which implies that 
\begin{equation*} 
	\CH=\sum_{r=1}^{\infty}\,\lambda_r \, \varphi_r \otimes \varphi_r,
\end{equation*}
where $\{\lambda_r\}_{r=1}^R$ are the eigenvalues of the covariance function $H(t,t')$.
Let $\widehat \CH$ denote the operator corresponding to the empirical covariance $\widehat H(t,t')$ defined in \eqref{hatH}. The key issue in proving Theorem \ref{thm: consistency of estimators} is to establish the uniform convergence rate of $\|\widehat \CH - \CH\|_{HS}$, as provided in \eqref{consis of H}. It can be shown that 
\begin{equation*}
	\widehat{H}\left(t,t'\right) - H\left(t,t'\right) = \Gamma_1 + \Gamma_2,
\end{equation*}
where
\begin{align*}
	\Gamma_1 :=&  \frac{1}{p (n-1)} \left\{ \sum\limits_{i=2}^{n-1}  \left\langle \BE_i\left(t\right), \BE_i\left(t'\right) \right\rangle_p \right\} - \frac{1}{p}\E \left\{ \left\langle \BE_i(t),\BE_i(t')  \right\rangle_p \right\},\\
	\Gamma_2 :=&  \frac{1}{2 p (n-1)}\left\{\left\langle \BE_1\left(t\right), \BE_1\left(t'\right) \right\rangle_p + \left\langle \BE_n\left(t\right), \BE_n\left(t'\right) \right\rangle_p\right.\\
	& + \left.\left\langle \BDelta\left(t\right), \BDelta\left(t'\right) \right\rangle_p + 2\left\langle \BDelta\left(t\right), \BE_{\tau^* + 1}\left(t'\right) -  \BE_{\tau^*}\left(t'\right)\right\rangle_p\right\}, 
\end{align*}
and $\BDelta = \{\Delta(\s_1),\dots,\Delta(\s_p)\}\trans \in (L^2(\CT))^p $.
Denote $\widehat\CC_{jk} = (n-1)^{-1}\sum_{i=2}^{n} \varepsilon_i(\s_j) \otimes \varepsilon_i(\s_k)  $ for any $j,k=1,\ldots,p$.
It follows from Assumption 1 and Theorem $8.1.2$ of \cite{Hsing2015Theoretical} that $\widehat\CC_{jk} - \CC_{jk}$ converges weakly to a mean-zero Gaussian random element in $\CB_{HS}\{ L^2(\CT)\}$, 
which implies that {$\widetilde\Delta_{jk} := \widehat\CC_{jk} - \CC_{jk}=O_p(n^{-1/2})$}. By definition we further set $\Gamma_1=p^{-1} \sum_{j=1}^p (\widehat\CC_{jj} - \CC_{jj})$, and it follows that
\begin{equation*}
	\E \left\|\Gamma_1\right\|_{HS}^2  \leq \frac{1}{p^2}\sum_{j=1}^p \E \left\| \widetilde\Delta_{jj}\right\|^2_{HS} + \frac{1}{p^2}\sum_{j\neq k} \E \left\|\widetilde\Delta_{jj} \widetilde\Delta_{kk}\right\|_{HS}.
\end{equation*}
The first term on the right hand is $O((pn)^{-1})$ since $\E \| \widetilde\Delta_{jj}\|^2_{HS} = O(n^{-1})$.
For the second term, we have $\E \|\widetilde\Delta_{jj} \widetilde\Delta_{kk}\|_{HS}\leq(\E \|\widetilde\Delta_{jj}\|_{HS}^2 \E \|\widetilde\Delta_{kk}\|_{HS}^2)^{1/2} =O(n^{-1})$ by Cauchy--Schwarz inequality.
It therefore follows that 
\begin{equation*}
	\|\Gamma_1\|_{HS} = O_p\left( \frac{1}{\sqrt{pn}}+\frac{1}{\sqrt{n}} \right). 
\end{equation*}
For $\Gamma_2$, note that by Assumption \ref{assump: moment} we have $\left\langle \BDelta, \BE_{\tau^* + 1} -  \BE_{\tau^*}\right\rangle_{L_p} = O_p(1)$ and $p^{-1}\|  \BDelta \|_{L_p} = O_p(1)$.
It follows that $\|\Gamma_2\|_{HS} = O_p(n^{-1})$, and then 
\begin{equation}\label{consis of H}
	\left\|\widehat\CH - \CH \right\|_{HS}=O_p\left(n^{-1/2}\right).
\end{equation}
We emphasize that this result holds under both the null and the alternative hypotheses.  Hence, by Lemmas 4.2 and 4.3 of \cite{bosq2000linear}, it follows that for each $r=1,\dots,R$, $|\hat\lambda_r-\lambda_r| =o_p(1)$ and
\begin{equation}\label{psi order}
	\left\| \hat\psi_r - \psi_r \right\|_\CT \leq 2\sqrt{2} c_R^{-1} \left\|\widehat\CH - \CH \right\|_{HS},
\end{equation}
where $c_R = \inf_{r=1,\dots,R} \left(\lambda_r - \lambda_{r+1}\right)$,
which implies that $\| \hat\psi_r - \psi_r \|_\CT=O_p(n^{-1/2})$.

We finally prove the element-wise convergence of $\widehat \BSigma_r$. 
Define $\BDelta_i^\varepsilon = \{\Delta_i^\varepsilon(\s_1),\dots,\Delta_i^\varepsilon(\s_p)\}\trans$ for $i=1,\dots,n-1$, where $\Delta_i^\varepsilon(\s_j) = \varepsilon_{i+1}(\s_j) - \varepsilon_i(\s_j)$.
Observe by \eqref{hat Sigma_r} that 
\begin{equation}\label{decom for sigma}
	\begin{aligned}
		\widehat \BSigma_r = & 
		\frac{1}{2(n-1)}\sum_{i=1}^{n-1} \left\langle \BDelta_i^\varepsilon,\hat\psi_r \right\rangle_\CT \left\langle \BDelta_i^\varepsilon,\hat\psi_r \right\rangle_\CT\trans - 
		\frac{1}{2(n-1)} \left\langle \BDelta_{\tau^*}^\varepsilon,\hat\psi_r \right\rangle_\CT \left\langle \BDelta_{\tau^*}^\varepsilon,\hat\psi_r \right\rangle_\CT\trans \\
		& + \frac{1}{2(n-1)}\left\langle  \left( \BX_{\tau^*+1}-\BX_{\tau^*} \right),\hat\psi_r \right\rangle_\CT \left\langle \left( \BX_{\tau^*+1}-\BX_{\tau^*} \right), \hat\psi_r \right\rangle_\CT\trans,  
	\end{aligned}
\end{equation}
and it can be shown that $\widehat \BSigma_r(j,k) =  \BTheta_1(j,k) + \BTheta_2(j,k) + \BTheta_3(j,k) + o_p(1)$ for each $j,k$, where
\begin{align*}
	\BTheta_1 &:= \frac{1}{2(n-1)}\sum_{i=1}^{n-1} \left\langle \BDelta_i^\varepsilon, \psi_r \right\rangle_\CT \left\langle \BDelta_i^\varepsilon, \psi_r \right\rangle_\CT\trans, \\
	\BTheta_2 &:= \frac{1}{2(n-1)}\sum_{i=1}^{n-1} \left\langle \BDelta_i^\varepsilon, \left( \hat\psi_r - \psi_r\right) \right\rangle_\CT \left\langle \BDelta_i^\varepsilon, \left( \hat\psi_r - \psi_r\right) \right\rangle_\CT\trans, \\
	\BTheta_3 &:= \frac{1}{(n-1)}\sum_{i=1}^{n-1} \left\langle \BDelta_i^\varepsilon, \psi_r \right\rangle_\CT \left\langle \BDelta_i^\varepsilon, \left( \hat\psi_r - \psi_r\right) \right\rangle_\CT\trans.
\end{align*}
Abbreviate $\Delta_i^\varepsilon(\s_j)$ as $\Delta_{ij}^\varepsilon$ for simplicity.
For $\BTheta_1$, the law of large numbers implies that $\BTheta_1(j,k) \stackrel{p}{\rightarrow} \E \{\langle \BDelta_{ij}^\varepsilon, \psi_r \rangle_\CT \left\langle \BDelta_{ik}^\varepsilon, \psi_r \right\rangle_\CT\}/2=\sigma_r(\s_j,\s_k)$, since $\langle \BDelta_{ij}^\varepsilon, \psi_r \rangle_\CT$ forms a 1-dependent, mean-zero moving-average sequence in $i$ with $\E \{\langle \BDelta_{ij}^\varepsilon, \psi_r \rangle_\CT \langle \BDelta_{ik}^\varepsilon, \psi_r \rangle_\CT\}  = 2\E \{ \xi_{ir}(\s_j) \xi_{ir}(\s_k)\}$. For $\BTheta_2$, it follows from Cauchy--Schwarz inequality that 
$\BTheta_2(j,k)\leq \frac{1}{2(n-1)}\sum_{i=1}^{n-1}\|\Delta_{ij}^\varepsilon\|_\CT\|\Delta_{ik}^\varepsilon\|_\CT\|\hat\psi_r - \psi_r\|_\CT^2$, which implies that $\BTheta_2(j,k) = o_p(1)$ by Assumption \ref{assump: moment} and \eqref{psi order}.
Similarly we can derive that $\BTheta_3(j,k) = o_p(1)$, which completes the proof for $\widehat \BSigma_r(j,k) = \BSigma_r(j,k) + o_p(1)$.

\subsection{Proof of Theorem \ref{thm: asy null of Qh}}\label{sec: proof null}
Recall the definitions that $\BSigma_r=\{\sigma_r(\s_j,\s_k) _{j,k=1,\dots,p}\}$, $\BP_r = \{ \rho_r(\s_j,\s_k)_{j,k=1,\dots,p}\}$, and let ${\bm{\xi}}^*_{ir}= ( \xi^*_{ir,1},\dots,\xi^*_{ir,p})\trans$ where  $\xi^*_{ir,j} = \lambda_{r,j}^{-1/2}\xi_{ir}(\s_j)$. 
It follows from Lemma \ref{lem: CLT} and the continuous mapping theorem that, for each $1\leq r \leq R$, 
\begin{equation*}
	n^{-1}\left(\sum_{1\leq i\leq nx } \bm{\xi}_{ir}^*-x \sum_{1\leq i \leq n} \bm{\xi}_{ir}^* \right)\trans{\BK}_{h} \left(\sum_{1\leq i\leq nx } \bm{\xi}_{ir}^*-x \sum_{1\leq I \leq n}  \bm{\xi}_{ir}^* \right) \stackrel{d}{\rightarrow} {\BB_{r}\trans(x) \BK_h \BB_{r}(x)},
\end{equation*}
where
$\BB_{r}(\cdot) = \{B_{1r}(\cdot),\dots,B_{pr}(\cdot)\}\trans$ is a $p$-dimensional Brownian bridge with covariance matrix $\BP_r$. Therefore, we have
\begin{equation}\label{weak conv true}
	\begin{aligned}
		&\sum_{r=1}^R  n^{-1} \left(\sum_{1\leq i\leq nx } \bm{\xi}_{ir}^*-x \sum_{1\leq i \leq n} \bm{\xi}_{ir}^* \right)\trans{\BK}_{h} \left(\sum_{1\leq i\leq nx } \bm{\xi}_{ir}^*-x \sum_{1\leq i \leq n}  \bm{\xi}_{ir}^* \right)\\
		\stackrel{d}{\rightarrow}& 
		\sum_{r=1}^R {\BB_{r}\trans(x) \BK_h \BB_{r}(x)},
	\end{aligned}
\end{equation}

Define that $\hat\xi_{ir}(\s_j) = \hat\lambda_{r,j}^{-1/2} \langle \varepsilon_{i}(\s_j), \hat\psi_r \rangle_\CT$ and
$\hat{\bm{\xi}}_{ir}=\{\hat\xi_{ir}(\s_1),\dots,\hat\xi_{ir}(\s_p)\}\trans$.
We then study the impact of replacing $ \bm{\xi}_{ir}^*$ in \eqref{weak conv true} by $\hat{\bm{\xi}}_{ir}$.
Note that under null hypothesis, we have
\begin{align}\label{eta null}
	\hat{\eta}_{\tau,r}(\s_j) &=  \hat\lambda_{r,j}^{-1/2}  \left\langle{\widehat\Delta}_{\tau}(\s_j), \hat{\psi}_{r} \right\rangle_\CT = 
	\frac{1}{\sqrt{n}} \left\{\sum_{i=1}^{\tau} \hat\xi_{ir}(\s_j) -  \frac{\tau}{n} \sum_{i=1}^{n} \hat\xi_{ir}(\s_j) \right\}.
\end{align}
Let $\bm{\lambda}_r=(\lambda_{r,1},\dots,\lambda_{r,p})\trans$,  $\hat{\bm{\lambda}}_r=(\hat\lambda_{r,1},\dots,\hat\lambda_{r,p})\trans$, and $\tilde\BE_{i,r} = \{ \lambda_{r,1}^{-1/2}\varepsilon_i(\s_1),\dots,\lambda_{r,p}^{-1/2} \varepsilon_i(\s_p)\}\trans$.
Observe that
\begin{align*}
	& \sup_{0<x<1} n^{-1/2} \left| \sum_{1\leq i \leq nx} \hat{\bm{\xi}}_{ir} - \sum_{1 \leq i \leq nx} \bm{\xi}_{ir}^* \right|  \\
	=& \sup_{0<x<1} n^{-1/2} \left| \sum_{1\leq i \leq nx} \hat{\bm{\lambda}}_{r}^{-1/2} \langle \BE_i,\hat\psi_r \rangle - \sum_{1\leq i \leq nx}  {\bm{\lambda}}_{r}^{-1/2}\langle \BE_i,\psi_r \rangle\right| \\
	\leq & \sup_{0<x<1} \left| \left(\hat{\bm{\lambda}}_{r}^{-1/2}-{\bm{\lambda}}_{r}^{-1/2}\right)\left\langle n^{-1/2}  \sum_{1\leq i \leq nx} \BE_i,\hat\psi_r \right\rangle  \right|\\
	&+\sup_{0<x<1} \left| {\bm{\lambda}}_{r}^{-1/2}\left\langle n^{-1/2}  \sum_{1\leq i \leq nx} \BE_i,\hat\psi_r - \psi_r \right\rangle \right|,
\end{align*}
where the first term is $o_p(1)$ due to the uniform consistency of $\hat{\bm{\lambda}}_r$ by Theorem \ref{thm: consistency of estimators} together with 
\begin{equation}\label{partial sum e_i}
	\sup_{0<x<1}  \left\| n^{-1/2} \sum_{1 \leq i \leq nx} \BE_i \right\|_p = O_p(1),
\end{equation}
and the second term is $o_p(1)$  due to the $l_2$ consistency of $\hat\psi_r$ in Theorem \ref{thm: consistency of estimators} together with 
\begin{equation}\label{partial sum sd e_i}
	\sup_{0<x<1}  \left\| n^{-1/2} \sum_{1 \leq i \leq nx} \tilde\BE_{i,r} \right\|_p = O_p(1).
\end{equation}
Here equations \eqref{partial sum e_i} and \eqref{partial sum sd e_i} follow from Assumptions \ref{assump: moment}--\ref{assump: eigengap} and the weak convergence in $D^{p} \{ [0,1], L^2(\CT)\}$ of the partial sum processes $\sum_{1 \leq i \leq nx} \BE_{i}$ and $\sum_{1 \leq i \leq nx} \tilde\BE_{i,r}$. Consequently, we obtain the result that $ \sup_{0<x<1} n^{-1/2} |\sum_{1\leq i \leq nx} \hat{\bm{\xi}}_{ir} - \sum_{1 \leq i \leq nx} \bm{\xi}_{ir}^*|=o_p(1) $, which in turn implies that 
\begin{equation}\label{gap xi}
	\left\| \left(\sum_{1\leq i\leq nx } \hat{\bm{\xi}}_{ir}-x \sum_{1\leq i \leq n} \hat{\bm{\xi}}_{ir}\right) - \left(\sum_{1\leq i\leq nx } \bm{\xi}_{ir}^*-x \sum_{1\leq i \leq n} \bm{\xi}_{ir}^*\right) \right\|_p = o_p(n^{-1/2})
\end{equation}

Combining \eqref{weak conv true} and \eqref{gap xi}, we have
\begin{equation}\label{weak conv est}
	\begin{aligned}
		&\sum_{r=1}^R  n^{-1} \left(\sum_{1\leq i\leq nx } \hat{\bm{\xi}}_{ir}-x \sum_{1\leq i \leq n} \hat{\bm{\xi}}_{ir} \right)\trans{\BK}_{h} \left(\sum_{1\leq i\leq nx } \hat{\bm{\xi}}_{ir}-x \sum_{1\leq i \leq n}  \hat{\bm{\xi}}_{ir} \right)\\ \stackrel{d}{\rightarrow} &
		\sum_{r=1}^R {\BB_{r}\trans(x) \BK_h \BB_{r}(x)},
	\end{aligned}
\end{equation}
which completes the proof of Theorem \ref{thm: asy null of Qh}.

\subsection{Proof of Theorem \ref{thm: asy alter of Qh}}\label{sec: proof alter}
Let $\Bmu_i=\{ \mu_i(\s_1),\dots,\mu_i(\s_p)\}\trans$ and recall the definition of $\BX_i$. Note that for $x\in(0,1)$,
\begin{equation}\label{diff sum X_i}
	\sum_{1\leq i \leq nx} \BX_i - x \sum_{1\leq i \leq n} \BX_i =\left( \sum_{1\leq i \leq nx} \BE_i - x \sum_{1\leq i \leq n} \BE_i \right) + \left( \sum_{1\leq i \leq nx} \Bmu_i - x \sum_{1\leq i \leq n} \Bmu_i \right),
\end{equation}
where the first term is $O_p(n^{1/2})$ by \eqref{partial sum e_i}.
Examining the second term we have
\begin{equation*}\label{diff sum mu}
	\sum_{1\leq i \leq nx} \Bmu_i - x \sum_{1\leq i \leq n} \Bmu_i = 
	\begin{cases}
		nx(1-\theta_0) \BDelta, & 0<x\leq \theta_0, \\
		n(1-x)\theta_0 \BDelta, & \theta_0< x< 1,
	\end{cases}
\end{equation*}
where $\theta_0$ and $\BDelta$ are defined in Section \ref{sec: cd proce}.

Define that $\hat{\Beta}_{r}(x) = \{{\hat{\eta}}_{r,1}(x),\dots,{\hat{\eta}}_{r,p}(x)\}\trans$
where $\hat{\eta}_{j,r}(x) = \hat\lambda_{r,j}^{-1/2}\langle\sum_{1\leq i \leq nx} X_i(\s_j) - x \sum_{1\leq i \leq n} X_i(\s_j) ,\hat\psi_r \rangle_\CT / \sqrt{n}$,
and 
$\BD_r=\mbox{diag}(\lambda_{r,1}^{-1/2} ,\dots,\lambda_{r,p}^{-1/2})$. 
It then follows from Theorem \ref{thm: consistency of estimators}, Slutsky's Theorem and functional continuous mapping theorem that for $0<x\leq \theta_0$,
\begin{equation}\label{gx1}
	\hat{\Beta}_{r}(x) = \sqrt{n} x(1-\theta_0)\BD_r\BDelta_r + O_p(1),
\end{equation}
which is uniformly true in $x\in(0,1)$.  
Similarly, for $x>\theta_0$, we have 
\begin{equation}\label{gx2}
	\hat{\Beta}_{r}(x) = \sqrt{n} \theta_0(1-x)\BD_r\BDelta_r + O_p(1).
\end{equation}
It then follows from \eqref{gx1} and \eqref{gx2} that
\begin{equation}\label{Q_h lim}
	\sup_{0<x<1} \left| \frac{1}{n}{\sum_{r=1}^R}\hat{\Beta}_r\trans(x) \widehat\BD_r\trans {\BK_h} \widehat\BD_r \hat{\Beta}_r(x) - g(x) \right| = o_p(1),
\end{equation}
where $\widehat\BD_r=\mbox{diag}\left(\hat\lambda_{r,1}^{-1/2},\dots,\hat\lambda_{r,p}^{-1/2}\right)$ and 
\begin{equation*}
	g(x) = \begin{cases}
		x^2(1-\theta_0)^2\sum\limits_{r=1}^{R} \BDelta_r\trans \BD_r\trans {\BK_h} \BD_r \BDelta_r, & 0< x \leq \theta_0, \\
		\theta_0^2(1-x)^2\sum\limits_{r=1}^{R} \BDelta_r\trans \BD_r\trans {\BK_h} \BD_r \BDelta_r, & \theta_0 < x < 1.
	\end{cases}
\end{equation*}
It then follows from Theorem 1 and the definition of $Q_h(\tau)$ that
$n^{-1} Q_h^{max} = \sup_{0<x<1} g(x) + o_p(1)$, 
and the conclusion for $Q_h^{sum}$ can be similarly obtained.

\subsection{Proof of Corollary \ref{thm: cp consis}}\label{sec: proof change est}
Denote $\bm\Gamma_r =  \BD_r\trans {\BK_h} \BD_r$ and $\widehat{\bm\Gamma}_r =  \widehat\BD_r\trans {\BK_h} \widehat\BD_r$.
Under the Assumptions in Theorem \ref{thm: asy null of Qh},
we have $g_R^* := \sum_{r=1}^{R} \BDelta_r\trans \bm\Gamma_r \BDelta_r > 0$,
thus the continuous function $g(x)$ has a unique maximum at $x = \theta_0$. Then the uniform convergence in \eqref{Q_h lim} implies that $| \hat\tau/n - \theta_0| = o_p(1)$
by noting the definition of $\hat\tau$. To further show $\hat\tau - \tau^* = O_p(1)$, we define that 
$Q_n(x) = \sum_{r=1}^R \hat{\Beta}_r\trans(x) \widehat{\bm\Gamma}_r \hat{\Beta}_r(x) $,
$\widehat\BDelta_r=\langle \BDelta, \hat\psi_r \rangle_\CT$ and
$L_{n,\tau} = -(\tau^*-\tau)(\tau+\tau^*)({n-\tau^*})^2$.
Following the similar arguments in the proof of Theorem 3.3(b) of \cite{aston2012detecting} and noting that $\widehat\Gamma_r$ are symmetric matrices, we can obtain that for $\tau \le \tau^*$, the decomposition of 
$Q_n(\tau/n)-Q_n(\tau^*/n)$ is dominated by $L_{n,\tau} \sum_{r=1}^R \widehat\BDelta_r\widehat{\bm\Gamma}_r\widehat\BDelta_r=L_{n,\tau}(g_R^*+o_p(1))$
due to Theorem \ref{thm: consistency of estimators}.
As a result, it can be shown that for some large $N$,
\begin{align*}
	\Pr(\hat\tau \le \tau^*-N) & = \Pr\left(\max_{\tau\leq \tau^*-N} Q_n\left({\tau}/{n}\right) - Q_n\left( {\tau^*}/{n} \right) \ge \max_{\tau>\tau^*-N} Q_n\left({\tau}/{n}\right) - Q_n\left( {\tau^*}/{n} \right) \right) \\
	& \le  \Pr\left(\max_{\tau\leq \tau^*-N} Q_n\left({\tau}/{n}\right)- Q_n\left( {\tau^*}/{n} \right) \ge 0 \right) \\
	& \le \Pr\left( \left(g_R^*+o_p(1)\right) \max_{\tau\leq \tau^*-N} L_{n,\tau} \ge 0 \right) \rightarrow 0,
\end{align*}
as $g_R^*>0$ and $ \max_{\tau\leq \tau^*-N} L_{n,\tau} \le -N\tau^*(n-\tau^*)^2<0$.
Analogous arguments can be applied to show that $\Pr(\hat\tau\ge \tau^*+N)\rightarrow 0$, which completes the proof.

\subsection{Proof of Theorem \ref{thm: FDR control}}\label{B5}
To streamline the notation and avoid ambiguity, we omit the superscripts ``O'' and ``E'' in
$\eta_{j,r}^{*O}$ and $\tilde{\eta}_{j,r}^{*E}$, yielding that $ W_j^* = \sum_{r=1}^R \eta_{j,r}^* \tilde{\eta}_{j,r}^* = {\Beta_j^*}\trans\tilde\Beta_j^*$. Similarly we abbreviate $\hat{\eta}_{\hat\tau, r}^O\left(\s_j\right)$ and $\tilde{\eta}_{\hat\tau, r}^E\left(\s_j\right)$ as $\hat{\eta}_{j,r}$ and $\tilde{\eta}_{j,r}$ for $W_j$, respectively. 
The first key result for the proof is the uniform symmetry property for $W_j$, i.e.,
\begin{equation}\label{sym property}
	\sup _{0 \leq t \leq t^*}\left|\frac{\sum_{j\in\mathcal{A}^c} \bI\left(W_j \geq t\right)}{\sum_{j\in\mathcal{A}^c} \bI\left(W_j \leq-t\right)}-1\right| = o_p(1)
\end{equation}
with the upper bound $t^*$ defined in \eqref{def t*}.
Note that Lemma \ref{lem: CLT} provides the analogous symmetry property for $W_j^*$, where the proof is based on Lemmas \ref{lem: sym for popu} and \ref{lem: conv for emp}. Meanwhile, Lemma \ref{lem: uni gap for W_j} provides the uniform convergence for
$\sum_{j\in\mathcal{A}^c} \bI(W_j \geq t)/\sum_{j\in\mathcal{A}^c} \bI(W_j^* \geq t)$ and 
$ \sum_{j\in\mathcal{A}^c} \bI(W_j \leq -t)/\sum_{j\in\mathcal{A}^c} \bI(W_j^* \leq -t ) $.
Then the result in \eqref{sym property} follows by noticing
\begin{equation*}
	\begin{aligned}
		\left|\frac{\sum_{j\in\mathcal{A}^c} \bI\left(W_j \geq t\right)}{\sum_{j\in\mathcal{A}^c} \bI\left(W_j \leq-t\right)}-1\right|
		\leq & \left| \frac{\sum_{j\in\mathcal{A}^c} \bI\left(W_j \geq t\right) - \sum_{j\in\mathcal{A}^c} \bI\left(W_j^* \geq t\right)}{\sum_{j\in\mathcal{A}^c} \bI\left(W_j \leq -t\right)} \right| \\
		&+ \left| \left(\frac{\sum_{j\in\mathcal{A}^c} \bI\left(W_j^* \geq t\right)}{\sum_{j\in\mathcal{A}^c} \bI\left(W_j^* \leq -t\right)} - 1\right) \cdot \frac{\sum_{j\in\mathcal{A}^c} \bI\left(W_j^* \leq -t\right)}{\sum_{j\in\mathcal{A}^c} \bI\left(W_j \leq -t\right)} \right|\\
		&+ \left| \frac{\sum_{j\in\mathcal{A}^c} \bI\left(W_j^* \leq -t\right)}{\sum_{j\in\mathcal{A}^c} \bI\left(W_j \leq -t\right)} - 1 \right|.
	\end{aligned}
\end{equation*}

By definition, our thresholding rule is equivalent to select index $j$ if $W_j \geq L$, where 
\begin{equation*}\label{thresh}
	L=\inf \left\{t\geq0: {1+\sum_j \bI ( W_j \leq-t )} \leq \alpha \max \left(\sum_j \bI( W_j \geq t), 1 \right)\right\}.
\end{equation*}
We need to establish an asymptotic bound for $L$ such that \eqref{sym property} can be applied.
Note that $\sum_{j\in\mathcal{A}^c}\bI\left( W_{j}^*\leq -t^* \right) \approx \alpha b_{p}$ by \eqref{t^* def}, where $A_n \approx B_n$ denotes that $A_n / B_n \xrightarrow{p} 1$.
By Lemmas \ref{lem: uni gap for W_j} and \ref{lem: lower bound W_j}, one then obtains that
$$\sum_j \bI\left(W_j<-t^*\right) \approx \sum_{j\in\mathcal{A}^c}\bI\left( W_{j}\leq -t^* \right) \approx \sum_{j\in\mathcal{A}^c}\bI\left( W_{j}^*\leq -t^* \right) \approx \alpha b_{p} \lesssim \alpha \sum_j \bI\left(W_j>t^*\right),$$ 
where $A_n \lesssim B_n$ means $\Pr(A_n \leq B_n) \rightarrow 1$. Hence, 
\begin{equation}\label{bound for L}
	\Pr\left( 0\leq L \leq t^* \right) \to 1,
\end{equation}
and consequently,
\begin{equation}\label{sym for emp W}
	\frac{\sum_{j\in\mathcal{A}^c} \bI\left(W_j \geq L\right)}{\sum_{j\in\mathcal{A}^c} \bI\left(W_j \leq-L\right)} \xrightarrow{p} 1.
\end{equation}
Finally, by the definition of FDP and the selection rule of $L$, we have
\begin{align*}\label{derive FDP}
	\text{FDP}_w(L)  =\frac{\sum_{j\in\mathcal{A}^c} \bI\left(W_j \geq L\right)}{1 \vee \sum_j \bI\left(W_j \geq L\right)} &= \frac{\sum_j \bI\left(W_j \leq-L\right)}{1 \vee \sum_j \bI\left(W_j \geq L\right)} \times \frac{\sum_{j\in\mathcal{A}^c} \bI\left(W_j \geq L\right)}{\sum_j \bI\left(W_j \leq-L\right)} \\
	&\leq \alpha \times \frac{\sum_{j\in\mathcal{A}^c} \bI\left(W_j \geq L\right)}{\sum_{j\in\mathcal{A}^c} \bI\left(W_j \leq-L\right)},
\end{align*}
which completes the proof for $\limsup\limits_{n,p\to\infty}\mbox{FDP}_w(L) \leq \alpha$ due to \eqref{sym for emp W}.

\subsection{Proof of Corollary \ref{thm: ident sr}}\label{B6}
This corollary is a direct conclusion derived from the proof of Theorem \ref{thm: FDR control}. Combining \eqref{bound for L} and \eqref{Wj t*}, we have
\begin{equation*}
	\operatorname{Pr}\left(W_j > L, \text{ for any } j\in\beta_p\right)\geq\Pr\left(W_j > t^*, \text{ for any } j\in\beta_p\right)\rightarrow 1,
\end{equation*}
which completes the proof.

\section{Additional lemmas}
The first is the standard Hoeffding’s inequality for sub-Gaussian variables, and the second is a moderate deviation result provided in \cite{cao2007moderate}.
\begin{lemma}\label{lem: Hoeffding}
	Let $\{X_i\}_{i=1}^n$ be independent centered sub-Gaussian random variables with parameter $\gamma^2_i$, i.e. $\E\{\exp ( s\, X_i) \} \leq \exp(s^2\gamma_i^2/2)$ for any $s\in\mathbb{R}$. 
	Then for any $\epsilon>0$, we have
	\begin{equation*}
		\Pr\left(\left|\sum_{i=1}^n X_i\right|>\epsilon\right)\leq 2\exp\left(\frac{-\epsilon^2}{2 \sum_{i=1}^n\gamma_i^2}\right).
	\end{equation*}
\end{lemma}

\begin{lemma}\label{lem: moderate}
	Let $X_1, \ldots, X_{n_1}$ be a sample of i.i.d. random variables with mean $\mu_1$ and variance $\sigma_1^2$, and $Y_1, \ldots, Y_{n_2}$ be another sample of i.i.d. random variables with mean $\mu_2$ and variance $\sigma_2^2$ that is independent of $\{X_i,1\leq i\leq n_1 \}$. Denote $T$ as the two-sample t-statistic 
	$
	T = \frac{\bar{X}-\bar{Y}-(\mu_1 - \mu_2)}{\sqrt{\sigma_1^2/n_1+\sigma_2^2/n_2}}.
	$
	Assume that $\E|X_1|^3 < \infty$, $\E|Y_1|^3 < \infty$ and that there exists $0<c_1 < c_2 <\infty$ such that $c_1 < n_1 / n_2 < c_2$. 
	Then
	\begin{equation*}
		\frac{\Pr(T \geq x)}{1-\Phi(x)} \rightarrow 1
	\end{equation*}
	uniformly in $x \in\left(0, o((n_1 + n_2)^{1/6})\right)$.
\end{lemma}

Define
$G(t) := p_{0}^{-1}\sum_{j\in\mathcal{A}^c} \Pr(W_{j}^* \geq t |{\mathcal{X}_E})$ and $G_{-}(t) := p_{0}^{-1}\sum_{j\in\mathcal{A}^c} \Pr(W_{j}^* \leq -t |\mathcal{X}_E)$,
where $p_0 = |\mA^c|$ denote the cardinality of null set. For any function $F$, let $F^{-1}(y) := \inf\{t \geq 0: F(t) \leq y\}$. Then set
\begin{equation}\label{def t*}
	t^* = G_{-}^{-1}(\alpha b_{p}/p_0),
\end{equation}
where $b_p$ is defined in Assumption \ref{assump: signal}.
Lemma \ref{lem: sym for popu} characterizes the symmetry property between $G(t)$ and $G_{-}(t)$.
\begin{lemma}\label{lem: sym for popu}
	Suppose Assumptions \ref{assump: eigengap} and \ref{assump: p}--\ref{assump: sr moment} hold. Then we have 
	\begin{equation*}
		\frac{G(t)}{G_{-}(t)} - 1 \to 0
	\end{equation*}
	uniformly for all $ t\in[0,t^*]$.
\end{lemma}

Lemma \ref{lem: conv for emp} establishes the uniform convergence of $\sum_{j\in\mathcal{A}^c}\bI(W_{j}^* \geq t)/p_0G(t)$ and $\sum_{j\in\mathcal{A}^c} \bI(W_j^* \leq -t)/p_0 G_{-}(t)$, 
\begin{lemma}\label{lem: conv for emp}
	Suppose Assumptions \ref{assump: eigengap}, \ref{assump: sr moment} and \ref{assump: dependence} hold. Then conditional on $\mathcal{X}_E$,
	\begin{align}\label{conv for emp 1}
		\sup\limits_{0\leq t \leq t^*}\left|\frac{\sum_{j\in\mathcal{A}^c} \bI(W_j^* \geq t)}{p_0 G(t)} - 1\right| &= o_p(1),\\
		\sup\limits_{0\leq t \leq t^*}\left|\frac{\sum_{j\in\mathcal{A}^c} \bI(W_j^* \leq -t)}{p_0 G_{-}(t)} - 1\right| &= o_p(1).\nonumber
	\end{align}
\end{lemma}

Lemma \ref{lem: uni gap for W_j} provides the uniform convergence of $\sum_{j\in\mathcal{A}^c} \bI(W_j \geq t)/\sum_{j\in\mathcal{A}^c} \bI(W_j^* \geq t)$ and ${\sum_{j\in\mathcal{A}^c} \bI(W_j \leq -t)}/{\sum_{j\in\mathcal{A}^c} \bI(W_j^* \leq -t)} $. 

\begin{lemma}\label{lem: uni gap for W_j}
	Suppose Assumptions \ref{assump: L2}--\ref{assump: eigengap}, \ref{assump: p}--\ref{assump: sr moment} and \ref{assump: dependence}--\ref{assump: kernel} hold, then
	\begin{equation*}
		\begin{aligned}
			\sup\limits_{0\leq t \leq t^*} \left| \frac{\sum_{j\in\mathcal{A}^c} \bI\left(W_j \geq t\right)}{\sum_{j\in\mathcal{A}^c} \bI\left(W_j^* \geq t\right)} - 1\right| = o_p(1), \\
			\sup\limits_{0\leq t \leq t^*} \left| \frac{\sum_{j\in\mathcal{A}^c} \bI\left(W_j \leq -t\right)}{\sum_{j\in\mathcal{A}^c} \bI\left(W_j^* \leq -t \right)} - 1\right| = o_p(1).
		\end{aligned}
	\end{equation*}
\end{lemma}

Lemma \ref{lem: lower bound W_j} further implies that, with probability tending to 1, $\sum_j \bI\left(W_j>t^*\right)$ is larger than $b_p$, where $b_p$ denotes the minimum number of identifiable effect sizes.
\begin{lemma}\label{lem: lower bound W_j}
	Suppose Assumptions \ref{assump: L2}--\ref{assump: eigengap}, \ref{assump: sr moment}--\ref{assump: signal} and \ref{assump: kernel} hold, then
	\begin{equation*}\label{signal number}
		\operatorname{Pr}\left(\sum_j \bI\left(W_j>t^*\right)\geq b_{p}\right)\rightarrow 1.
	\end{equation*}
\end{lemma}

\section{Proofs for the lemmas}
\subsection{Proof of Lemmas \ref{lem: CLT}-\ref{lem: bound for est}}
\begin{proof}[Proof of Lemma \ref{lem: CLT}] 
	Note that under the weak separability structure in Section \ref{sec: weak sepa}, we have $\cov\{\xi_{ir}(\s_j),\xi_{i'r'}(\s_k)\}=\bI(i=i')\bI(r=r')\sigma_r(\s_j,\s_k)$. Then the result in Lemma \ref{lem: CLT} follows directly from the functional central limit theorem \citep[e.g.][]{bosq2000linear} with respect to the Skorokhod space $D^{pR}[0,1]$.
\end{proof}
\begin{proof}[Proof of Lemma \ref{lem: bound for eta*}]
	(a) This result follows immediately from the fact that the uncorrelatedness of $\eta_{j,r}^*$ across $r$ and by the direct calculation
	\begin{equation*}
		\var\left(\eta_{j,r}^*\right) = \var\left\{\frac{\tau^*(n-\tau^*)}{n\sqrt{n}}\left(\frac{1}{\tau^*}\sum_{i=1}^{\tau^*}\xi_{ir,j}^* - \frac{1}{n-\tau^*}\sum_{i=\tau^*+1}^n\xi_{ir,j}^*\right)\right\} = \theta_0(1-\theta_0)(1-2\theta_0).
	\end{equation*}
	(b) We shall show that the assertion holds for each $\eta_{j,r}^*$ and $\delta_{j,r}^*$ with some large $C_\eta>0$, i.e.
	\begin{equation*}
		\operatorname{Pr}\left(\left|\eta_{j,r}^* -\sqrt{n} \theta_0(1-\theta_0)\delta_{j,r}^* \right| > \sqrt{C_\eta\log p} \right) = o(1/p),
	\end{equation*}
	and the case for $\Beta_j^*$ is straightforward by using the Bonferroni inequality since $R$ is fixed. For each $r$, define that $Z_{ij,r} =  (1-\theta_0)\xi_{ir,j}^*$ for $i \le \tau^*$
	and $Z_{ij,r} = - \theta_0\xi_{ir,j}^*$ for $i>\tau^*$. By definition, we have
	$\eta_{j,r}^* - \sqrt{n} \theta_0(1-\theta_0)\delta_{j,r}^* = n^{-1/2}\sum_{i=1}^{n}Z_{ij,r}$,   
	and thus it suffices to show that 
	$ \operatorname{Pr}\left( \left| \sum_{i=1}^{n}Z_{ij,r}\right| > \sqrt{C_\eta n\log p}\right) = o(1/p) $
	for each $j$ and $r$. 
	This can be obtained directly from Lemma \ref{lem: Hoeffding} and Assumption \ref{assump: sr moment} with 
	\begin{equation}\label{bound for Zij}
		\operatorname{Pr}\left( \left| \sum_{i=1}^{n}Z_{ij}\right| > \sqrt{C_\eta n\log p}\right)\leq 2\exp\left(\frac{-C_\eta\log p}{2 c\gamma^2}\right) = o(1/p),
	\end{equation}
	where $c = \min\{\theta_0^2,(1-\theta_0)^2\}$.
	
	To prove the result for $\tilde\Beta_j^*$, we denote
	$\bxi_j^* = (\xi_{j,1}^*,\dots,\xi_{j,R}^*)\trans$ where $\xi_{j,r}^*=\frac{1}{\tau^*}\sum_{i=1}^{\tau^*}\xi_{ir,j}^*-\frac{1}{n-\tau^*}\sum_{i=\tau^*+1}^{n}\xi_{ir,j}^*$,
	and $w_{k,j} =  K_h(\s_k-\s_j)/ \sum_k K_h(\s_k-\s_j)$
	for each $j$.
	By definition, we then have
	$\Beta_j^* = \sqrt{n}\theta_0\left(1-\theta_0\right)\left(\Bdelta_j^* +\bxi_j^*\right)$, 
	$\tilde\Beta_j^* = \sqrt{n}\theta_0\left(1-\theta_0\right){\sum_{k} w_{k,j}\left(\Bdelta_k^* +\bxi_k^*\right)}$,
	and
	\begin{align*}
		\|\tilde\Beta_j^*-\Beta_j^*\|&\leq\sqrt{n}\theta_0(1-\theta_0)\left(\left\|{\sum_{k}w_{k,j}(\Bdelta_k^*-\Bdelta_j^*)}\right\| + \left\|{\sum_{k} w_{k,j}(\bxi_k^*-\bxi_j^*)}\right\|\right)\\
		&\leq\sqrt{n}\theta_0(1-\theta_0)\left(\max_{k:|k-j|\leq h}\|\Bdelta_k^*-\Bdelta_j^*\| + \max_{k}(\|\bxi_k^*\| + \|\bxi_j^*\|)\right),
	\end{align*}
	It then follows from Assumption \ref{assump: kernel} and the same argument in \eqref{bound for Zij} that
	\begin{equation}\label{uni gap for K_eta}
		\operatorname{Pr}\left(\left\|\tilde\Beta_j^*-\Beta_j^*\right\|> \sqrt{C\log p}\right) = o(1/p)
	\end{equation}
	for some large $C$, which consequently implies that 
	\begin{equation*}
		\operatorname{Pr}\left(\left\|\tilde\Beta_j^* -\sqrt{n} \theta_0(1-\theta_0)\Bdelta_j^* \right\| > \sqrt{C\log p} \right) = o(1/p).
	\end{equation*}
\end{proof}

\begin{proof}[Proof of Lemma \ref{lem: sym for true W}]
	The asymptotic symmetry result in Lemma \ref{lem: sym for true W} follows directly from Lemmas \ref{lem: sym for popu} and \ref{lem: conv for emp}.
	Hence, it remains only to show that $t^*\leq C_\eta \log p$.
	According to the definition of $t^*$ and Lemma \ref{lem: conv for emp},
	\begin{equation}\label{t^* def}
		\alpha b_{p} / p_0 = G_{-}\left(t^*\right)=\frac{1}{p_{0}} \sum_{j\in\mathcal{A}^c} \bI\left(W_{j}^* \leq -t^*\right)\{1+o(1)\}.
	\end{equation}
	On the other hand, it follows from Lemma \ref{lem: bound for eta*}(b) that
	\begin{align*}
		&\Pr(\sum_{j\in\mA^c}\bI\left(W_j^*\leq -C_\eta\log p\right)>0)\\
		\leq&\sum_{j\in\mA^c}\Pr\left(W_j^*\leq-C_\eta\log p\right)\\
		\leq&\sum_{j\in\mA^c}\Pr\left(\|\Beta_j^*\|\|\tilde\Beta_j^*\|\geq C_\eta\log p\right)\\
		\leq&\sum_{j\in\mA^c}\Pr\left(\|\tilde\Beta_j^*\|> \sqrt{C_\eta\log p}\right) + \sum_{j\in\mA^c}\Pr\left(\|\Beta_j^*\|> \sqrt{C_\eta\log p}\right)= o(1),
	\end{align*}
	which implies
	$\Pr\left(\sum_{j\in\mA^c}\bI\left(W_j^*\leq-C_\eta\log p\right)=0\right)\to 1$. Hence combining with \eqref{t^* def}, one obtains $t^*\leq C_\eta\log p$.
\end{proof}

\begin{proof}[Proof of Lemma \ref{lem: bound for est}] 
	(a) Applying the similar arguments as in the proof of Theorem 4 in \cite{zapata2021partial}, we have
	\begin{equation*}
		\Pr\left(\| \hat\CH - \CH \|_{HS} \geq \epsilon \right) \leq C'_2 \exp\left(- C'_1 \varpi_r^{-2} n \epsilon^2 \right),
	\end{equation*}
	and
	\begin{equation*}
		\Pr\left(\| \hat\CC_{jj} - \CC_{jj} \|_{HS} \geq \epsilon \right) \leq C'_2 \exp\left(- C'_1 \varpi_r^{-2} n \epsilon^2 \right),
	\end{equation*}
	where $\varpi_1 = 2\sqrt{2}(\lambda_1-\lambda_2)^{-1}$ and $\varpi_r = 2\sqrt{2} \max \{ (\lambda_{r-1}-\lambda_r)^{-1},(\lambda_{r}-\lambda_{r+1})^{-1}\} $ for $r \ge 2$.
	Moreover, applying the result (S4) in \cite{zapata2021partial}, for each $j=1,\dots,p$,
	\begin{equation*}
		| \hat{\lambda}_{r,j}-\lambda_{r,j} | \le 2 M \varpi_r \| \hat\CH - \CH \|_{HS} + \| \hat\CC_{jj} - \CC_{jj} \|_{HS},
	\end{equation*}
	Therefore, we can obtain the concentration inequality
	\begin{equation}\label{ConInq for lambda}
		\Pr\left(\left|\hat{\lambda}_{r,j}-\lambda_{r,j}\right| \geq \epsilon \right) \leq C_2 \exp\left(- C_1 \varpi_r^{-2} n \epsilon^2 \right),
	\end{equation}
	It then follows that for any $c>1$,  
	\begin{align*}
		\Pr\left(\max_{j}\left|\hat{\lambda}_{r,j}-\lambda_{r,j}\right|\geq\sqrt{\frac{c\log p}{n}}\right) &\leq \sum_{j}\Pr\left(\left|\hat{\lambda}_{r,j}-\lambda_{r,j}\right|\geq\sqrt{\frac{c\log p}{n}}\right)\\
		&\leq C p \exp\left(-c\log p\right) = o(1),
	\end{align*}
	which implies that
	\begin{equation*}
		\max_{j}\left|\hat{\lambda}_{r,j}^{-1/2}-\lambda_{r,j}^{-1/2}\right| = \max_{j}\left|\frac{\hat{\lambda}_{r,j}-\lambda_{r,j}}{\hat{\lambda}_{r,j}^{-1/2}\lambda_{r,j}^{-1/2}\left(\hat{\lambda}_{r,j}^{-1/2}+\lambda_{r,j}^{-1/2}\right)}\right| = O_p\left(\left(\log p /n\right)^{1/2}\right).
	\end{equation*}
	(b) Assume that $\hat\tau\leq \tau^*$ w.l.o.g., we have
	\begin{align*}
		&\left|\left\langle \widehat\Delta_{\hat\tau}(\s_j) -  \widehat\Delta_{\tau^*}(\s_j),\psi_r \right\rangle_\CT\right|\\
		=& \frac{1}{\sqrt{n}}\left|\left\langle\left(\sum_{i=1}^{\hat\tau}X_i(\s_j) - \frac{\hat\tau}{n}\sum_{i=1}^{n}X_i(\s_j)\right) - \left(\sum_{i=1}^{\tau^*}X_i(\s_j) - \frac{\tau^*}{n}\sum_{i=1}^{n}X_i(\s_j)\right),\psi_r\right\rangle_\CT\right|\\
		\leq & \frac{1}{\sqrt{n}}\left|\left\langle\sum_{i=\hat\tau +1}^{\tau^*}X_i(\s_j),\psi_r\right\rangle_\CT\right| + \frac{\tau^*-\hat\tau}{n\sqrt{n}}\left|\left\langle\sum_{i=1}^{n}X_i(\s_j),\psi_r\right\rangle_\CT\right|\\
		=&\frac{1}{\sqrt{n}}\left|\sum_{i=\hat\tau+1}^{\tau^*}\left\{\mu_{ir}(\s_j)+\xi_{ir}(\s_j)\right\}\right|+\frac{\tau^*-\hat\tau}{n\sqrt{n}}\left|\sum_{i=1}^n\left\{\mu_{ir}(\s_j)+\xi_{ir}(\s_j)\right\}\right| := W_{j1}+W_{j2}
	\end{align*}
	where $\mu_{ir}(\s_j) = \langle\mu_i(\s_j),\psi_r\rangle_\CT$.
	Note that $\tau^* - \hat\tau = O_p(1)$ by Corollary \ref{thm: cp consis}, applying Lemma \ref{lem: Hoeffding} to $\xi_{ir}(\s_j)$ under Assumption \ref{assump: sr moment}, we have
	\begin{equation*}
		\Pr\left(\left|\sum_{i=\hat\tau+1}^{\tau^*}\xi_{ir}(\s_j)\right|>\sqrt{C\log p}\right)\leq 2\exp\left(\frac{-C_\tau\log p}{2(\tau^*-\hat\tau)M\gamma^2}\right)=o(1/p)
	\end{equation*}
	for some large constant $C_\tau$, and
	\begin{equation}\label{ConIneq for xi}
		\Pr\left(\left|\sum_{i=1}^{n}\xi_{ir}(\s_j)\right|>\sqrt{Cn\log p}\right)\leq 2\exp\left(\frac{-C\log p}{2 M\gamma^2}\right)=o(1/p),
	\end{equation}
	which implies 
	$
	\Pr(\max_{j}|\sum_{i=\hat\tau+1}^{\tau^*}\xi_{ir}(\s_j)|>\sqrt{C\log p})=o(1)$
	and 
	$ \Pr(\max_{j}|\sum_{i=1}^{n}\xi_{ir}(\s_j)|>\sqrt{Cn\log p})=o(1)$, respectively.
	It then follows that
	\begin{equation*}
		\max_j W_{j1} =O_p\left(\left(\log p /n\right)^{1/2}\right) \mbox{ and } \max_j W_{j2} = O_p\left(\sqrt{\log p}/n\right),
	\end{equation*}
	provided that the mean function $\Bmu_{i}(\cdot)$ is uniformly bounded, which yields the result in (b). \\
	(c) Recall that $\eta_{j,r}^* = \lambda_{r,j}^{-1/2} \langle \widehat\Delta_{\tau^*}(\s_j),\psi_r \rangle_\CT $ and $\hat\eta_{j,r} = \hat\lambda_{r,j}^{-1/2} \langle \widehat\Delta_{\hat\tau}(\s_j),\hat\psi_r \rangle_\CT$, where for notation simplicity we write $\Delta_{\tau^*,j}:=\Delta_{\tau^*}(\s_j)$ and $\Delta_{\hat\tau,j}:=\Delta_{\hat\tau}(\s_j)$.
	Moreover, we omit the subscripts ``$r$'' in $\hat\eta_{j,r}$, $\eta_{j,r}^*$, $\delta_{j,r}^*$ and $\lambda_{r,j}$, and ``$\CT$'' in $\langle \cdot,\cdot\rangle_\CT$ without confusion.
	For each $j=1,\dots, p$, we have
	\begin{align*}
		\left|\hat\eta_{j}-\eta_{j}^*\right|  =& \lambda_j^{-1/2} \left|\left\langle \widehat\Delta_{\hat\tau}(\s_j) -  \widehat\Delta_{\tau^*}(\s_j), \psi_r \right\rangle\right|  
		+ \lambda_j^{-1/2} \left|\left\langle \widehat\Delta_{\hat\tau}(\s_j), \hat\psi_r-\psi_r \right\rangle\right|\\
		&+  \left|\left( \hat\lambda_j^{-1/2} - \lambda_j^{-1/2} \right) \left\langle \widehat\Delta_{\hat\tau}(\s_j),\hat\psi_r \right\rangle\right| := L_{j1} + L_{j2} + L_{j3}
	\end{align*}
	According to Lemma \ref{lem: bound for est}(b), $\max_j L_{j1}= O_p\left((\log p /n)^{1/2}\right)$ since $\lambda_j$ is uniformly bounded from Assumptions \ref{assump: eigengap} and \ref{assump: sr moment}.
	To obtain the order for $ L_{j2}$ and $L_{j3}$, we first show that for any $j\in \mA^c_h$,
	\begin{equation}\label{ConIneq for Delta_r}
		\Pr\left(\left| \left\langle \widehat\Delta_{\tau^*}(\s_j) ,\psi_r \right\rangle \right|>\sqrt{C_1\log p}\right)=o(1/p),
	\end{equation}
	and
	\begin{equation}\label{ConIneq for Delta_r hat}
		\Pr\left(\left| \left\langle \widehat\Delta_{\tau^*}(\s_j) ,\hat\psi_r \right\rangle \right|>\sqrt{C_1\log p}\right)=o(1/p),
	\end{equation}
	for some large $C_1$.
	Note that $ \langle \widehat\Delta_{\tau^*}(\s_j) ,\psi_r \rangle=\sqrt n \theta_0(1-\theta_0)\lambda_j\delta_j^* + n^{-1/2}\{\sum_{i=1}^{\tau^*}\xi_{ir}(\s_j)-\frac{\tau^*}{n}\sum_{i=1}^{n}\xi_{ir}(\s_j)\}$.
	Then \eqref{ConIneq for Delta_r} follows by Assumption \ref{assump: kernel} and that
	$ \Pr (|\sum_{i=1}^{\tau^*}\xi_{ir}(\s_j)-\frac{\tau^*}{n}\sum_{i=1}^{n}\xi_{ir}(\s_j)|>\sqrt{C_1n\log p} )=o(1/p)$, which can be shown using similar arguments as \eqref{ConIneq for xi}.
	To obtain \eqref{ConIneq for Delta_r hat}, we further show that
	\begin{equation}\label{ConIneq for Delta diffr}
		\Pr\left(\left| \left\langle \widehat\Delta_{\tau^*}(\s_j) ,\hat\psi_r-\psi_r \right\rangle \right|>{\frac{C_2\log p}{\sqrt n}}\right)=o(1/p),
	\end{equation}
	for some large $C_2$.
	Note that $\max_j \E \|\widehat\Delta_{\tau^*}(\s_j) \|^2 \leq M_1$ for some large constant $M_1$ that depends on $M$ specified in Assumption \ref{assump: sr moment}, and the similar arguments for \eqref{ConIneq for xi} yield 
	\begin{equation*}
		\Pr\left( \|\hat\psi_r-\psi_r \|> \sqrt{\frac{C\log p}{n}} \right ) = o(1/p).
	\end{equation*}
	Then \eqref{ConIneq for Delta diffr} follows by applying Cauchy--Schwarz inequality, which leads to
	$\max_j L_{j2}= O_p(\log p /n^{1/2})$ based on Lemma \ref{lem: bound for est}(b). 
	Similarly, we can obtain $\max_j L_{j3}= O_p(\log p /n^{1/2})$ from \eqref{ConIneq for Delta_r hat} and Lemma \ref{lem: bound for est}(a), which completes the proof for $\max_{j\in\mA^c_h}|\hat\eta_{j}-\eta_{j}^*| = O_p(\log p /n^{1/2})$.\\
	(d) Observe that $|W_j - W_j^*| \leq \sum_{r=1}^R |\hat\eta_{j,r}(\tilde{\eta}_{j,r} - \tilde{\eta}_{j,r}^*) + \tilde\eta_{j,r}^* (\hat\eta_{j,r} - {\eta}_{j,r}^*)|$. 
	For each $r = 1,\dots,R$, it follows from Lemma \ref{lem: bound for est}(c) that
	\begin{equation*}
		\max_{j\in\mA^c} |\hat\eta_{j,r} - {\eta}_{j,r}^*| = {O_p\left(\log p/n^{1/2}\right)},
		\tag{I.1}
	\end{equation*}
	and
	\begin{equation*}
		\max_{j\in\mA^c}|\tilde{\eta}_{j,r} - \tilde{\eta}_{j,r}^*| \leq\max_{j\in\mA^c_h}|\hat\eta_{k,r} - {\eta}_{k,r}^*| =  O_p\left(\log p /n^{1/2}\right).
		\tag{I.2}
	\end{equation*}
	On the other hand, it follows from Lemma \ref{lem: bound for eta*}(b) that 
	\begin{equation*}
		\Pr\left(\max_{j\in\mA^c}\left|\eta_{j,r}^*\right|>\sqrt{C_\eta\log p}\right)\leq\sum_{j\in\mA^c}\Pr\left(\left|\eta_{j,r}^*\right|>\sqrt{C_\eta\log p}\right)=o(1),
	\end{equation*}
	which implies that $\max_{j\in \mA^c}\left|\eta_{j,r}^*\right| = O_p\left(\left(\log p\right)^{1/2}\right)$. 
	Therefore,
	\begin{equation*}
		\max_{j\in\mA^c}|\hat\eta_{j,r}|\leq\max_{j\in\mA^c}|\hat\eta_{j,r}-\eta_{j,r}^*| + \max_{j\in\mA^c}|\eta_{j,r}^*| = O_p\left(\left(\log p\right)^{1/2}\right),
		\tag{I.3}
	\end{equation*}
	and
	\begin{equation*}
		\max_{j\in\mA^c}|\tilde\eta_{j,r}^*|\leq \max_{j\in\mA^c}|\eta^*_{j,r}| + \max_{j\in\mA^c}|\tilde\eta^*_{j,r} - \eta^*_{j,r}|
		= O_p\left(\left(\log p\right)^{1/2}\right),
		\tag{I.4}
	\end{equation*}
	where $\max_{j\in\mA^c}|\tilde\eta^*_{j,r} - \eta^*_{j,r}|
	= O_p\left(\left(\log p\right)^{1/2}\right)$ is due to \eqref{uni gap for K_eta} in the proof of Lemma \ref{lem: bound for eta*}(b).
	Combining (I.1)-(I.4), we conclude that 
	\begin{equation}\label{uni bound for Wj}
		\max_{j\in\mA^c}\left| W_j^* - W_j \right|=  O_p\left((\log p)^{3/2}/n^{1/2}\right) = o_p \left( (\log p)^{-1/2} \right)
	\end{equation}
	by Assumption \ref{assump: p}. 
\end{proof}

\subsection{Proof of Lemmas \ref{lem: sym for popu}-\ref{lem: lower bound W_j}}
\begin{proof}[Proof of Lemma \ref{lem: sym for popu}]
	By the definitions of $W_j^*$, $G(t)$ and $G_{-}(t)$, we have
	\begin{equation*}
		{G(t)}/{G_{-}(t)} - 1 =  \{p_0 G_{-}(t)\}^{-1} \sum_{j\in\mA^c} \widetilde{P}_j(t),
	\end{equation*}
	where $\widetilde{P}_j(t)=\Pr\left({\Beta_j^*}\trans\tilde\Beta_j^*\geq t|\mathcal{X}_E\right) - \Pr\left({\Beta_j^*}\trans\tilde\Beta_j^*\leq -t|\mathcal{X}_E\right)$.
	On one hand, for $\mA_t:=\{j: \|\tilde\Beta_j^*\| < t/\sqrt{C_\eta\log p}\}$ where $C_\eta$ is a large constant specified in Lemma \ref{lem: bound for eta*}(b), we have
	\begin{align*}
		\frac{\sum_{j\in\mA^c \cap \mA_t }\widetilde{P}_j(t)}{p_0G_{-}(t)} &\leq \frac{\sum_{j\in\mA^c}\Pr\left({\Beta_j^*}\trans\tilde\Beta_j^*\geq t|\mathcal{X}_E\right)}{p_0G_{-}(t)}\\
		&\leq \frac{\sum_{j\in\mA^c}\Pr\left(\|\Beta_j^*\|\geq \sqrt{C_\eta\log p}|\mathcal{X}_E\right)}{\alpha b_p} = o\left( 1 \right),
	\end{align*}
	where the second inequality follows immediately from the fact that $G_{-}(t)$ is decreasing and $t\leq G_{-}^{-1}(\alpha b_{p}/p_0)$.
	
	On the other hand, for the case $\|\tilde\Beta_j^*\| \geq t/\sqrt{C\log p}$, we note that for $j\in\mA^c$,
	\begin{equation*}
		{\Beta_j^*}\trans\tilde\Beta_j^* = \frac{1}{\sqrt{n}}\left(\sum_{i=1}^{\tau^*}{\bxi_{i,j}^*}\trans\tilde\Beta_j^* - \frac{\tau^*}{n}\sum_{i=1}^{n}{\bxi_{i,j}^*}\trans\tilde\Beta_j^*\right),  
	\end{equation*}
	where $\bxi_{i,j}^* = (\xi_{i1,j}^*,\dots,\xi_{iR,j}^*)\trans$. 
	Denote $\tilde\Beta_j^*=\bu$ given $\mathcal{X}_E$, we can derive that $\E({\Beta_j^*}\trans\bu|\mathcal{X}_E) = 0$ and
	$\cov({\Beta_j^*}\trans\bu|\mathcal{X}_E) = {\bu\trans\BSigma_j^*\bu}$,
	where $\BSigma_j^*$ is defined in Lemma \ref{lem: bound for eta*}(a). It then follows by Lemma \ref{lem: moderate} and Assumptions \ref{assump: p}--\ref{assump: sr moment} that 
	\begin{align*}
		\Pr\left({\Beta_j^*}\trans\tilde\Beta_j^*\geq t|\mathcal{X}_E\right) = \{ 1-\Phi\left({t}/{\bu\trans\BSigma_j^*\bu} \right)\} \left\{1+o_p(1)\right\},
	\end{align*}
	where $o_p(1)$ is uniformly in $j$.
	Similarly, we can obtain that
	$\Pr({\Beta_j^*}\trans\tilde\Beta_j^*\leq - t|\mathcal{X}_E) = \Phi({-t}/{\bu\trans\BSigma_j^*\bu})\left\{1+o_p(1)\right\}.
	$
	It then follows that
	$ \sum_{j\in\mA^c \cap \mA_t^c }\widetilde{P}_j(t) = o\left( 1 \right)$,
	which leads to that ${G(t)}/{G_{-}(t)}-1=o(1)$ by combining the result for $j\in\mA_t$.
	
\end{proof}
\begin{proof}[Proof of Lemma \ref{lem: conv for emp}]
	We only prove the first formula, and the second one holds similarly. 
	Note that $G(t)$ is a decreasing and continuous function. Following the discretization approach in \cite{du2021false}, 
	we let $z_0<z_1<\cdots<z_{h_n} \leq 1$ and $t_i=G^{-1}\left(z_i\right)$, where $z_0=a_p / p_0$, $a_p=\alpha b_{p}$, $z_i=a_p / p_0+w_p \exp(i^\zeta) /p_0$, $h_n=\{\log((p_0-a_p) / w_p)\}^{1 / \zeta}$ with $w_p / a_p \rightarrow 0$ and $0<\zeta<1$. 
	Note that $G(t_i) / G(t_{i+1})=1+o(1)$ uniformly in $i$. 
	It is therefore enough to derive the convergence rate of
	\begin{equation*}
		D_n=\sup _{0 \leq i \leq h_n}\left|\frac{\sum_{j\in\mathcal{A}^c}\left\{\bI\left(W_j^*>t_i\right)-\operatorname{Pr}\left(W_j^*>t_i\mid \mathcal{X}_E\right)\right\}}{p_0 G\left(t_i\right)}\right|.
	\end{equation*}
	Define that
	\begin{align*}
		D(t) & :=\mathbb{E}\left[\left(\sum_{j\in\mathcal{A}^c}\left\{\bI\left(W_j^*>t\right)-\operatorname{Pr}\left(W_j^*>t\mid \mathcal{X}_E\right)\right\}\right)^2 \mid \mathcal{X}_E\right]\\
		&=\sum_{j\in\mathcal{A}^c} \sum_{k \in \mathcal{A}^c}\left\{\operatorname{Pr}\left(W_j^*>t, W_k^*>t \mid \mathcal{X}_E\right)-\operatorname{Pr}\left(W_k^*>t \mid \mathcal{X}_E\right) \operatorname{Pr}\left(W_j^*>t \mid \mathcal{X}_E\right)\right\},
	\end{align*}
	and $\mathcal{M}_j=\{k \in \mathcal{A}^c:|\sum_{r=1}^R \rho_{r}(\s_j,\s_k)| \geq C_\rho(\log n)^{-2-\nu}\}$, where $C_\rho$ and $\nu$ are defined in Assumption \ref{assump: dependence}.
	It then follows that
	\begin{equation}\label{Dt}
		\begin{aligned}
			D(t) \leq &l_p p_0 G(t)+\sum_{j\in\mathcal{A}^c} \sum_{k \in \mathcal{M}_j^c}\left\{\operatorname{Pr}\left(W_k^*>t, W_j^*>t \mid \mathcal{X}_E\right)\right.\\
			&\left.-\operatorname{Pr}\left(W_k^*>t \mid \mathcal{X}_E\right) \operatorname{Pr}\left(W_j^*>t \mid \mathcal{X}_E\right)\right\}.
		\end{aligned}
	\end{equation}
	For the second term of the right-hand side of \eqref{Dt}, we notice that for 
	$j\in\mA^c$ and $k\in\mathcal{M}_j^c$,
	\begin{equation*}
		\left|\frac{\cov\left\{\left(W_j^*,W_k^*\right)|\mathcal{X}_E\right\}}{\sqrt{\var\left(W_j^*|\mathcal{X}_E\right)\var\left(W_k^*|\mathcal{X}_E\right)}}\right|\leq C(\log n)^{-2-v}
	\end{equation*}
	for some constant $C$ depend on $C_\rho$ and $\theta_0$. To see this, denote $\tilde\eta_{j,r}^* = u_{j,r}$ given $\mathcal{X}_E$. 
	For each $j\in\mathcal{A}^c$ and $k \in \mathcal{M}_j^c$, it follows from Lemma \ref{lem: bound for eta*}(a) that
	\begin{align*}
		&\var\left(W_j^*|\mathcal{X}_E\right) =\sum_{r=1}^R\var\left({\eta_{j,r}^*}u_{j,r}\right) =  \theta_0(1-\theta_0)(1-2\theta_0)\sum_{r=1}^R u_{j,r}^2,\\
		&\cov\left\{\left(W_j^*,W_k^*\right)|\mathcal{X}_E\right\} = \sum_{r=1}^R u_{j,r}u_{k,r}\cov\left(\eta_{j,r}^*,\eta_{k,r}^*\right)=c_{\theta_0}\sum_{r=1}^Ru_{j,r}u_{k,r}\rho_r(\s_j,\s_k),
	\end{align*}
	where $c_{\theta_0}>0$ is some constant depends on $\theta_0$. Consequently, applying Lemma 1 in \cite{cai2016large} we obtain that
	\begin{equation*}
		\left|\frac{\operatorname{Pr}\left(W_k^*>t, W_j^*>t \mid \mathcal{X}_E\right)-\operatorname{Pr}\left(W_k^*>t \mid \mathcal{X}_E\right) \operatorname{Pr}\left(W_j^*>t \mid \mathcal{X}_E\right)}{\operatorname{Pr}\left(W_k^*>t \mid \mathcal{X}_E\right) \operatorname{Pr}\left(W_j^*>t \mid \mathcal{X}_E\right)}\right| \leq (\log n)^{-1-\nu_1}
	\end{equation*}
	uniformly holds, where $\nu_1=\min (\nu, 1 / 2)$. 
	It then follows from \eqref{Dt} that
	\begin{equation}\label{D_n}
		\begin{aligned}
			\operatorname{Pr}\left(D_n \geq \epsilon \mid \mathcal{X}_E\right)
			\leq &\sum_{i=0}^{h_n} \operatorname{Pr}\left(\left|\frac{\sum_{j\in\mathcal{A}^c}\left[\bI\left(W_j^*>t_i\right)-\operatorname{Pr}\left(W_j^*>t_i \mid \mathcal{X}_E\right)\right]}{p_0 G\left(t_i\right)}\right| \geq \epsilon\right) \\
			\leq &\frac{1}{\epsilon^2} \sum_{i=0}^{h_n} \frac{D\left(t_i\right)}{p_0^2 G^2\left(t_i\right)} 
			\leq \frac{1}{\epsilon^2}\left\{l_p \sum_{i=0}^{h_n} \frac{1}{p_0 G\left(t_i\right)}+h_n (\log n)^{-1-\nu_1}\right\} .
		\end{aligned}
	\end{equation}
	Observing that
	\begin{equation*}
		\sum_{i=0}^{h_n} \frac{1}{p_0 G\left(t_i\right)}=\frac{1}{a_p}+\sum_{i=1}^{h_n} \frac{1}{a_p+w_p e^{i \zeta}} \leq w_p^{-1}\{1 + O(1)\},
	\end{equation*}
	we can then conclude that $D_n=o_p(1)$ when $l_p / b_{p} \rightarrow 0$ by noting that (a) $\zeta$ can be arbitrarily close to 1 such that $h_n(\log n)^{-1-\nu_1}\rightarrow 0$, 
	and (b) $w_p$ can be made arbitrarily large as long as $w_p / a_p \rightarrow 0$. This completes the proof of (\ref{lem: conv for emp}).
\end{proof}

\begin{proof}[Proof of Lemma \ref{lem: uni gap for W_j}]
	Note that the result in Lemma \ref{lem: conv for emp} implies that for any $\epsilon>0$,
	\begin{equation*}\label{liu1}
		\sup\limits_{0\leq t \leq t^*}\Pr\left(\left|\frac{\sum_{j\in\mathcal{A}^c} \bI(W_j^* \geq t)}{p_0 G(t)} - 1\right|\geq \epsilon\right) = o(1),
	\end{equation*}
	\begin{equation*}\label{liu2}
		\int_{0}^{t^*}\Pr\left(\left|\frac{\sum_{j\in\mathcal{A}^c} \bI(W_j^* \geq t)}{p_0 G(t)} - 1\right|\geq \epsilon\right)dt= o(t^*/h_n).
	\end{equation*}
	Then, it follows from \eqref{uni bound for Wj} and the arguments in the proof of Theorem 3.1 in \cite{liu2013gaussian} that
	\begin{equation*}\label{gap W}
		\sup\limits_{0\leq t\leq t^*} \left| \frac{\sum_{j\in\mathcal{A}^c} \bI\left(W_j \geq t\right)}{p_0 G(t)} - 1\right| = o_p(1),
	\end{equation*}
	which is a counterpart of \eqref{conv for emp 1} based on $W_j$. 
	Then the first equation in Lemma \ref{lem: uni gap for W_j} can be obtained using again \eqref{conv for emp 1}, and the second follows similarly.
\end{proof}

\begin{proof}[Proof of Lemma \ref{lem: lower bound W_j}]
	We first show that for each $j \in \beta_{p}$,  $\operatorname{Pr}(W_{j}^* \leq C_\eta\log p)=o(1/p)$. 
	According to the proof of Lemma \ref{lem: bound for eta*}(b), we have
	$\Beta_j^* = \sqrt{n}\theta_*(\Bdelta_j^* +\bxi_j^*)$ and
	$\tilde\Beta_j^* = \sqrt{n}\theta_*{\sum_{k} w_{k,j}(\Bdelta_k^* +\bxi_k^*)}$,
	where $\theta_* := \theta_0\left(1-\theta_0\right)$.
	Then for each $j \in \beta_{p}$,
	\begin{align*}
		&\operatorname{Pr} (W_{j}^*\leq C_\eta\log p )\\ \leq &\Pr\left(n\theta_*^2\sum_{k}w_{k,j}\left({\Bdelta_j^*}\trans\Bdelta_k^*+{\Bdelta_j^*}\trans\bxi_k^*+{\bxi_j^*}\trans\Bdelta_k^*+{\bxi_j^*}\trans\bxi_k^*\right) \leq C_\eta\log p\right)\\
		= & \Pr\left(\sum_{k}w_{k,j}\left({\Bdelta_j^*}\trans\bxi_k^*+{\bxi_j^*}\trans\Bdelta_k^*+{\bxi_j^*}\trans\bxi_k^*\right) \leq \frac{C_\eta\log p}{n\theta_*^2} - \sum_{k}w_{k,j}{\Bdelta_j^*}\trans\Bdelta_k^*\right)\\
		\leq & \Pr\left(\sum_{k}w_{k,j}\left(\|\Bdelta_j^*\|\|\bxi_k^*\|+\|\bxi_j^*\|\|\Bdelta_k^*\|+\|\bxi_j^*\|\|\bxi_k^*\|\right) \geq \sum_{k}w_{k,j}{\Bdelta_j^*}\trans\Bdelta_k^* - \frac{C_\eta\log p}{n\theta_*^2}\right),
	\end{align*}
	where the last inequality is due to Cauchy--Schwarz inequality.
	Define $d_j = \sum_{k}w_{k,j}{\Bdelta_j^*}\trans\Bdelta_k^* - {C_\eta\log p}/{n\theta_*^2}$.
	Then it suffices to show that 
	\begin{equation*}
		\Pr\left(\sum_{k}w_{k,j}\|\Bdelta_j^*\|\|\bxi_k^*\| \geq \frac{d_j}{3}, \, j \in\beta_p\right) = o(1/p),
	\end{equation*}
	and similar results for other two terms regarding  $\|\bxi_j^*\|\|\Bdelta_k^*\|$ and $\|\bxi_j^*\|\|\bxi_k^*\|$ can be easily obtained. Note that $\Pr( \max_{k}\|\bxi_k^*\|\geq C\sqrt{\log p/n})=o(1/p)$ by the proof in  Lemma \ref{lem: bound for eta*}(b), it then suffices to show $\sqrt{n}d_j/\left(\|\Bdelta_j^*\|\sqrt{\log p}\right)\rightarrow \infty$ for any $j \in\beta_p$. To see this,
	\begin{align*}
		\frac{\sqrt{n}d_j}{\|\Bdelta_j^*\|\sqrt{\log p}}
		&\geq \frac{\sqrt{n}\sum_{k}w_{k,j}{\Bdelta_j^*}\trans\left(\Bdelta_j^*+\Bdelta_k^*-\Bdelta_j^*\right)}{\|\Bdelta_j^*\|\sqrt{\log p}} - \frac{C_\eta\sqrt{\log p}}{\theta_*^2\|\Bdelta_j^*\|\sqrt{n}}\\
		&\geq \frac{\sqrt{n}\|\Bdelta_j^*\|}{\sqrt{\log p}} - \frac{\sqrt{n}\max_{k\in\beta_p^h}\|\Bdelta_k^* - \Bdelta_j^*\|}{\sqrt{\log p}} - \frac{C_\eta\sqrt{\log p}}{\theta_*^2\|\Bdelta_j^*\|\sqrt{n}}\to\infty 
	\end{align*}
	as $n,p\to\infty$ by Assumptions \ref{assump: signal} and \ref{assump: kernel},
	where $\beta_p^h = \beta_p\cup\{k:|k-j|\leq h \text{ where } j\in\beta_p\}$. 
	Therefore, we conclude that $\operatorname{Pr}(W_{j}^*\leq C_\eta \log p)=o(1/p)$ for each $j\in\beta_p$, and similarly $\operatorname{Pr}(W_{j}\leq C_\eta \log p)=o(1/p)$ using similar arguments for proving \eqref{uni bound for Wj} and Lemma \ref{lem: bound for est}(c).
	Since $t^* \leq C_\eta \log p$, it then follows that 
	\begin{equation}\label{Wj t*}
		\Pr (W_{j} \leq t^*, \mbox { for some } j \in \beta_{p}) \leq \sum_{j \in\beta_p} \Pr (W_{j} \leq t^*) \rightarrow 0,
	\end{equation}
	and consequently, 
	\begin{equation*}
		\operatorname{Pr}\left(\sum_j \bI\left(W_j>t^*\right) \geq b_{p}\right)\geq\operatorname{Pr}\left(\sum_{j\in\beta_p} \bI\left(W_j>t^*\right) \geq b_{p}\right)\rightarrow 1.
	\end{equation*}
	
\end{proof}

\section{Additional numerical results}
\subsection{additional simulation results}\label{E1}
\begin{figure}
	\centering
	\includegraphics[width=1\linewidth]{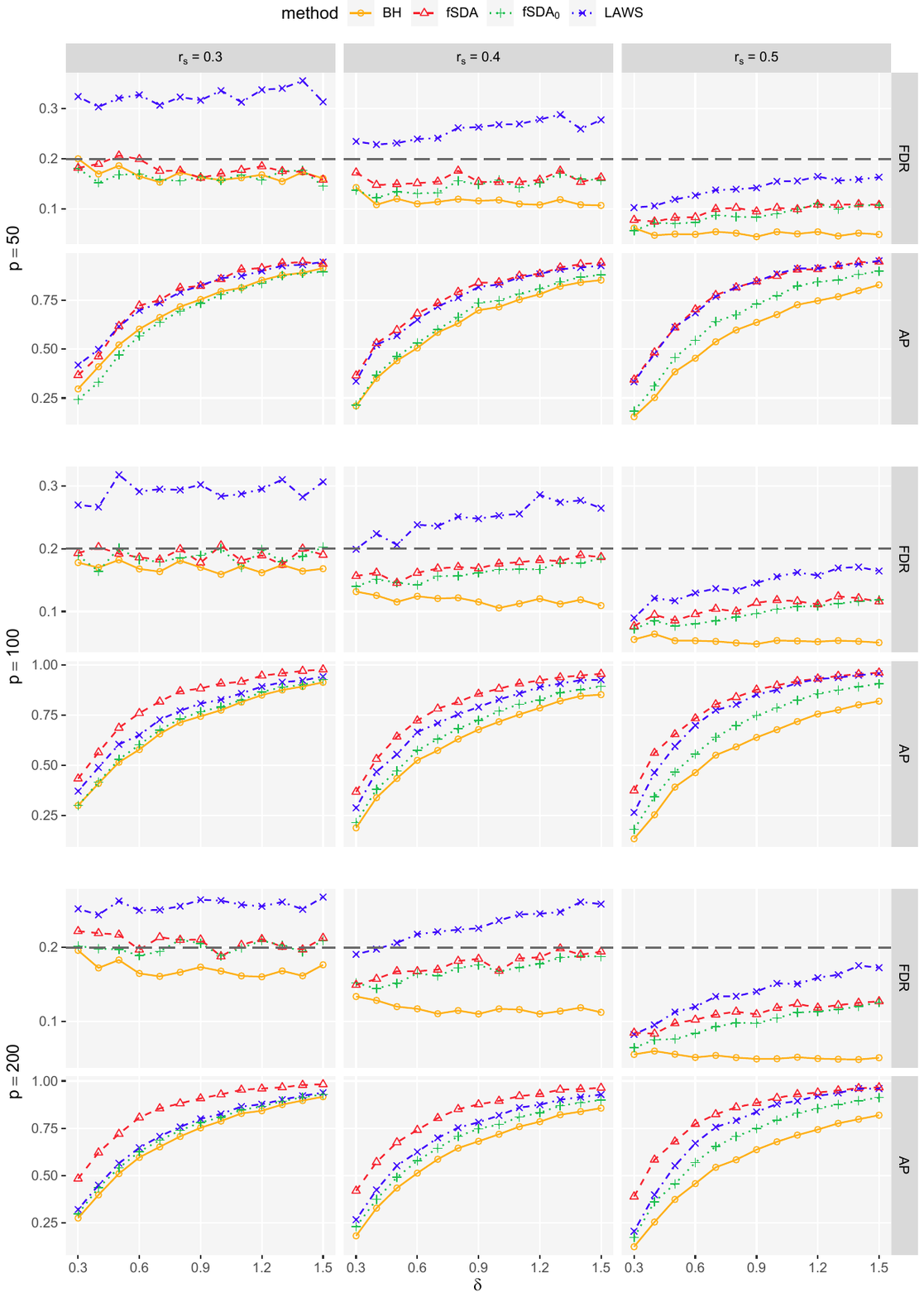}
	\caption{FDR and AP comparison for varying $r_s$ with $n=200$. The dashed line represents the nominal level $\alpha=0.2$.}
	\label{simu: fdr2}
\end{figure}

First, we analyze the simulation results for the case $n=200$. Table \ref{power n200} presents the empirical power results for $n=200$ under various spatial signal scenarios, with the highest power in each scenario highlighted in bold, and Figure \ref{simu: fdr2} reports the multiple testing results for $p=50, 100$, and $200$ with $n=200$. It is evident that, as the sample size increases, both the change-point detection and support recovery procedures exhibit substantially higher power compared to the $n=100$ case. Meanwhile, the results of empirical FDR remain similar to those in the main paper, where our methods still outperform competing approaches. Additionally, Table \ref{tab: consistency1} assesses the performance of the proposed change-point estimation approach in terms of the mean and standard deviation when $r_s=0.6$ and $n=100$. $Q_h$ exhibits the smallest estimation errors across nearly all settings.

\begin{figure}
	\centering
	\includegraphics[width=1\linewidth]{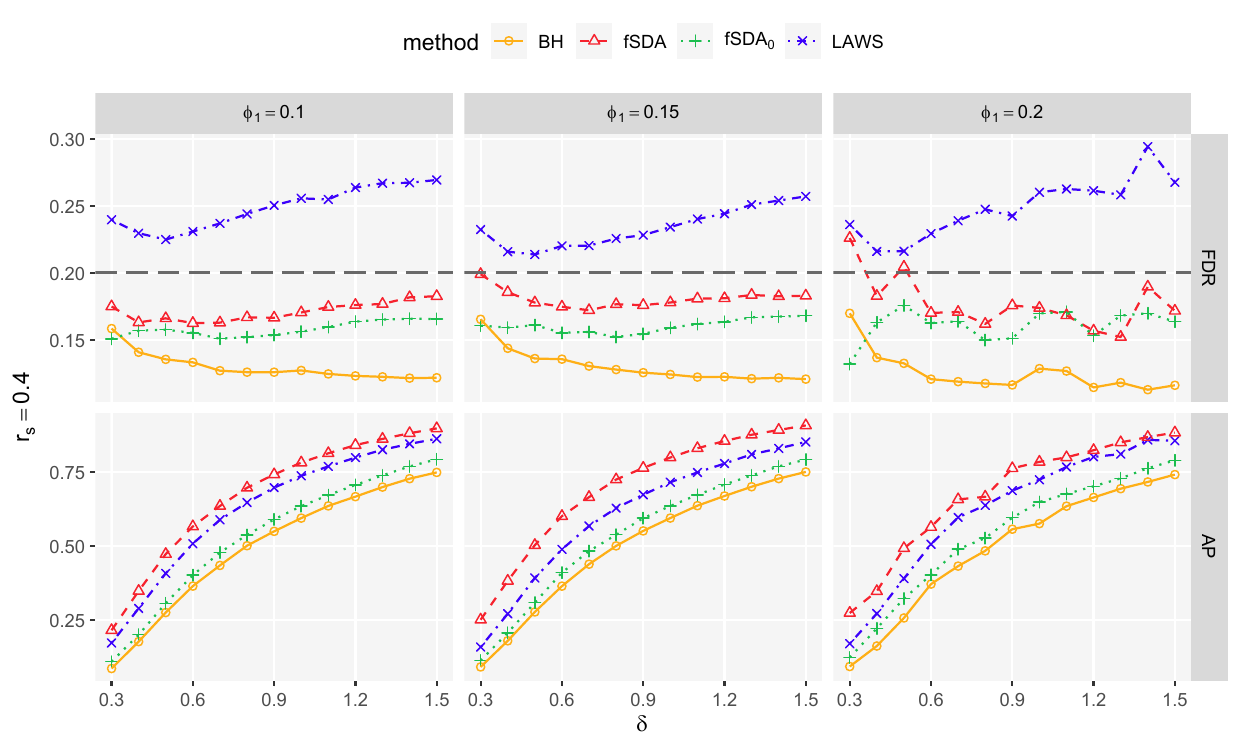}
	\caption{FDR and AP comparison for varying $\phi_1$ with $n=100, p=100$. The dashed line represents the nominal level $\alpha=0.2$.}
	\label{simu: fdr3}
\end{figure}

Next, we investigate the effect of varying spatial covariance on the performance of the change-point detection and support-recovery procedures. By adjusting the range parameter \(\phi_1\), we control the degree of spatial dependence, with larger values of \(\phi_1\) indicating stronger spatial correlation. Table \ref{tab: size phi} reports the empirical size under different levels of spatial dependence. For \(n=100\) and \(p=100\), all methods except \(\widehat{\Lambda}_2\) maintain empirical size close to the nominal level, regardless of the strength of spatial dependence. 
Table \ref{tab: power phi} presents the empirical power results under various scenarios of spatial correlations, with the highest power in each scenario highlighted in bold. Although the power gradually decreases with increasing spatial dependence, our method $Q_h$ maintains superior detection power across all correlation levels.
Figure \ref{simu: fdr3} illustrates the impact of varying spatial correlation on FDR and average power. We observe that fSDA exhibits slightly inflated FDR beyond the nominal level under strong spatial correlation with weak signals, while still demonstrating substantially higher power than competing methods. Furthermore, LAWS fails to maintain proper FDR control, whereas the remaining two methods show relatively conservative performance.

\begin{table}
	\caption{Empirical power (\%) of the change-point tests using various methods with $n=200$.} 
	\label{power n200}
	\centering
	\begin{tabular}{c c c ccc ccc}
		\toprule
		& \multirow{2}{*}{Method} & \multirow{2}{*}{Type} & \multicolumn{3}{c}{$r_s = 0.4$} & \multicolumn{3}{c}{$r_s = 0.6$} \\
		\cmidrule(r){4-6} \cmidrule(l){7-9}
		& & & $\delta = 0.2$ & $\delta = 0.3$ & $\delta = 0.4$ & $\delta = 0.2$ & $\delta = 0.3$ & $\delta = 0.4$ \\
		\midrule
		
		& \multirow{2}{*}{$Q_0$} & sum & 16.5 & 36.5 & 51.5 & 14.5 & 28.0 & 60.0 \\
		
		& & max & 10.5 & 21.5 & 38.5 & 8.0 & 15.5 & 45.0 \\
		
		\multirow{2}{*}{$p=20$} & \multirow{2}{*}{$Q_h$} & sum & \textbf{21.5} & 36.5 & \textbf{55.0} & \textbf{19.0} & \textbf{51.0} & 73.5 \\
		
		& & max & 15.0 & \textbf{37.5} & \textbf{55.0} & 17.5 & 48.0 & \textbf{74.0} \\
		
		& \multirow{2}{*}{Gro} & $\widehat{\Lambda}_1$ & 17.0 & 34.0 & 53.0 & 15.0 & 27.5 & 61.5 \\
		
		& & $\widehat{\Lambda}_2$ & 10.5 & 17.5 & 29.5 & 12.5 & 14.0 & 28.5 \\
		\midrule
		
		& \multirow{2}{*}{$Q_0$} & sum & 27.0 & 42.0 & 76.5 & 24.5 & 47.0 & 79.5 \\
		
		& & max & 14.5 & 24.5 & 51.5 & 13.0 & 26.5 & 63.0 \\
		
		\multirow{2}{*}{$p=50$} & \multirow{2}{*}{$Q_h$} & sum & \textbf{32.5} & \textbf{58.0} & \textbf{81.5} & \textbf{43.5} & \textbf{79.5} & 90.5 \\
		
		& & max & 30.0 & 52.0 & 77.0 & 37.0 & 74.5 & \textbf{91.0} \\
		
		& \multirow{2}{*}{Gro} & $\widehat{\Lambda}_1$ & 27.0 & 46.0 & 76.5 & 28.0 & 48.0 & 80.5 \\
		
		& & $\widehat{\Lambda}_2$ & 20.0 & 25.5 & 36.0 & 16.0 & 27.5 & 42.0 \\
		\midrule
		
		& \multirow{2}{*}{$Q_0$} & sum & 42.5 & 74.0 & 94.0 & 49.5 & 93.0 & \textbf{100} \\
		
		& & max & 42.0 & 57.5 & 72.5 & 42.0 & \textbf{95.5} & \textbf{100} \\
		
		\multirow{2}{*}{$p=100$} & \multirow{2}{*}{$Q_h$} & sum & \textbf{65.5} & 93.0 & \textbf{97.5} & 76.5 & \textbf{95.5} & \textbf{100} \\
		
		& & max & 60.5 & \textbf{93.5} & 97.0 & \textbf{77.0} & 95.0 & \textbf{100} \\
		
		& \multirow{2}{*}{Gro} & $\widehat{\Lambda}_1$ & 46.0 & 72.5 & 92.5 & 56.0 & 92.5 & \textbf{100} \\
		
		& & $\widehat{\Lambda}_2$ & 26.5 & 43.0 & 59.0 & 31.0 & 59.0 & 86.0 \\
		\bottomrule
	\end{tabular}
\end{table}

\begin{table}
	\caption{Simulation study on the consistency of the change-point detection procedures with the mean and standard deviation (in parentheses) of $|\hat{\tau} - \tau^*|$'s under different settings with $r_s = 0.6$ and $n=100$. }
	\label{tab: consistency1}
	\centering
	\begin{tabular}{cccccccc}
		\toprule
		\multirow{2}{*}{Method} & \multicolumn{3}{c}{$\delta = 0.4$} & & \multicolumn{3}{c}{$\delta = 1$}\\
		\cmidrule(r){2-4} \cmidrule(r){6-8}
		& $p=20$ & $p=50$ & $p=100$ & & $p=20$ & $p=50$ & $p=100$\\
		\midrule
		
		$Q_0$ & $3.57_{(3.71)}$ & $\mathbf{2.13}_{(2.56)}$ & $2.02_{(2.44)}$ & & $0.52_{(1.04)}$ & $0.18_{(0.48)}$ & $0.05_{(0.24)}$\\
		$Q_h$ & $\mathbf{3.49_{(3.69)}}$ & $2.17_{(\mathbf{2.55})}$ & $\mathbf{1.74_{(2.27)}}$ & & $\mathbf{0.51_{(1.01)}}$ & $\mathbf{0.17_{(0.46)}}$ & $\mathbf{0.04_{(0.23)}}$\\
		$\widehat{\Lambda}_1$ & $4.25_{(3.91)}$ & $2.68_{(3.05)}$ & $2.62_{(2.84)}$ & & $0.55_{(1.17)}$ & $0.20_{(0.57)}$ & $0.06_{(0.27)}$\\
		$\widehat{\Lambda}_2$ & $5.34_{(5.12)}$ & $4.64_{(4.51)}$ & $4.20_{(4.13)}$ & & $1.41_{(2.37)}$ & $0.69_{(1.18)}$ & $0.27_{(0.59)}$\\
		\bottomrule
	\end{tabular}
\end{table}

\begin{table}
	\caption{Empirical size (\%) of the change-point tests with different $\phi_1$.}
	\label{tab: size phi}
	\centering
	\begin{tabular}{c c ccc}
		\toprule
		\multirow{2}{*}{ Method } & \multirow{2}{*}{Type} & \multicolumn{3}{c}{$n=100, p=100$}\\
		\cmidrule(r){3-5}
		& & $\phi_1=0.1$ & $\phi_1=0.15$ & $\phi_1=0.2$ \\
		\midrule
		\multirow{2}{*}{$Q_0$} & sum & 4.5 & 5.7 & 5.2 \\
		& max & 4.4 & 5.3 & 3.2 \\
		\multirow{2}{*}{$Q_h$} & sum & 5.5 & 5.0 & 4.7 \\
		& max & 6.2 & 6.3 & 5.2 \\
		\multirow{2}{*}{Gro} & $\widehat{\Lambda}_1$ & 5.7 & 6.3 & 4.3 \\
		& $\widehat{\Lambda}_2$ & 9.8 & 9.3 & 9.2 \\
		\bottomrule
	\end{tabular}
\end{table}

\begin{table}
	\caption{Empirical power (\%) of the change-point tests with different $\phi_1$ when $n=100, p=100$.} 
	\label{tab: power phi}
	\centering
	\begin{tabular}{c c c ccc ccc}
		\toprule
		& \multirow{2}{*}{Method} & \multirow{2}{*}{Type} & \multicolumn{3}{c}{$\delta=0.2$} & \multicolumn{3}{c}{$\delta=0.4$} \\
		\cmidrule(r){4-6} \cmidrule(l){7-9}
		& & & $\phi_1 = 0.1$ & $\phi_1 = 0.15$ & $\phi_1 = 0.2$ & $\phi_1 = 0.1$ & $\phi_1 = 0.15$ & $\phi_1 = 0.2$ \\
		\midrule
		
		& \multirow{2}{*}{$Q_0$} & sum & 24.5 & 15.5 & 14.0 & 79.5 & 80.0 & \textbf{70.5} \\
		& & max & 14.0 & 12.0 & 10.5 & 80.0 & 72.5 & 63.0 \\
		\multirow{2}{*}{$r_s = 0.4$} & \multirow{2}{*}{$Q_h$} & sum & \textbf{36.0} & \textbf{22.5} & \textbf{15.5} & 91.5 & \textbf{83.5} & 65.0 \\
		& & max & 30.0 & 16.0 & 9.5 & \textbf{92.0} & 79.5 & 53.0 \\
		& \multirow{2}{*}{Gro} & $\widehat{\Lambda}_1$ & 22.5 & 13.5 & 10.0 & 78.0 & 63.0 & 48.0 \\
		& & $\widehat{\Lambda}_2$ & 19.5 & 15.0 & \textbf{15.5} & 42.0 & 30.5 & 26.0 \\
		\bottomrule
	\end{tabular}
\end{table}

\subsection{additional empirical data results}\label{E2}
To evaluate the space-time separability structure of the considered dataset,
we perform hypothesis testing based on the strong separability test proposed by \cite{aston2017tests} and the weak separability test from  \cite{luotest}. Both tests are appropriate for replicated spatiotemporal functional data, with the latter also applicable for assessing the partial separability of multivariate functional data. To account for potential mean shifts in the precipitation data, the cross-covariance estimation in both tests is adapted by using the estimated mean functions defined in \eqref{mean diff}, for the years before and after the change-point, rather than the typical sample mean function for all years. The test result provided in Table \ref{tab: separable} shows clear evidence for rejecting the hypothesis of strong separability. However, the assumption of weak separability is not rejected across different FVE levels, which provides adequate support for the subsequent functional change-point modeling approach.

\begin{table}
	\caption{The $p$-values of different types of separability tests for China precipitation data. }
	\label{tab: separable}
	\centering
	\begin{tabular}{ccccc}
		\toprule
		& \multicolumn{3}{c}{weak separability} &\multirow{2}{*}{strong separability} \\
		\cmidrule(r){2-4} 
		& FVE=80\% & FVE=90\% & FVE=95\% \\
		\midrule
		& 0.402 & 0.286 & 0.253 & 0.016\\
		\bottomrule
	\end{tabular}
\end{table}

\begin{figure}
	\centering
	\includegraphics[width=0.7\linewidth]{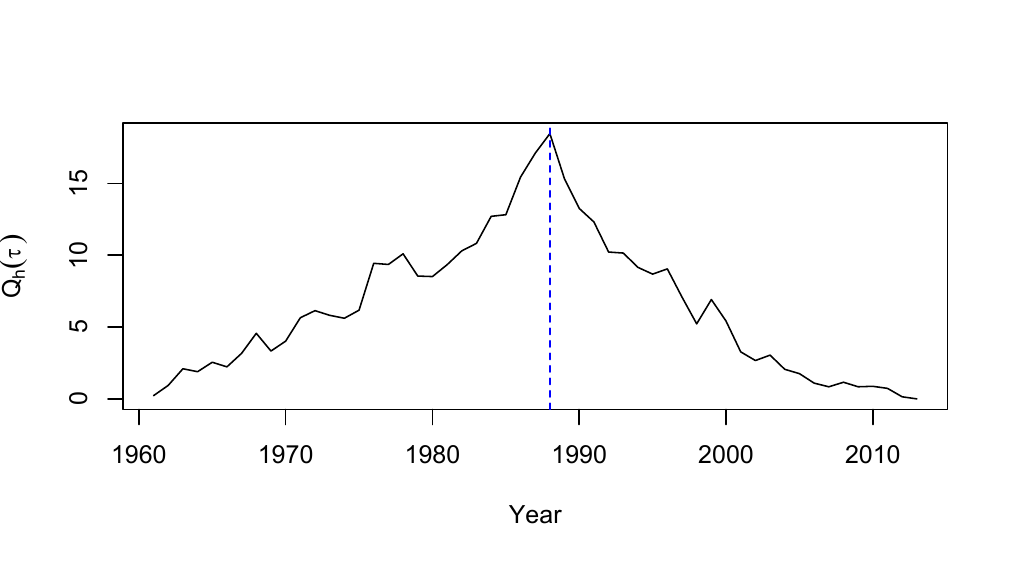}
	\caption{The function $Q_h(\tau)$ against $1961 \le \tau \le 2013$ with the dashed line marking the estimated change-point.}
	\label{fig: real Qh}
\end{figure}

Figure \ref{fig: real Qh} visualizes the function $Q_h(\tau)$ across different $\tau$ values, indicating that the estimated change-year for China's precipitation data is $\hat\tau=1988$.

\end{appendix}

\bibliographystyle{imsart-nameyear}
\bibliography{ref}
\end{document}